\documentclass[aps, showpacs, pra, superscriptdaddress,  eqsecnum]
 {revtex4-1}
\usepackage{hyperref}
\usepackage{amsmath }
\usepackage{amssymb}
\usepackage{epsfig}
\usepackage{natbib}
\usepackage{color}
\usepackage{mathrsfs}
\newcommand*{\be}{\begin{equation}}
\newcommand*{\ee}{\end{equation}}
\newcommand*{\bea}{\begin{align}}
\newcommand*{\eea}{\end{align}}
\newcommand{\PT}{$\mathcal{PT}$}
\newcommand{\C}{{\frak C}}
\newcommand{\A}{{\sf A}}
\newcommand{\B}{{\sf B}}
\newcommand{\F}{{\sf F}}
\newcommand{\G}{{\sf G}}
\newcommand{\iA}{{\mathscr A}}
\newcommand{\iB}{{\mathscr B}}

\newcommand{\sech}{\, \mathrm{sech}}

\DeclareMathAlphabet\mathbfcal{OMS}{cmsy}{b}{n}

\begin{document}

\title[]{Solitons  in  \PT-symmetric ladders of 
 optical waveguides}

\author{N V Alexeeva$^{1,2,3}$,
I  V  Barashenkov$^{3,4,5}$,  Y S Kivshar$^{2,6}$
}
 \affiliation{
 $^1$ Department of Mathematics, University of Cape Town, Rondebosch 7701, South Africa \\
 $^2$ Nonlinear Physics Centre, Australian National University, Canberra ACT 0200,  Australia\\
 $^3$ Department of Physics, University of Bath, Claverton Down, 
Bath BA2 7AY, UK \\
 $^4$ National Institute for Theoretical Physics, Western Cape and 
Centre for Theoretical and Mathematical Physics,
	University of Cape Town, Rondebosch 7701, South Africa \\
	$^5$ Joint Institute for Nuclear Research, Dubna 141980, Russia \\	
$^6$ ITMO University, St. Petersburg 197101, Russia}

\begin{abstract}
We consider a \PT-symmetric ladder-shaped optical array consisting of a chain of  waveguides  with gain coupled to a parallel chain of waveguides with loss.
All waveguides have the focusing Kerr nonlinearity.
The array supports two co-existing solitons, an in-phase and an antiphase one,
and each of these can be centred either on a lattice site or midway between two neighbouring sites.
We show that both bond-centred (i.e. intersite) solitons are  unstable regardless of their amplitudes
and parameters of the chain.  The  site-centred 
{\it in-phase\/}  soliton is 
 stable when its amplitude lies below
 a threshold that depends on the 
 coupling and   gain-loss coefficient.
 The threshold is  lowest when the  gain-to-gain and loss-to-loss coupling constant in each chain   is close to the interchain gain-to-loss coupling coefficient.
 The {\it antiphase\/} soliton in the strongly-coupled chain or in a chain close to the  \PT-symmetry breaking point, is stable  when its amplitude lies
 above a critical value and unstable otherwise.  
 The instability growth  rate of solitons with small amplitude is exponentially small in this parameter regime; hence the small-amplitude solitons, though unstable, 
 have exponentially long lifetimes. On the other hand, 
the  antiphase  soliton in the weakly or moderately coupled chain and away from 
 the \PT-symmetry breaking point, is unstable  when its amplitude falls in one or two finite bands.
 All amplitudes outside those bands are stable. 

\end{abstract}

\maketitle

\section{Introduction}

In soliton-bearing optical systems, 
 weak dissipative losses are typically compensated by the application of a
resonant pump.  
 The damping and driving terms break  the  scaling-,
 phase- and Galilean invariances of the underlying nonlinear Schr\"odinger equation.
 As a result, 
the amplitude, phase, and velocity  of the damped-driven solitons are 
found to be fixed by the driver's strength and dissipation coefficient \cite{DD}.

Dissipative solitons in 
systems with competing linear and nonlinear gain and loss --- systems modelled by the 
Ginsburg-Landau equations --- are equally inflexible.
 The balance of nonlinearity and diffraction or dispersion 
singles out a one-parameter family of solitons, while the competition of gain and loss 
fixes a particular member of the family \cite{CGL}.
This inflexibility can be a disadvantage in applications where one would like the 
parameters of the soliton to be determined just by the initial conditions.

\begin{figure}[h]
 \includegraphics*[width=0.4 \linewidth]{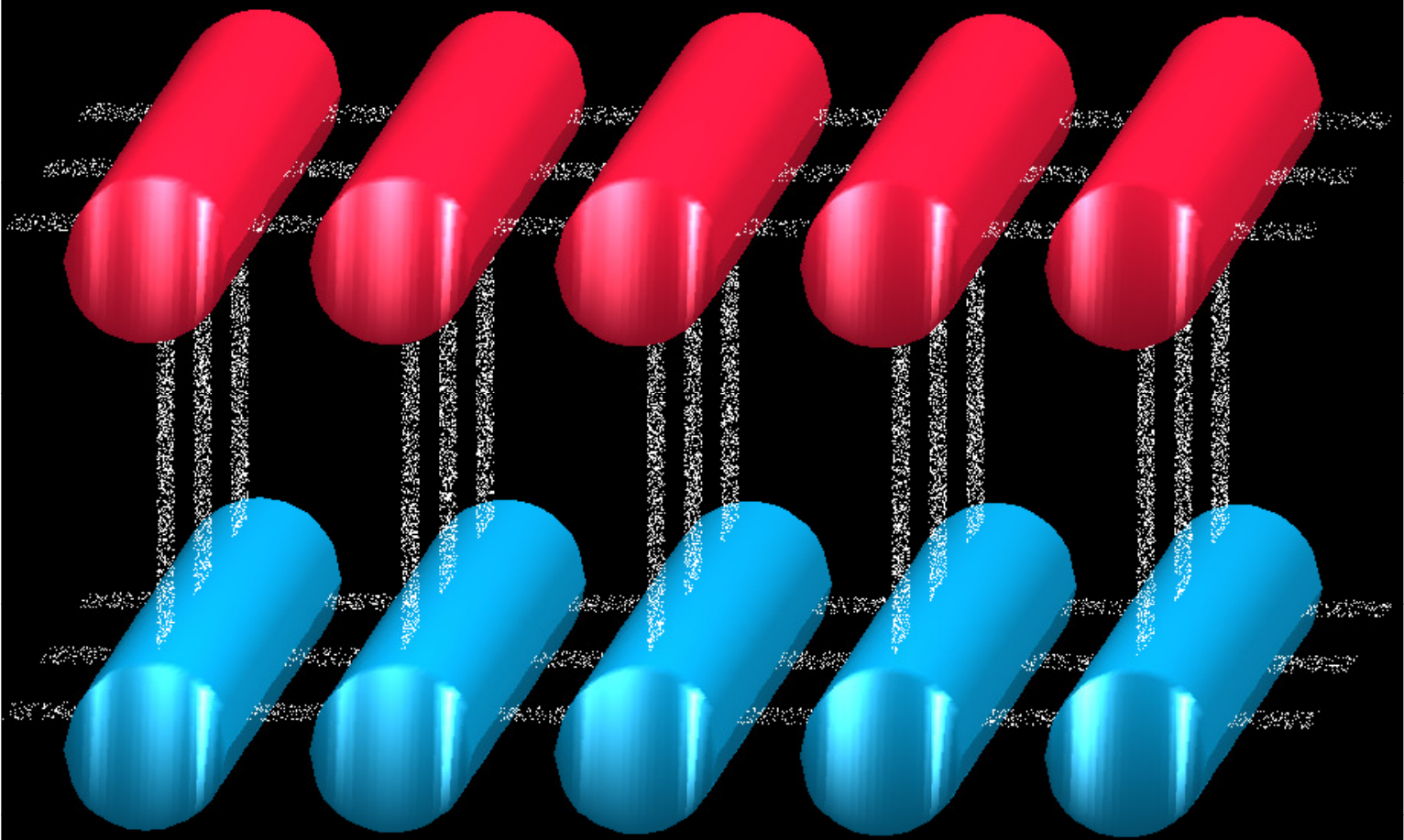}
  \caption{A \PT-symmetric ladder of coupled optical waveguides with gain (red) and loss (blue). 
  Each guide is coupled to its left and right nearest neighbours in the horizontal plane. 
  We are occasionally referring to this coupling as ``horizontal" in the text.
  In addition, each guide with gain is coupled to its nearest guide with loss
  --- this will be referred to as the ``vertical coupling".
  }
 \label{ladder}
 \end{figure}

 Yet there are  ways of supplying energy in a soliton-bearing system with 
 loss, free from the above drawback. 
The pertinent class of approaches 
  exploits the concept of the
 parity-time (\PT-)  symmetry, originally developed in the context of quantum mechanics \cite{Bender}.
One  \PT-symmetric recipe consists in the 
 manufacturing of two identical replicas of the same system and
 arranging for the energy supply in one copy and energy drain in the other one.
 With a judicious choice of coupling,
 the combined  system with gain and loss will contain, as its scalar reduction,
  the unperturbed 
 nonlinear Schr\"odinger equation. 
As a result, this \PT-symmetric system will support families of solitons with continuously variable amplitudes, phases, and velocities.

In fact the parametric flexibility is but one of the whole range of unusual properties 
 of  optical 
 $\mathcal{PT}$-symmetric structures. Other 
 behaviours afforded by the symmetric application of gain and loss and
  unattainable with standard arrangements, include
 the
unconventional beam refraction \cite{Musslimani,Zheng,Regensburger}, Bragg scattering \cite{Berry_Longhi},
nonreciprocal light propagation \cite{Nonreciprocal_propagation,WGM,Shramkova} and  Bloch oscillations \cite{Longhi}, 
  loss-induced  transparency \cite{Guo},
  single-mode lasing \cite{single-mode_laser}, coherent perfect absorption of light \cite{Chong},
  loss-induced onset of lasing \cite{loss_lase}
   and
conical diffraction \cite{Ramezani}. 
The \PT-symmetric systems are expected to promote an
   efficient control of light, including
all-optical low-threshold  \cite{RKEC,SXK,Lin} and asymmetric \cite{EPswitch}  switching
and unidirectional
invisibility \cite{RKEC,Feng,Regensburger,Sanchez}.
There is also  a rapidly growing interest 
in the context of plasmonics \cite{plasmonics}, quantum optics of atomic gases \cite{Lambda},  optomechanical systems \cite{OM,Kepesides}
and metamaterials \cite{Lazarides,Shramkova}.

The early experimental realisations of the optical $\mathcal{PT}$ symmetry  were  in a directional coupler 
consisting of two waveguides with gain and loss \cite{Guo,Rueter,Kottos}
and in a pair of coupled  whispering-gallery mode resonators \cite{WGM}. 
The  work that followed focussed on {\it chains\/} of $\mathcal{PT}$-symmetric couplers 
\cite{Regensburger,Regensburger2,Miri1,Christo_2016}.
The corresponding theoretical studies progressed from
a single  Schr\"odinger dimer \cite{RKEC,SXK,Miri2,standimer} 
and oligomer \cite{oligomer,Liam,Turitsyn,Izrailev,Kepesides},
 to
  $\mathcal{PT}$-symmetric dimer arrays 
\cite{binary,SMDK,KPZ,staggered,Chernyavsky,Miri3}. 
The subsequent analyses 
 included the effects of  diffraction of
spatial beams and dispersion of  temporal pulses,  that is, included an additional spatial or temporal dimension
\cite{SMDK,Driben1,Driben2,ABSK,BSSDK,Driben3}. $\mathcal{PT}$-symmetric necklaces of
dispersive waveguides
(dispersive 
$\mathcal{PT}$-oligomers)  were examined in \cite{Liam}. 
For comprehensive reviews, see \cite{reviews}.

In this paper, we consider an infinite chain of $\mathcal{PT}$-symmetric dimers 
coupled
to their nearest neighbours.
The coupling of the neighboring dimers is arranged  in such a way that,
apart from its ``internal" counterpart with gain,  
 each waveguide with loss 
interacts only with its two lossy neighbours
(Fig \ref{ladder}). 
Likewise,  each  waveguide with gain is 
coupled only to its two neighbours with gain --- besides its counterpart with loss with who they make up 
the \PT-symmetric entity
\cite{SMDK}:
\begin{align}
 i {\dot u_n}+ \C \Delta u_n + 2 |u_n|^2 u_n+v_n & = i \gamma u_n, \nonumber \\
 i {\dot v_n} + \C \Delta v_n + 2|v_n|^2v_n + u_n & = - i \gamma v_n.
\label{B1}
\end{align}
Here $\Delta$ is the second difference operator acting on bi-infinite sequences $..., u_{-2}, u_{-1}, u_0, u_{1}, u_{2}, ...$ according to  the rule
\[
\Delta u_n = u_{n+1}-2u_n + u_{n-1} \quad (n= 0,\pm 1, \pm 2, ....).
\]
In the basis of
 infinite-component vectors 
 \[
 {\bf u}= (..., u_{-2}, u_{-1},   u_0,   u_1,  u_2, ...)^T,  \]
  the operator $\Delta$ is represented by a tridiagonal matrix.

 The $u_n$ is the nondimensional complex mode amplitude in the $n$th waveguide with 
 gain, and $v_n$ is the amplitude in the $n$th waveguide with loss.
 The overdot indicates the derivative with respect to $z$, the propagation coordinate.
 The sign of the cubic terms corresponds to the self-focusing Kerr nonlinearity; the 
 coefficients in front of these terms have been set to 2 by scaling the mode
 amplitudes.
Finally, $\C>0$ is a dimer-to-dimer coupling constant
(the ``horizontal" coupling in the language of Fig \ref{ladder}), and 
$\gamma \geq 0$ is the gain-loss coefficient.

The ladder-shaped chain \eqref{B1} supports four families of  soliton excitations \cite{SMDK}.
These are distinguishable by the  phase difference between $u_n$ and $v_n$, and
by the centring of the soliton relative to lattice sites.
In the continuum limit, their stability has been classified in \cite{Driben1,ABSK}.
Ref \cite{Susanto} considered the anticontinuum limit, under an additional assumption of $\gamma \to 0$.
In the present paper we study the discrete solitons in the entire range of their 
amplitudes, for all values of the coupling $\C$ and all physically admissible values of $\gamma$.

We will demonstrate that the soliton stability problem can be decomposed into an eigenvalue
problem for perturbations belonging to the same symmetric manifold as 
the stationary soliton  itself, and perturbations that take the evolution
out of the symmetry. 
This decomposition alone is sufficient to classify the stability of the bond-centred solitons
--- which  happen to be all unstable against a perturbation in the symmetric manifold.
In contrast, the instability of the site-centred solitons can only be nucleated by 
nonsymmetric perturbations. In the latter case, the advantage of our decomposition is  that it reduces
the underlying two-component eigenvalue problem to a much simpler one-component problem.

We will show that stability properties of the in-phase and antiphase site-centred soliton are completely
determined just by two combinations of $\C$, $\gamma$ and the soliton's amplitude.
The stability domain of the in-phase soliton admits a simple characterisation in terms of a 
single function relating these two self-similar combinations. As a result, for each   $\C$
and $\gamma$ we will be able to determine a critical value of the amplitude above which the
soliton becomes unstable.

Stability properties of the  antiphase  soliton
in  a  strongly-coupled chain or in a weakly or moderately coupled chain close to its  \PT-symmetry breaking point, follow an 
opposite pattern. Here, the soliton instability sets in {\it below\/} a critical amplitude. 
Reducing the inter-dimer coupling or turning down  the gain-loss coefficient of the chain, changes the topography of this soliton's parameter space. 
Namely, the weakly or moderately coupled array away from the \PT-symmetry breaking threshold
displays one or two bands of unstable amplitudes, with all antiphase solitons outside those bands being stable.

The outline of the paper is as follows. Section \ref{CLS} summarises stability properties of two
solitons arising in the continuum limit of the array \eqref{B1}.  In section  \ref{DS}, we introduce four 
discrete solitons: the in-phase and  antiphase site-centred soliton,
 and the in- and out-of-phase bond-centred one. 
 The subsequent section \ref{SF} lays out the general framework of our stability analysis
 and draws some general conclusions.
 
 In section \ref{HFS} we classify the stability of the in-phase solitons while
 section \ref{LFS} focusses on the antiphase variety. 
 Some technical results on eigenvalues  playing the central role in our analysis have been
relegated to eight appendices.
 Finally,  section \ref{Conc} summarises conclusions of this study
 and contrasts them with the earlier results of \cite{SMDK} and \cite{Susanto}.

\section{Continuum limit: summary}
\label{CLS}

If the characteristic wavelength in the chain is much greater than $\C^{-1/2}$, 
 the \PT-symmetric lattice \eqref{B1} reduces to  a system of two  partial
differential equations:
\begin{align}
i {\dot u}+  u_{xx} + 2 |u|^2 u +v  & = i \gamma u, 
\nonumber \\
i {\dot v} +  v_{xx}  + 2|v|^2v + u & = - i \gamma v.
\label{Z1}
\end{align}
This  \PT-symmetric
system of coupled nonlinear Schr\"odinger equations was considered in \cite{SMDK,Driben1,ABSK,BSSDK}. 
Results  of the previous analyses can be summarised as follows.

(1)
For each $\gamma <1$, there are two continuous families of coexisting solitons.
Each family is characterised by its own value of
 the phase difference between the $u$- and $v$-component. 
Representatives of the two families with equal amplitudes have unequal propagation constants  \cite{SMDK}.

(2)
The soliton stability and internal dynamics are  determined 
by a single self-similar combination $\eta= 2 \sqrt{1-\gamma^2}/A^2$ of the 
gain-loss coefficient $\gamma$ and  the soliton's amplitude,
 $A$ \cite{ABSK}.

(3)
The in-phase solitons with amplitudes smaller than $\mathcal A_0= \sqrt{\frac23} (1-\gamma^2)^{1/4}$ are stable,
and with amplitudes greater than $\mathcal A_0$, unstable
\cite{Driben1,ABSK}.  All antiphase solitons are unstable; however the lifetimes of 
the antiphase solitons with small amplitudes are exponentially long
\cite{ABSK,Dima_Deconinck2}.

 In this paper, we will extend these considerations to localised
 solutions of  the discrete system \eqref{B1}.
 
\section{Discrete solitons}
\label{DS}

\begin{figure}
      \includegraphics*[width=0.4\linewidth]{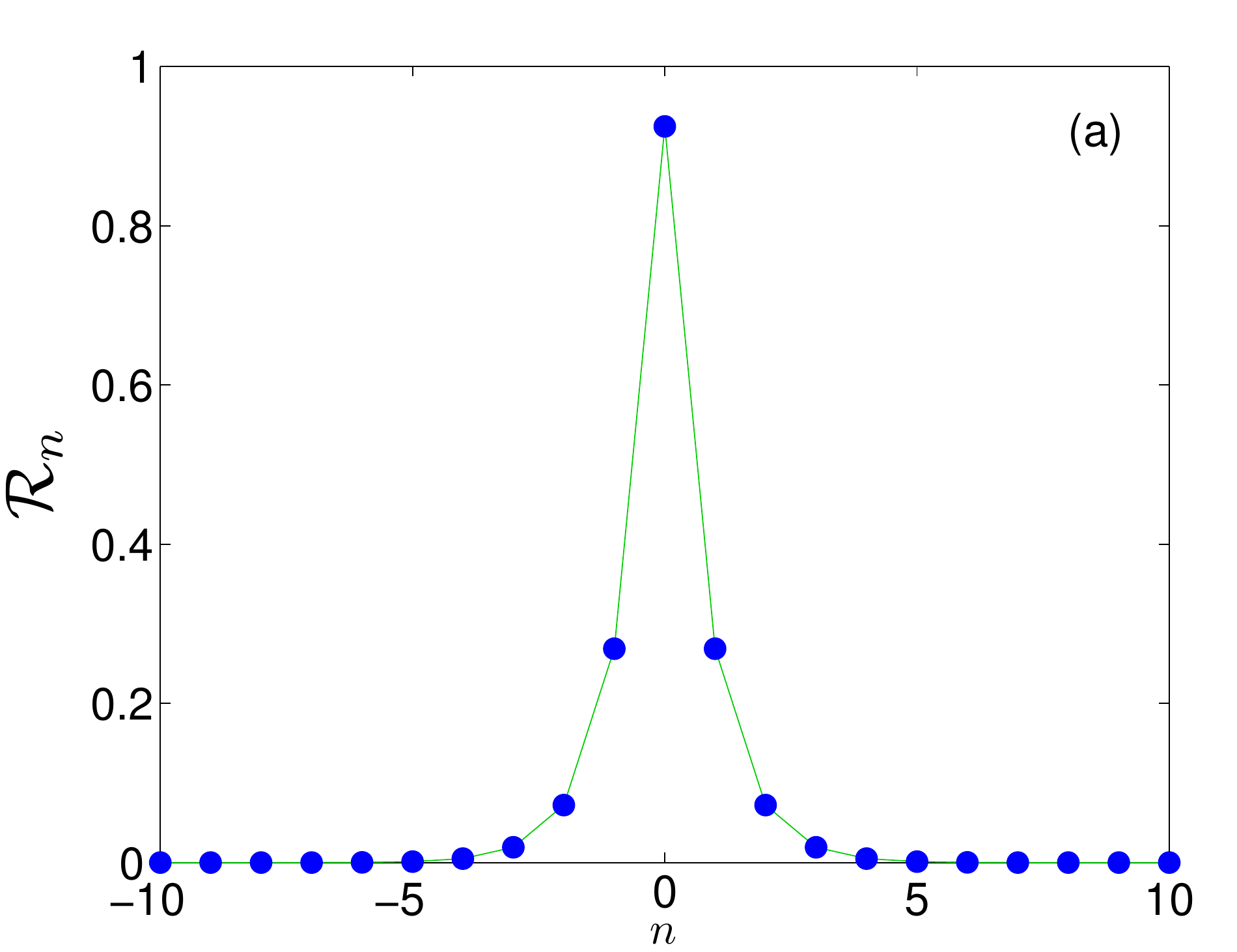}    
 \hspace*{5mm}  \includegraphics*[width=0.4\linewidth]{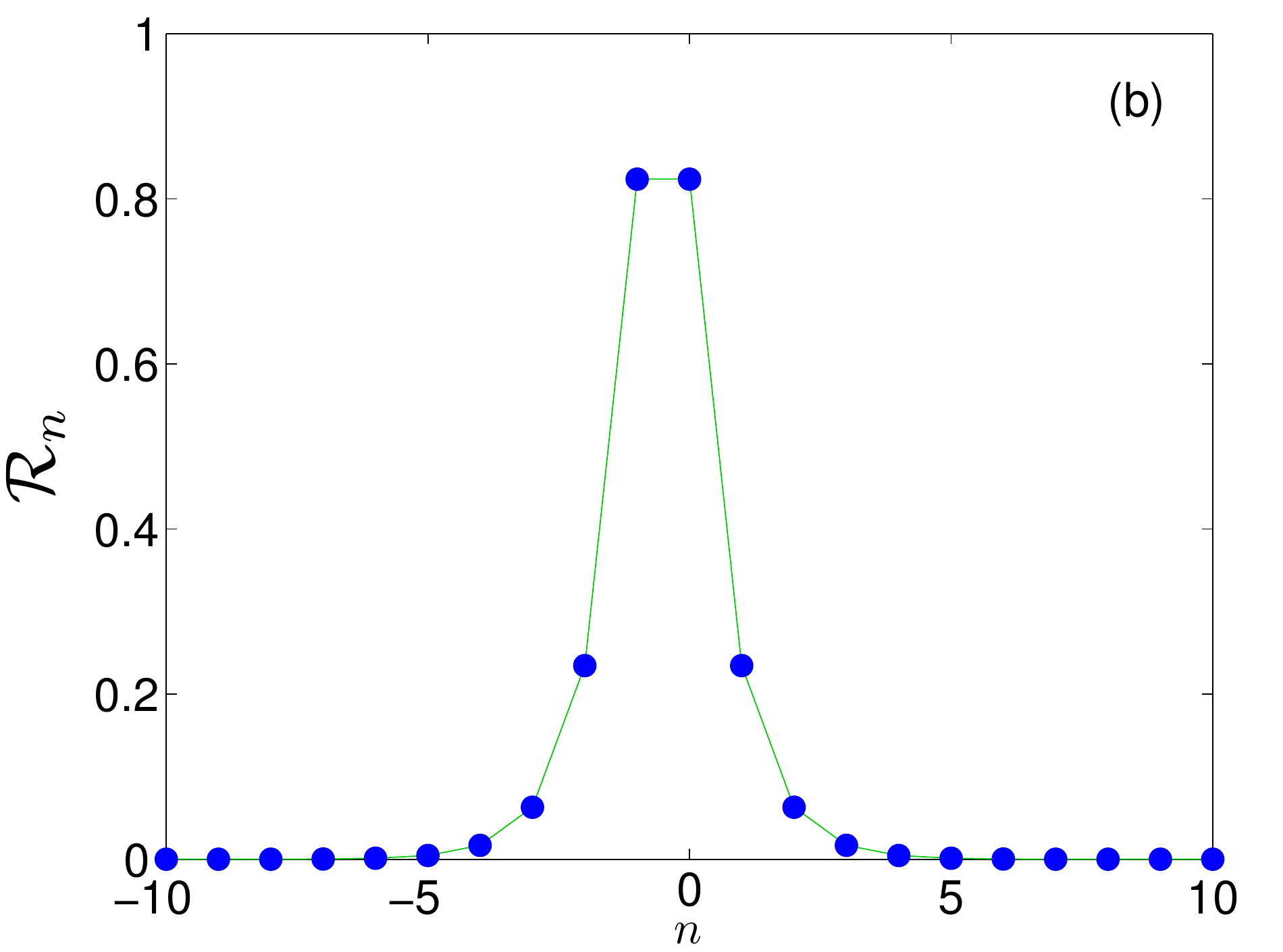}                 
                        \caption{The two single-hump solutions  of equation \eqref{B3} with the same $h$.
                        (In these plots, $h= \sqrt{2}$.)  The 
                         left panel shows the form of  $|u_n|$ and $ |v_n|$ for the
                        site-centred soliton (both for the in-phase and antiphase one).
                         The right panel illustrates the magnitude profile of the bond-centred soliton --- again, for the in- and out-of-phase one alike.
                         }
 \label{R_n}
 \end{figure}

The dispersion relations for the
small-amplitude harmonic waves in the system \eqref{B1} are
\[
\omega = \pm \omega_0 -4\C \sin^2 \frac{k}{2}, \quad \omega_0=\sqrt{1-\gamma^2}.
\]
In what follows, we assume that the trivial solution, $u_n=v_n=0$, is stable: 
$0 \leq \gamma \leq 1$.

For future convenience, we perform the change of variables
\be
u_n= e^{i \theta} a_n + b_n, 
\quad
v_n = e^{i \theta} b_n -a_n,
\label{A40}
\ee
where 
$\theta = \arcsin \gamma$.
This transformation diagonalises the linear part of 
equations \eqref{B1} \cite{BSSDK,Liam}:
\begin{align}
i {\dot a}_n+ \C \Delta a_n -\omega_0 a_n +
\frac{
|u_n|^2u_n -e^{-i \theta}
 |v_n|^2v_n}{ \omega_0} =0,
\nonumber
\\
i{\dot b}_n+ \C \Delta b_n + \omega_0  b_n +
\frac{e^{-i \theta} |u_n|^2u_n+ |v_n|^2 v_n}{\omega_0} =0,
\label{A6}
\end{align}
where  $u_n$ and $v_n$ denote the linear combinations \eqref{A40}.

The system \eqref{B1} has two invariant 
manifolds where the evolution is conservative \cite{ABSK}.
Indeed,  the equations \eqref{A6} admit a reduction  $a_n=0$, $b_n=\phi_n$
to the scalar
 discrete nonlinear Schr\"odinger equation
 \be
 i {\dot \phi_n} + \C \Delta \phi_n + \omega_0  \phi_n + 2|\phi_n|^2 \phi_n=0.
 \label{B2}
\ee
(Note that under this reduction, $u_n=\phi_n$ and $v_n= e^{i \theta} \phi_n$.)
A different, independent, scalar reduction arises by letting 
$b_n=0$, $a_n=\phi_n$. The equations \eqref{A6} become
\be
 i {\dot \phi_n} + \C \Delta \phi_n - \omega_0  \phi_n + 2|\phi_n|^2 \phi_n=0.
 \label{B200}
\ee
(This time, $u_n= e^{i \theta} \phi_n$ and $v_n= -\phi_n$.)

 In this paper, we are focussing on localised solutions
 of \eqref{B2} and \eqref{B200}: $ |\mathcal \phi_n| \to 0$ as $|n| \to \infty$. 
 The simplest solutions of this sort are separable; these  are given by 
 \[
 \phi_n= e^{i \beta z} A \mathcal{R}_n.
 \]
Here $A>0$ is an arbitrary amplitude and
$\mathcal{R}_n$ solves
\be
\frac{1}{h^2}    \Delta \mathcal{R}_n  - \mathcal{R}_n+ 2 {\mathcal R}_n^3=0,
\label{B3}
\ee
where $h$ is the amplitude-to-coupling ratio:
\be
h= \frac{A}{\sqrt \C}. 
\label{h}
\ee  
In what follows, we are referring to the corresponding solutions of \eqref{A6} as solitons.

Equation \eqref{B3} determines the profile of the localised solution 
and $h$ has the meaning of discretisation stepsize in this equation.
 In the case of the reduction \eqref{B2},
 the propagation constant $\beta$ is given by
 \be
  \beta= A^2 + \omega_0,
  \label{OmegaP}
 \ee
 while the equation \eqref{B200} requires
\be
  \beta= A^2 - \omega_0.
  \label{OmegaM}
 \ee

\begin{figure}[t]
 \includegraphics*[width=0.6  \linewidth]{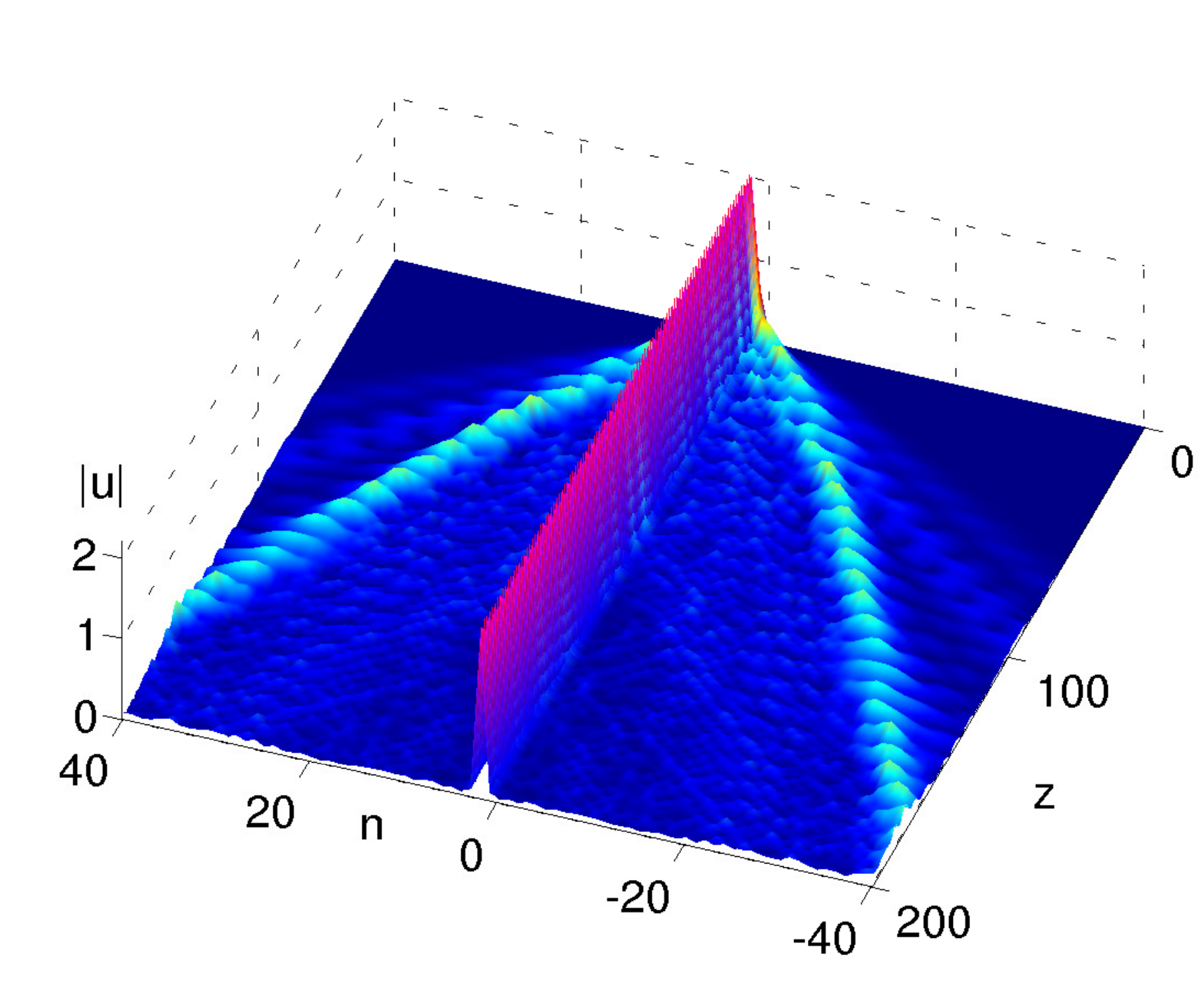}
  \caption{
  The formation of a  soliton  out of a gaussian initial condition $u_n(0)=   e^{i (\theta+\chi)} A_{\rm gs}   e^{-n^2}$, $v_n(0)=-A_{\rm gs}  e^{-n^2}$.
  Here $\chi$ is a phase mismatch  taking the initial condition out of the scalar reduction \eqref{B200}. 
  In this plot the parameters of the lattice are $\C=0.3$ and $\gamma=0.93$; the amplitude of the gaussian $A_{\rm gs}=2$
  and the mismatch $\chi=0.3$.  Note small-amplitude oscillations of the soliton's amplitude --- the non-oscillatory soliton would result from the gaussian with $\chi=0$.
  }
 \label{Gauss}
 \end{figure}

 We note that equations \eqref{B2}-\eqref{B200}
admit no localised separable solutions with {\it complex\/} $\mathcal R_n$ 
(except for the trivial situation where all $\mathcal R_n$ share the same, $n$-independent, phase factor) \cite{HT}.
Therefore, equation \eqref{B3} is the most general equation for localised profiles.

 Each   localised  
 solution of \eqref{B3} gives rise to two families of separable solutions  of the original system \eqref{B1},
  with $\C= (A/h)^2$.
  The first family corresponds to the $a_n=0$ reduction of Eq.\eqref{A6}.
  When the gain-loss coefficient $\gamma \to 0$, 
  the $u_n$ and $v_n$ components of any solution in this family become equal: $v_n  \to u_n$.
  In contrast, the components of any separable solution corresponding to the $b_n=0$ reduction 
  become equal in magnitude but opposite in sign: $v_n  \to -u_n$. 
  Because of this simple and noticeable difference, 
   the localised separable solutions with  $a_n=0$ will
 be called the {\it in-phase\/} solitons, and those with $b_n=0$ 
 the {\it  antiphase\/} ones.
(By continuity,  we will use this terminology  even in the  $\gamma \to 1$
limit where the difference between the two families becomes negligible.)

   Note that
 the   soliton
 solution of Eq.\eqref{A6} with $a_n=0$ and amplitude $A$, has a larger propagation constant than 
 the $(b_n=0)$-soliton  of the same amplitude.
 For this reason,  the in-phase separable solution was 
  referred to as the {\it high-frequency\/} soliton in \cite{ABSK}, and the one with $b_n=0$ 
as the {\it  low-frequency\/} one.

Equation \eqref{B3} has a variety of localised solutions, 
in particular two fundamental, i.e., single-hump,  nonlinear modes.
One of these is centred on the site with $n=0$ (the so-called Sievers-Takeno mode)  and the other one 
midway between the sites with $n=0$ and $n=1$ (the Page mode)  \cite{KC,LST,KK,PKF,Herrmann}. We will be referring to the former
 as
the `site-centred' mode,  and the latter as `bond-centred'. 
Both solutions exist for all $0 < h < \infty$ \cite{ABK}; both
 satisfy $\mathcal{R}_n>0$ for all $n$ \cite{Herrmann}. 
 We depict these  solutions in Fig. \ref{R_n}.

Thus there are four  fundamental  solitons in the dimer array \eqref{B1}: the in-phase and antiphase
site-centred soliton, and the in-phase and antiphase
bond-centred one. 
The distribution of the waveguide mode  magnitudes in the site- and bond-centred soliton is shown in Fig \ref{R_n}. 
The phases of $u_n$ and $v_n$ remain constant  as $n$  grows
or decreases;
hence there is no flow of energy along the array. 
On the other hand, the phase difference between  $u_n$ and $v_n$ reflects a nonzero 
 energy flux from the gaining to the losing waveguide of each dimer.

Small $h$ correspond to  the continuum limit
(small soliton amplitudes or strongly coupled sites). 
As $h \to 0$, the site-centred and bond-centred solutions satisfy
$\mathcal{R}_n \to \sech   (hn)$ and 
$\mathcal{R}_n \to \sech [h(n-~\frac12)]$,
respectively. Large $h$ pertain to the anti-continuum limit
 (large amplitudes or weak dimer-to dimer coupling).

Before proceeding to stability analysis,  we note that discrete solitons 
and their breather-like perturbations emerge from a broad class of  initial
conditions.
Fig \ref{Gauss} shows the formation of a soliton from a gaussian with generic amplitude and phase.

\section{Stability framework} 
\label{SF}
\subsection{The in-phase soliton}

Consider the in-phase soliton first. Letting
\begin{align*} 
a_n =e^{i \beta z}  \left[ \mathrm{Re} (p^{(1)}_n e^{\mu z})+ i \, \mathrm{Re} (p_n^{(2)}e^{\mu z})\right],
 \\
b_n=  e^{i \beta z}  \left[ A \mathcal{R}_n+
 \mathrm{Re} (q^{(1)}_n e^{\mu z})+ i \, \mathrm{Re} (q_n^{(2)}e^{\mu z})\right],
\end{align*}
where $\mathcal{R}_n$ is a solution of \eqref{B3}, $\beta$ is as in \eqref{OmegaP},
and $p_n^{(1,2)}$, $q_n^{(1,2)}$, and $\mu$ are complex,
we linearise equations \eqref{A6} in $p_n^{(1,2)}$ and $q_n^{(1,2)}$.
This gives an eigenvalue problem
\begin{subequations}  
\label{B7}
\begin{align}
\mathcal{L}{\vec p}_n + \omega_0  {\vec p}_n = \mu J {\vec p}_n,  \label{A70}
\\
\mathcal{L}  {\vec q}_n - \omega_0  {\vec q}_n  + 4 \gamma A^2 \mathcal{R}_n^2 \sigma_1 {\vec p}_n  = \mu J {\vec q}_n,  \label{A80}
\end{align}
\end{subequations} 
where
the operator
\[
\mathcal{L}= \left( \begin{array}{cc}
-\C  \Delta  + \beta -6A^2 \mathcal{R}_n^2 & 0 \\ 0 & -\C \Delta + \beta -2A^2 \mathcal{R}_n^2
\end{array} \right)
\]
acts on  two-component complex sequences
\[
{\vec {p}}_n= \left( \begin{array}{c} p_n^{(1)} \\ p_n^{(2)} \end{array} \right), \quad
{\vec {q}}_n= \left( \begin{array}{c} q_n^{(1)}  \\ q_n^{(2)} \end{array} \right).
\]
The $J$ in the right-hand sides of \eqref{A70}-\eqref{A80} denotes a unimodular skew-symmetric matrix:
\[
J= \left( \begin{array}{lr} 0 & -1 \\
1 & 0 \end{array} \right),
\]
and $\sigma_1$ in \eqref{A80} is the conventional Pauli matrix.

The purpose of the stability analysis is to determine whether 
 the  eigenvalue problem \eqref{B7} has eigenvalues  $\mu$ with nonzero real part.
 If it has, the soliton is unstable; if it has not, the soliton is classified as linearly stable.

The continuous spectrum of $\mu$'s occupies two opposite pairs of bands on the 
imaginary axis:  $\mu= \pm i \omega_{1,2}$, where
\begin{align}
 \beta+ \omega_0 \leq \omega_1 \leq \beta+ \omega_0 +4\C,  \label{Y110} \\
 \beta- \omega_0 \leq \omega_2  \leq \beta- \omega_0 +4\C. \label{Y111}
\end{align}
Outside the domain of intersection of the bands \eqref{Y110} and \eqref{Y111},  the problem \eqref{B7} may have discrete eigenvalues. 

To find the discrete eigenvalues, one does  not have to solve 
the full system \eqref{B7}. Indeed, one class of eigenvectors
is selected simply by  letting ${\vec p}_n=0$; this choice selects  perturbations belonging to the
scalar   Schr\"odinger  equation \eqref{B2}. The associated ${\vec q}_n$ satisfies
\be
\mathcal{L}  {\vec q}_n - \omega_0  {\vec q}_n   = \mu J {\vec q}_n,  \label{A85}
\ee
 the linearised eigenvalue problem for the scalar  discrete nonlinear Schr\"odinger equation.

It is well  known that for the {\it bond}-centred soliton, the operator $J^{-1} (\mathcal{L}-\omega_0 I)$ in 
 \eqref{A85} has positive  eigenvalues \cite{LST,KK,PKF}
--- and so
the bond-centred soliton of the scalar nonlinear Schr\"odinger equation \eqref{B2} is unstable.
 This observation implies
that the   bond-centred in-phase soliton of the vector equation \eqref{B1} is unstable for all $\C$, $\gamma$, and $A$.

On the other hand, the {\it site}-centred soliton of the scalar Schr\"odinger equation \eqref{B2} is stable for all $A$ and $\C$ \cite{LST,KK,PKF}.
The operator   $J^{-1} (\mathcal{L}-\omega_0 I)$  in this case has only pure imaginary eigenvalues:
two zeros,  two pairs of opposite nonzero imaginary eigenvalues and two bands of continuous spectrum
on the imaginary axis:
\be
\mu= \pm i \omega_2, 
\quad 
A^2  \leq \omega_2  \leq A^2+ 4\C.
\label{cs}
\ee
This does not yet guarantee that 
 the corresponding soliton of the vector equations \eqref{B1} is stable;
the stability of the latter will be determined by  eigenvectors of \eqref{B7} with ${\vec p}_n \neq 0$.
These correspond to linear evolutions that do not satisfy the scalar reduction \eqref{B2}. 

The nontrivial sequences ${\vec p}_n \neq 0$ can be found as eigenvectors  of  the eigenvalue problem  \eqref{A70}. 
This is a significant simplification as the dimension of the matrix $ \mathcal{L}+ \omega_0 I$ in 
\eqref{A70} is half the dimension of the full triangular matrix 
\[
\left(
\begin{array}{cc}
 \mathcal{L}+ \omega_0 I  & 0 \\
 4 \gamma A^2 \mathcal{R}_n^2 \sigma_1 &  \mathcal{L}- \omega_0 I
\end{array} 
\right)
\]
in 
\eqref{B7}.
Once  the ${\vec p}_n$ and the associated eigenvalue $\mu$  have been found,  
Eq.\eqref{A80} becomes a nonhomogeneous equation with the right-hand side determined by ${\vec p}_n$:
\be
\frak N {\vec q}_n =-4 \gamma A^2 \mathcal{R}_n^2 \sigma_1  {\vec p}_n.
 \label{A10}
 \ee
 Here $\frak N$ is a nonhermitian operator defined by 
\[
 \frak N= \mathcal{L} - \omega_0 I - \mu J.
 \]
 The two-component eigenvector  
 \be
\left( \begin{array}{c} {\vec p}_n \\ {\vec q}_n \end{array} \right)
\label{pq}
\ee 
of the full ``triangular" eigenvalue problem \eqref{B7} 
exists if the equation \eqref{A10} has a solution 
 with
 \[
 \sum_{n=-\infty}^\infty  \left(   |q_n^{(1)} |^2  + |q_n^{(2)} |^2 \right) < \infty.
 \]
 
 Whether the finite-norm solution ${\vec q_n}$ exists or not, 
 depends on whether the kernel space of 
 the adjoint operator
\[
\frak N^\dagger = \mathcal{L}  - \omega_0 I + \mu^{*} J
\]
is empty, and if not --- whether  the right-hand side in \eqref{A10} is orthogonal to the kernel.
  For the given eigenvalue $\mu$ in \eqref{A70},
  the operator $\frak N^\dagger$ has  a nonempty kernel  space
  only if there is a  sequence ${\vec y_n}$ such that 
  \[
  (\mathcal{L}- \omega_0 I) {\vec y_n}=-\mu^{*} J {\vec y_n},
  \]
  that is, if 
   $-\mu^*$ 
 coincides with one of the  (pure imaginary) eigenvalues 
 of the symplectic operator $J^{-1} (\mathcal L - \omega_0 I)$, 
 or falls  in its continuous spectrum  occupying two bands  in \eqref{cs}.
 Therefore if  $\mu$ has a nonzero real part --- the situation that  concerns us here ---
 the kernel space of $\frak N^\dagger$ is empty and
  the two-component eigenvector \eqref{pq} does exist.

Thus the stability analysis of the in-phase soliton reduces to the solution of the eigenvalue problem \eqref{A70}. 
All unstable  eigenvalues of \eqref{A70} (if exist) are automatically eigenvalues of the 
full  eigenvalue problem \eqref{B7}.

\subsection{The antiphase soliton}

In the case of the antiphase soliton, we 
let
\begin{align*} 
a_n =e^{i \beta z}  \left[ A \mathcal{R}_n+
\mathrm{Re} (p^{(1)}_n e^{\mu z})+ i  \, \mathrm{Re} (p_n^{(2)}e^{\mu z})\right],
 \\
b_n=  e^{i \beta z}  \left[
 \mathrm{Re} (q^{(1)}_n e^{\mu z})+ i  \, \mathrm{Re} (q_n^{(2)}e^{\mu z})\right],
\end{align*}
where, as before, 
  $\mathcal{R}_n$ is a solution of \eqref{B3} and
$p_n^{(1,2)}$, $q_n^{(1,2)}$, $\mu$ are complex.
This time, however,  $\beta$ is given by \eqref{OmegaM}
rather than \eqref{OmegaP}.
Linearising equations \eqref{A6} in $p_n^{(1,2)}$ and $q_n^{(1,2)}$
gives 
\begin{subequations}  
\label{B70}
\begin{align}
\mathcal{L}{\vec q}_n - \omega_0  {\vec q}_n = \mu J {\vec q}_n,  \label{A700}
\\
\mathcal{L}  {\vec p}_n + \omega_0  {\vec p}_n  - 4 \gamma  A^2 \mathcal{R}_n^2 \sigma_1 {\vec q}_n  = \mu J {\vec p}_n.
  \label{A800}
\end{align}
\end{subequations}

The analysis of the eigenvalue problem \eqref{B70} is similar to
the analysis for the problem \eqref{B7}. Letting ${\vec q}_n=0$ gives 
the linearised eigenvalue problem for the scalar nonlinear Schr\"odinger equation \eqref{B200}:
\be
(\mathcal{L}   + \omega_0 I) {\vec p}_n   = \mu J {\vec p}_n.
\label{X10}
\ee
There are positive eigenvalues in the case of the bond-centred soliton of the scalar
equation \eqref{B200}; this implies instability of the bond-centred antiphase soliton of 
the vector equation \eqref{B1}.

 In the case of the site-centred soliton,
the spectrum of \eqref{X10}
 consists of two zeros, two pairs of opposite nonzero imaginary eigenvalues and two bands of continuous spectrum on the imaginary axis:
$\mu= \pm i \omega_1$, where
\[
A^2 \leq \omega_1 \leq A^2+ 4\C.
\]
Therefore,  in order to detect instabilities of the site-centred soliton
one has to turn to eigenvectors with ${\vec q}_n \neq 0$. 

The eigenvalues pertaining to ${\vec q}_n \neq 0$ are determined by solving
the ``reduced" eigenvalue problem \eqref{A700}. The ${\vec p}_n$-component
of the eigenvector $({\vec p}_n, {\vec q}_n)$ is then recovered from the solution of the
nonhomogeneous equation \eqref{A800}.
The question that concerns us here is whether there are any eigenvalues  with nonzero real part;
for those $\mu$,
the nonhomogeneous equation \eqref{A800} has a finite-norm solution
(see the argument in the previous subsection).
Therefore  the full triangular matrix in \eqref{B70} has an unstable eigenvalue 
 as long as the half-size matrix in \eqref{A700} has one. 

\subsection{Symplectic eigenvalue problem}
\label{symp}

Since the general instability of the in- and out-of-phase bond-centred solitons has been established,
we  only need to consider the site-centred ones in what follows.

Starting with the in-phase soliton and
forming (infinite-component) vectors ${\bf p}^{(1)}$ and ${\bf p}^{(2)}$ out of the elements 
of the sequences $p_n^{(1)}$ and $p_n^{(2)}$,  respectively,  we rename ${\bf p}^{(1)}={\bf g}$ and ${\bf p}^{(2)}={\bf f}$ for notational   convenience.
 The eigenvalue problem \eqref{A70} is then written in the matrix form:
 \be
 \label{B8}
 \left( \begin{array}{cc}
 L_+ + \sigma I & 0 \\
 0 & L_-+ \sigma  I
 \end{array}
 \right)
 \left(
 \begin{array}{c}
 {\bf g} \\ {\bf f} \end{array}
 \right) = 
 \lambda J 
  \left(
 \begin{array}{c}
 {\bf g} \\ {\bf f} \end{array}
 \right). 
 \ee
Here $L_+$ and $L_-$ are symmetric matrices with elements 
\begin{align}
(L_+)_{mn}= - \Delta_{mn} +  h^2 (1-6 \mathcal{R}^2_n) \delta_{mn},  \label{B10} \\
(L_-)_{mn}= -\Delta_{mn} + h^2 (1-2 \mathcal{R}^2_n) \delta_{mn},  \label{B10A}
\end{align}
where
\[
\Delta_{mn}= \delta_{m, n+1} + \delta_{m, n-1} - 2 \delta_{mn},
\]
$h$ is as in \eqref{h},
and $\mathcal{R}_n$ is the site-centred single-hump solution of Eq. \eqref{B3}
(the Sievers-Takeno mode).
In \eqref{B8}, we have denoted
 $\lambda=\mu/\C$
and  introduced a parameter 
$\sigma= 2 \omega_0/\C>0$.

Turning to the antiphase soliton
and treating the sequences $q_n^{(1)}$ and $q_n^{(2)}$
 as  infinite-component vectors ${\bf g}$ and ${\bf f}$, respectively,
 the eigenvalue problem  \eqref{A700} is represented in the same form \eqref{B8} --- but with 
$\sigma= -2 \omega_0/\C<0$.

Therefore,  equation \eqref{B8} with $\sigma$ varying from negative to positive 
values  defines the stability problem 
 both for the in- and the out-of-phase  soliton. Here
 \be 
\sigma= \left\{
\begin{array}{lr}
\phantom{-} 2 \omega_0/\C,  & \mbox{in-phase \ soliton}; \\
- 2 \omega_0/\C,  & \mbox{antiphase \ soliton}.
\end{array}
\right.
\label{Z10}
\ee

 Note that the single-hump site-centred solution ${\mathcal R}_n$
is uniquely specified  by the value of the parameter $h$. This means that the stability properties of the soliton with amplitude $A$, in the 
system \eqref{B1} with coupling $\C$ and gain-loss coefficient $\gamma$,  are completely determined by {\it two\/} combinations of those  three parameters:
 $\sigma$ and $h$.
 
The combination $\sigma$ in \eqref{Z10}  is a structural parameter. It is fixed by the coupling constant $\C$
and the gain/loss coefficient $\gamma$ --- but does not depend on  the soliton's amplitude. On the other hand, $h$
in \eqref{h}  is the  amplitude of a particular solution
under scrutiny    normalised by the coupling: $h= A/ \sqrt{\C}$. 
 Solitons with different amplitudes and in systems with different couplings and
gain-loss coefficients have the same stability properties as long as they share the values of $\sigma$ and $h$. 

It is fitting to note that a self-similarity of this sort was previously encountered in the 
continuum limit of the system \eqref{B1}. There, the stability of solitons was determined by a 
single similarity parameter  $\eta= 2 \sqrt{1-\gamma^2}/A^2$   \cite{ABSK}.

According to Eq.\eqref{B3},  the matrix $L_-$ has a zero eigenvalue,
 with the eigenvector $\mathcal{R}_n$.
 Keeping in mind that $\mathcal{R}_n>0$ for all $n$, equation \eqref{B3} can be cast in the form
  \be
h^2 (1-2{\mathcal R}_n^2)= \frac{\Delta \mathcal{R}_n}{\mathcal{R}_n}.
\label{F1}
\ee
Making use of the identity \eqref{F1}, 
 the quadratic form $\langle {\bf y}| L_- | {\bf y} \rangle$ with an arbitrary ${\bf y}=\{y_n\}$ admits the following representation:
\[
\sum_{nm} y_n (L_-)_{nm} y_m=
 \sum_{n} \frac{ (y_n \mathcal{R}_{n+1} - \mathcal{R}_n y_{n+1})^2} {\mathcal{R}_n \mathcal{R}_{n+1}}.
\] 
This representation implies that  the matrix $L_-$ does not have any negative eigenvalues.

On the other hand, 
 one can readily check that  the matrix $L_+$ has  a negative  eigenvalue. A simple upper bound for it is given by 
 equation \eqref{a3} in Appendix A. The negative eigenvalue is single.
 
 In fact, the operators \eqref{B10} and \eqref{B10A} are not unfamiliar in the soliton stability literature.
 These infinite-dimensional matrices occurred in the studies of 
 the   {\it scalar\/} discrete nonlinear Schr\"odinger equation --- where
some of their properties were established. In particular, it was
  shown that the operator $L_+$ corresponding to the site-centred soliton has a single
 negative eigenvalue
 \cite{LST,KK}. 
 (Had there been two negative eigenvalues, the site-centred scalar  soliton would have been unstable.
 In actual fact, stability is a firmly established property 
 of that solution  \cite{LST,KK,PKF}.)

The operator 
\be
J^{-1} \left(
\begin{array}{cc}
L_+ +{\sigma} I & 0  \\ 0 & L_-+{\sigma} I \end{array}
\right)
\label{symplec}
\ee
in \eqref{B8},
 is symplectic, that is, 
generates a hamiltonian flow.
The spectrum of symplectic operators 
consists of pairs of  pure-imaginary values,  real pairs, and
complex  quadruplets. If $\lambda$ is a real or pure imaginary point
of spectrum, then $-\lambda$ is another one;  if a complex $\lambda$ is in the spectrum, 
then so are $\lambda^*$, $-\lambda$,
 and $-\lambda^*$ \cite{Arnold}.

\begin{figure}[t]
   \includegraphics*[width=0.3\linewidth]{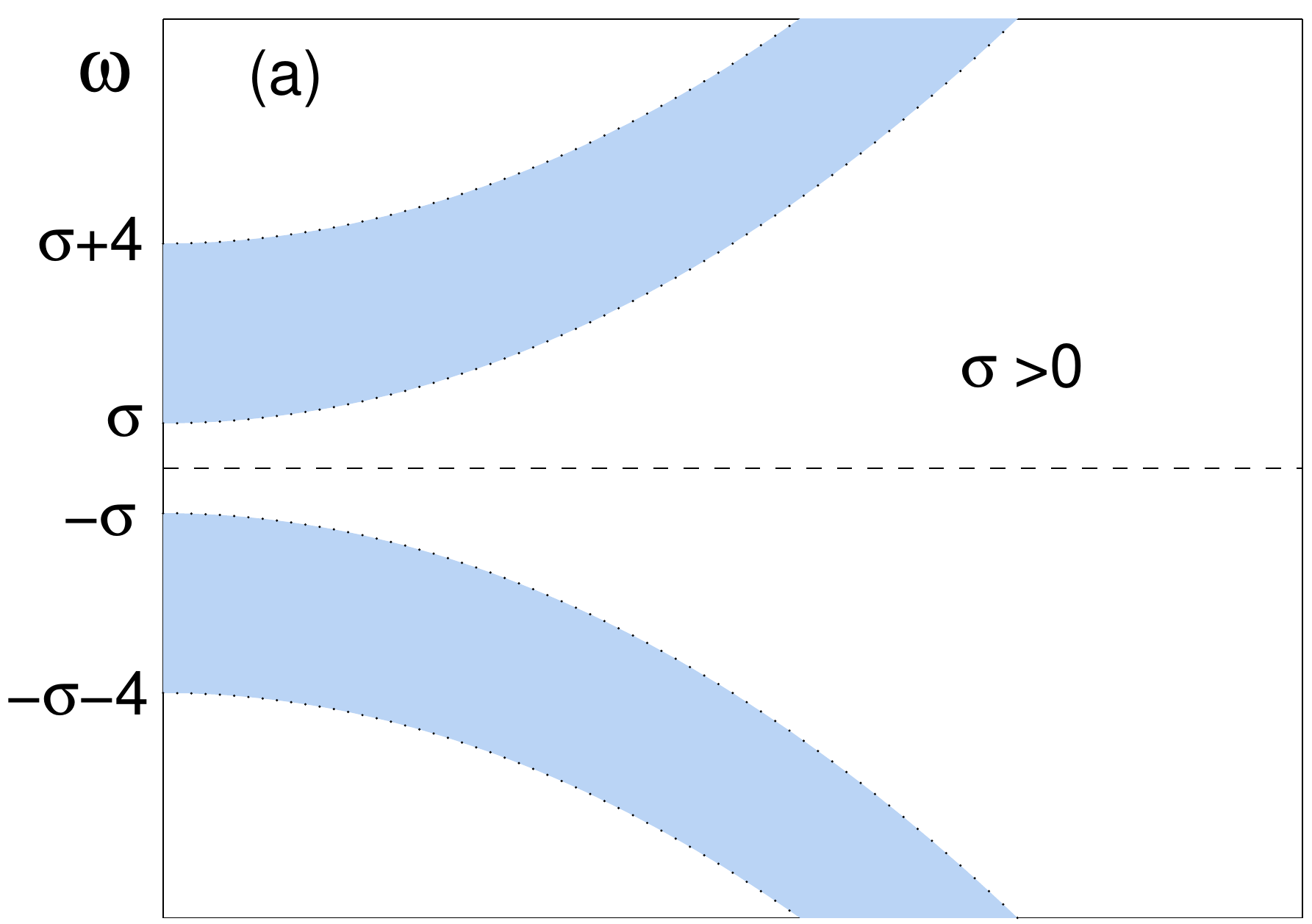}  \hspace*{5mm}
  \includegraphics*[width=0.3\linewidth]{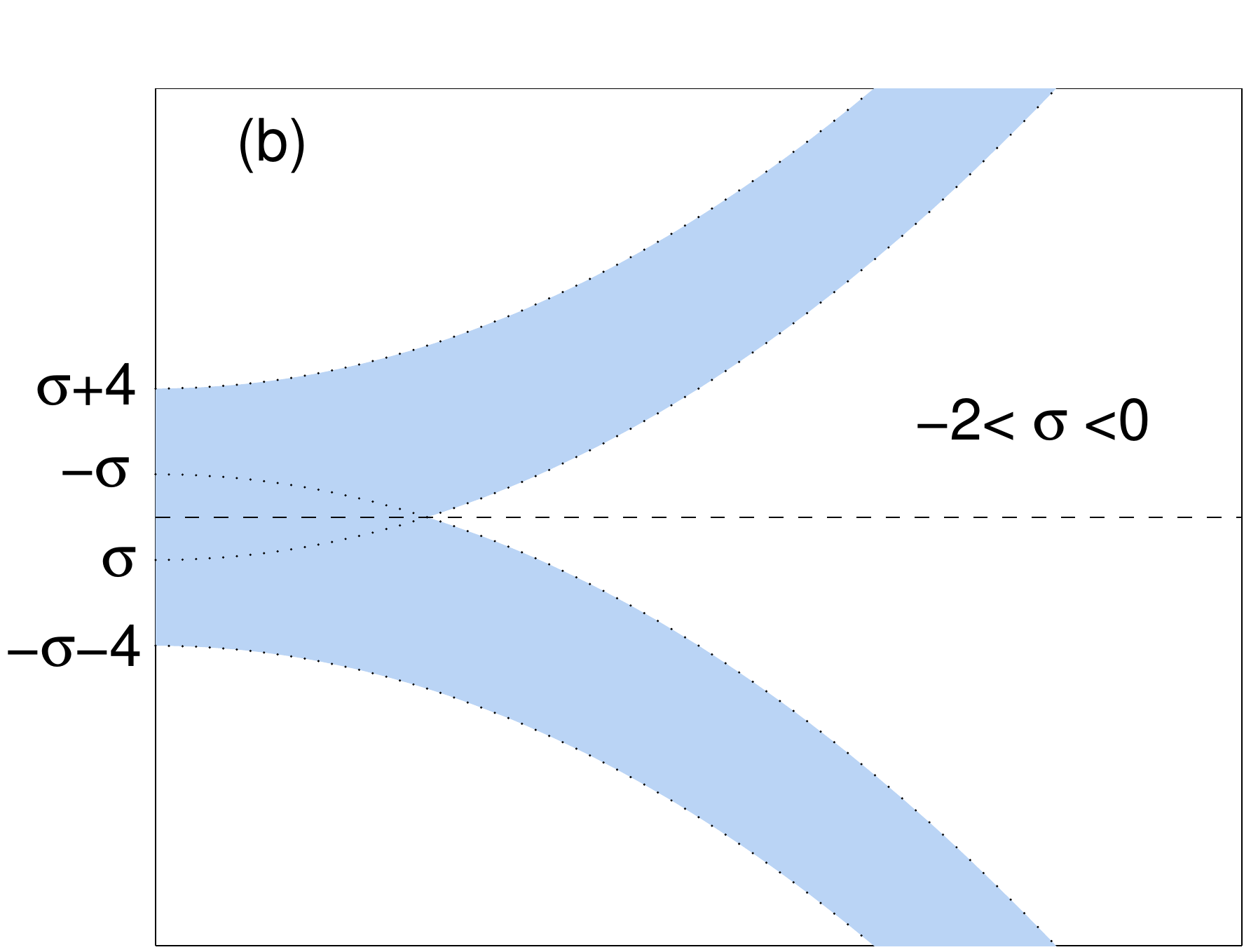}  \\
    \includegraphics*[width=0.3\linewidth]{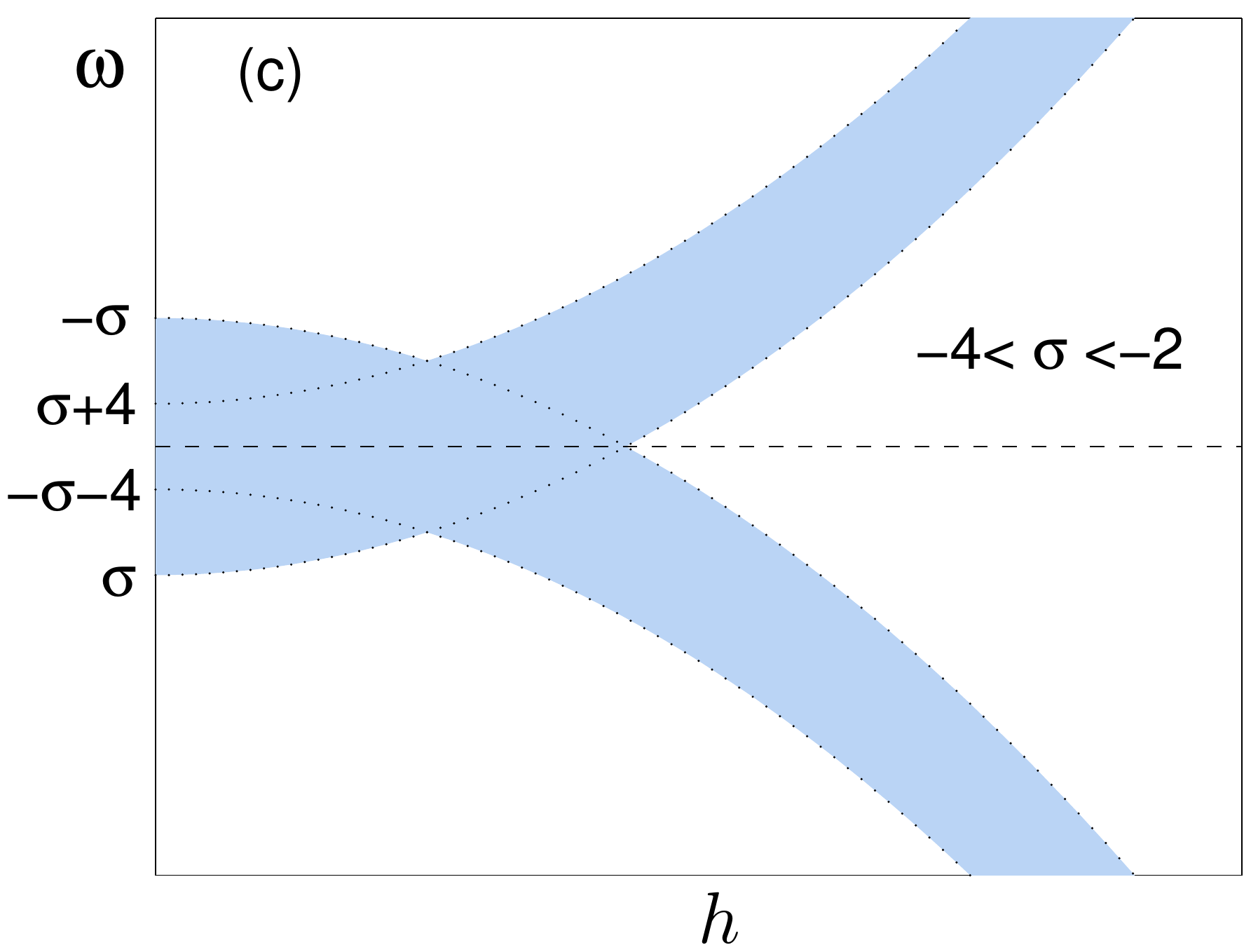} \hspace*{5mm}  
  \includegraphics*[width=0.3\linewidth]{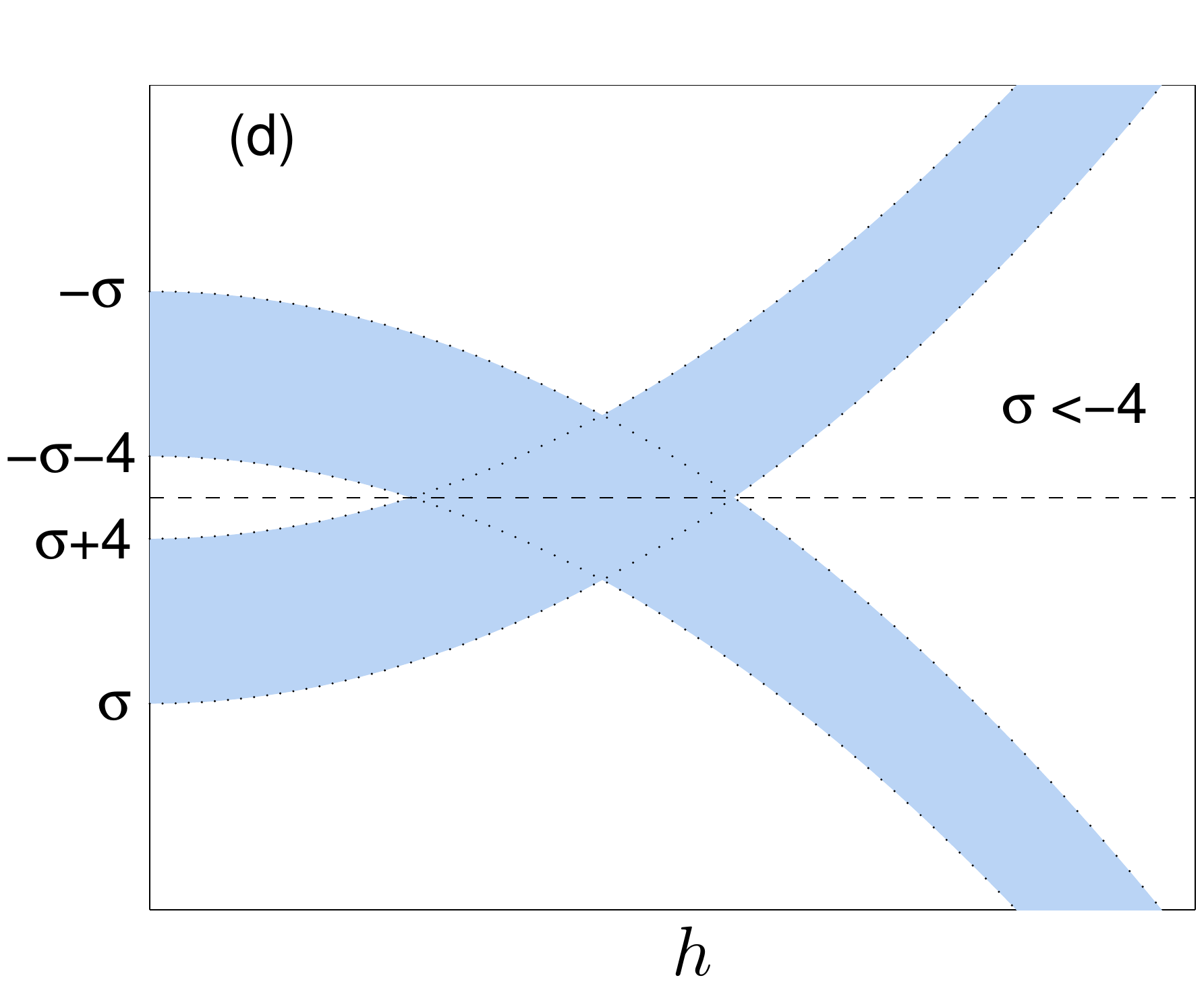}
     \caption{Two bands of continuous spectrum of the operator \eqref{symplec} on the $(h, \omega)$-plane, in four different intervals of $\sigma$.   }
 \label{cont}
 \end{figure}

The continuous spectrum of $\lambda$ lies on the imaginary axis and  consists of two bands:
$\lambda= \pm i \omega$, where $\sigma+ h^2 \leq  \omega \leq  \sigma+ 4+ h^2 $.
When $\sigma>0$ (the in-phase soliton), the bands are separated by a gap,
whereas when $\sigma<0$ (the antiphase soliton), the bands overlap.
See Fig \ref{cont}.

When $\sigma=0$,  equation  \eqref{B8} coincides with the eigenvalue problem
for the  scalar nonlinear Schr\"odinger equation, 
with no dissipative or driving terms.  It is important to keep in mind, however, that  the  problem  \eqref{B8} with  ${\sigma}=0$ 
does not result from the  coupled waveguides with $\gamma=0$.
The value ${\sigma}=0$  corresponds to  $\gamma =  1$ rather than $\gamma=0$. The significance of this value is  that 
 as $\gamma$ 
is increased through 1, the in-phase and antiphase solitons  merge and disappear. As for 
the ``no-gain, no-loss" ($\gamma=0$) pair of waveguides, it is represented by a nonzero $\sigma$ (${\sigma}=\pm 2/\C$) and is not special as far as the 
eigenvalue problem \eqref{B8} is concerned.

The  scalar discrete nonlinear Schr\"odinger equation has two zero eigenvalues 
in its linearised spectrum, stemming from its U(1) phase  invariance 
 and the transformation to the linearly growing phase.
Accordingly, the eigenvalue problem \eqref{B8} with $\sigma=0$
has two zero eigenvalues.
As ${\sigma}$ deviates from zero,  the two eigenvalues move out of the origin.

Finally we note that the eigenvalue problem \eqref{B8} with general $\sigma$  occurred previously in a context unrelated to the  \PT-symmetry. (Specifically,  
it appeared
in the analysis of the transverse stability of the
one-dimensional soliton in the scalar  two-dimensional  lattice \cite{PY}.)
The authors of \cite{PY} have established some useful 
 properties of eigenvalues  in the anticontinuum limit $h \to \infty$. 
 We make contact with those results in what follows. 

\section{In-phase soliton}
\label{HFS}
\subsection{Stability domain}

In this and the next section the word ``soliton" refers to the 
site-centred soliton (the Sievers-Takeno mode). 

In the case of the in-phase soliton ($\sigma>0$), the eigenvalue problem
\eqref{B8}  is amenable to simple analysis.
The smallest eigenvalue of the operator $L_-+\sigma I$ equals ${\sigma}$; therefore,
the matrix $L_-+{\sigma}I$ is positive definite and admits an inverse.
Hence the problem \eqref{B8} can be written  as a generalised
eigenvalue problem for the vector $\bf g$:
\be
(L_++{\sigma I}){\bf g} = -\lambda^2 (L_-+{\sigma I})^{-1} {\bf g}.
\label{A12}
\ee
The operator on the left in \eqref{A12} is symmetric, and the one on the right
is symmetric and positive definite. The lowest eigenvalue
$-\lambda^2$ is given by the minimum of the Rayleigh quotient
\be
-\lambda^2 = \min \frac{\langle {\bf g} |L_++{\sigma I}| {\bf g} \rangle}{\langle {\bf g} |(L_-+{\sigma I})^{-1}  |{\bf g}  \rangle}.
\label{A13}
\ee

Let $E_0=E_0(h)$ denote the (single) negative eigenvalue of $L_+$. 
The minimum in \eqref{A13}  is nonnegative  (hence the soliton is stable) if the smallest eigenvalue of the 
operator  in the numerator is nonnegative: $E_0(h)  +\sigma \geq  0$.

 \begin{figure}
      \includegraphics*[width=0.32\linewidth]{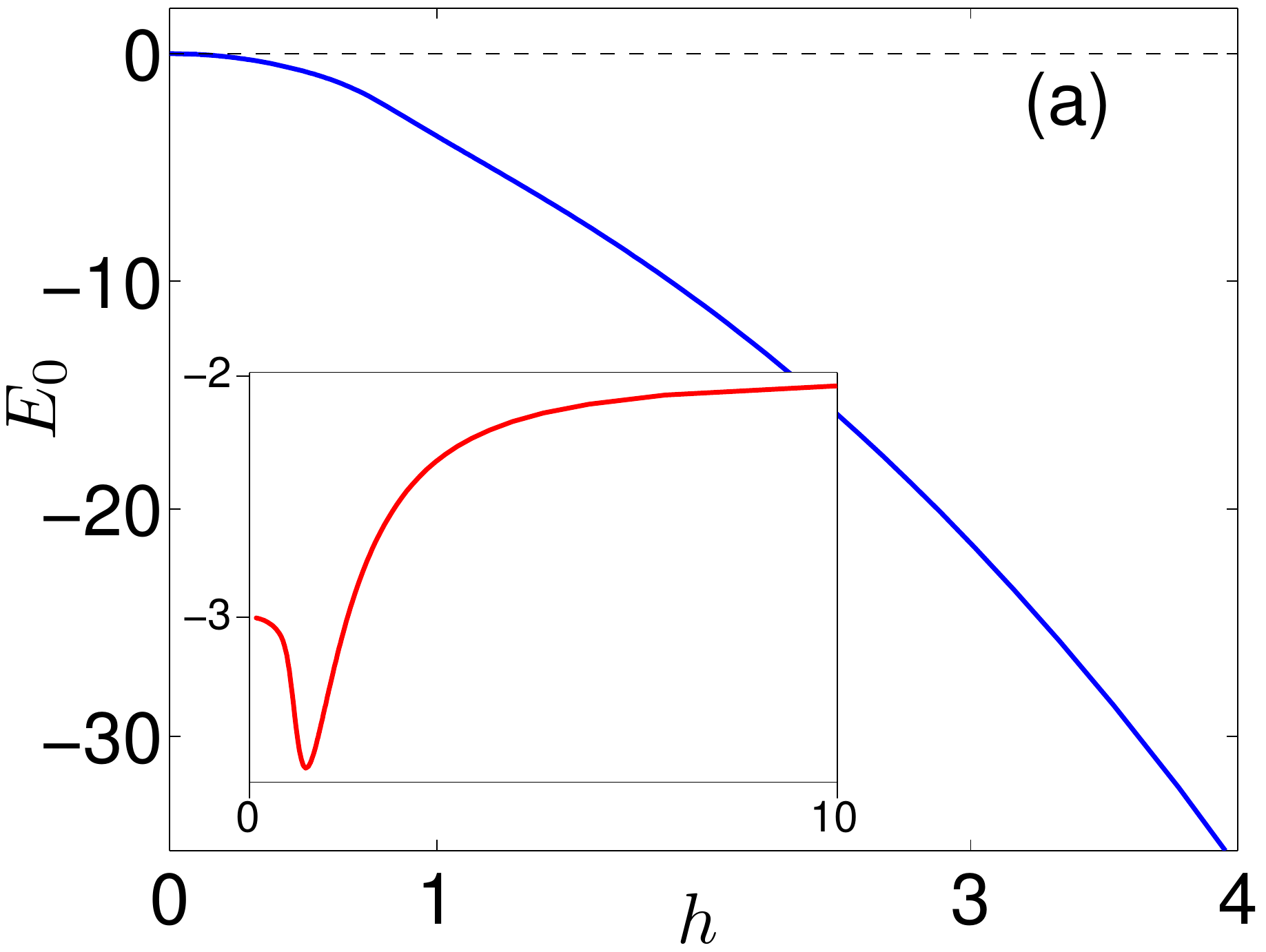}    \hspace*{4mm}        
         \includegraphics*[width=0.30\linewidth]{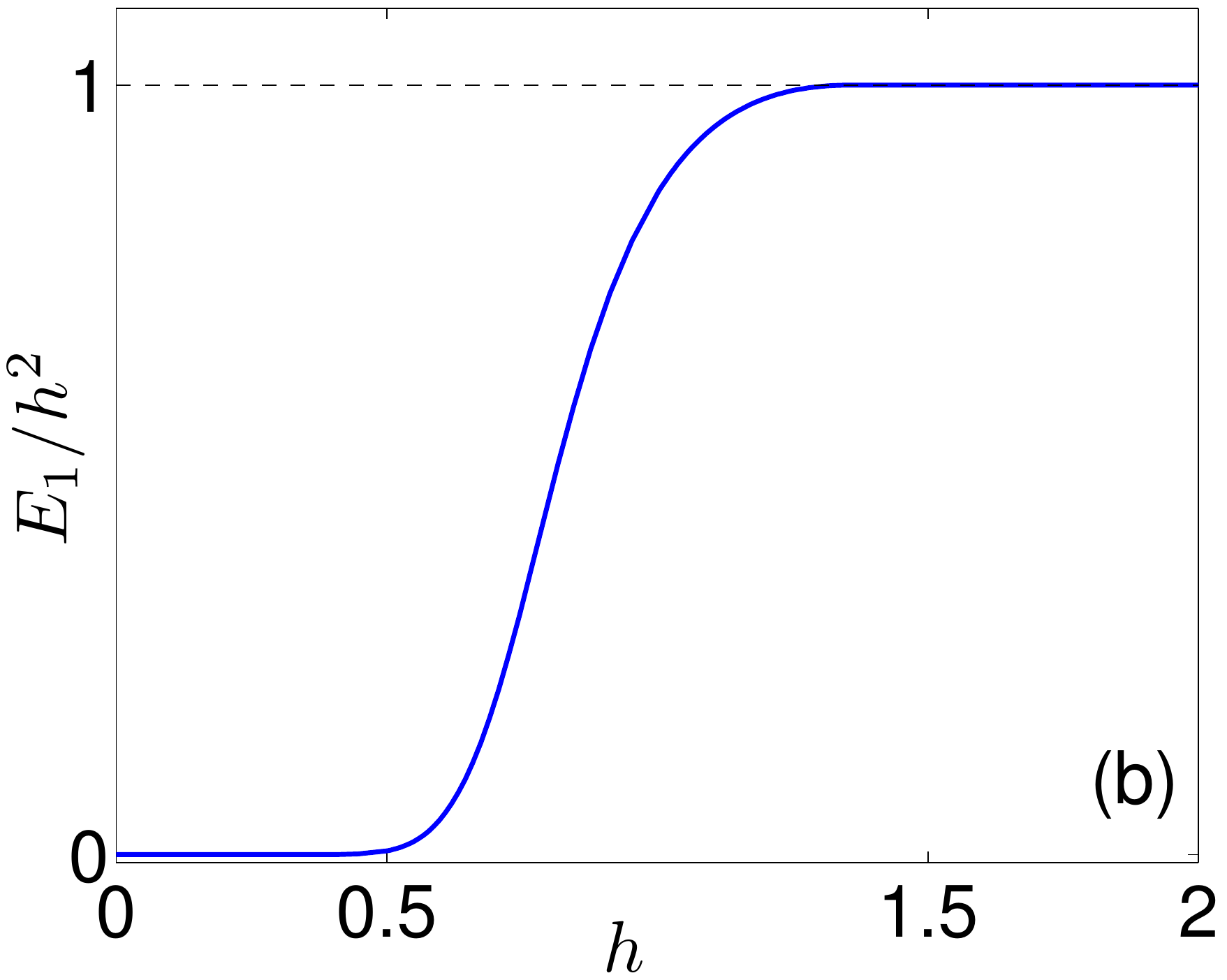}         \hspace*{5mm}    
     \includegraphics*[width=0.30\linewidth]{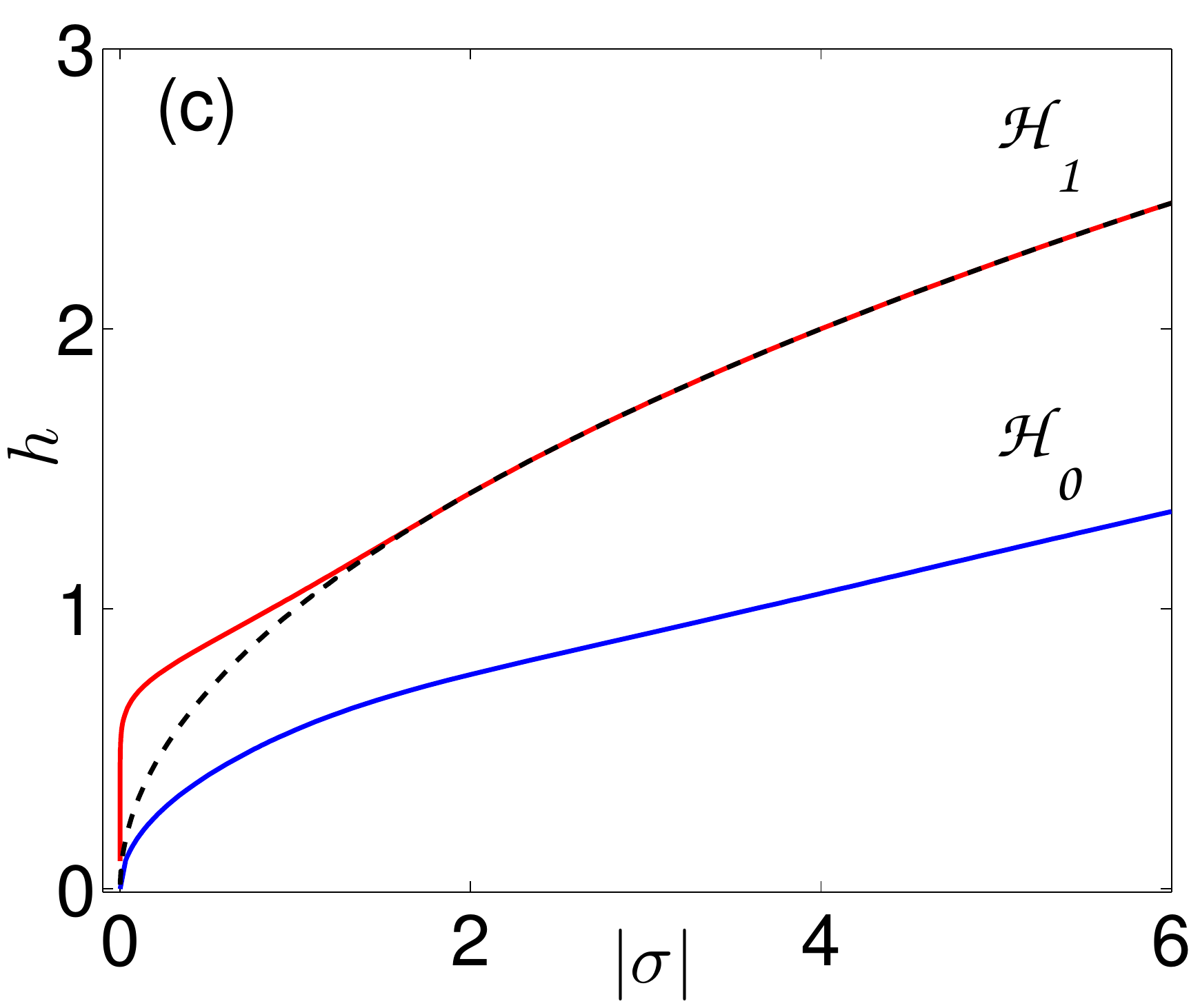}    
         \caption{The negative (a) and positive (b)  eigenvalue of the matrix $L_+$
   computed numerically. 
   In (b), the eigenvalue  $E_1$ is shown relative to $h^2$ (the lower endpoint of the continuous
   spectrum). In a similar way,
   the inset in (a) gives $E_0/h^2$.
     Panel  (c)
   displays the functions $\mathcal H_1(|\sigma|)$ (the inverse of $E_1(h)$)  and $\mathcal H_0(|\sigma|)$ (inverse of $-E_0(h)$).
   The dashed line  here is $h=|\sigma|^{1/2}$.
   }
 \label{L1E1}
 \end{figure} 

The eigenvalue $E_0$, calculated numerically for a sequence of $h$ values, is plotted in Fig \ref{L1E1} (a).  
It is clear from the figure that 
the  function $\sigma=-E_0(h)$ grows, monotonically, from zero to infinity as $h$ varies over the positive part of the axis. 
Denoting $h= \mathcal H_0(\sigma)$ its inverse function, the stability condition for the in-phase soliton
can be written as 
\be
h \leq   \mathcal H_0 (\sigma).     \label{hf}
\ee

The function $\mathcal H_0(\sigma)$, determined numerically, is plotted in Fig.\ref{L1E1}(c).
Equation \eqref{a3} of the Appendix \ref{App1} provides a simple upper bound for this function:
\be
\mathcal H_0(\sigma) < \sqrt{\sigma/2}, \quad 0<\sigma < \infty.   \label{ub}
\ee
Using equations \eqref{b1} and \eqref{a7}  in the corresponding Appendices, we obtain its small- and large-$\sigma$ asymptotic behaviours:
 \begin{align} 
 \mathcal H_0(\sigma) = \sqrt{\frac{\sigma}{3}}\left[ 1- \frac{\sigma}{90} +\frac{7}{16200}\sigma^2+ O(\sigma^3)  \right], \quad \sigma \to 0;   \label{CL} 
\\
\mathcal H_0 (\sigma) =\sqrt{\frac{\sigma}{2}}
\left[1 -\frac{2}{\sigma} + \frac{10}{3} \frac{1}{\sigma^2}  + O\left(\frac{1}{\sigma^3}\right) \right],
\quad \sigma \to \infty.   \label{ACL}
\end{align}

The stability domain admits a clear characterisation in terms of the soliton's amplitude and the
original parameters of the model \eqref{B1}.
Recalling that 
$h= A/ \sqrt{\C}$ and
$\sigma= 2 \sqrt{1-\gamma^2}/\C$, we infer from \eqref{hf} that 
 the in-phase soliton is stable if its amplitude satisfies
\be
A \leq {\mathcal A}_0(\gamma,\C),    \label{thresh}
\ee
 and unstable otherwise. Here $\mathcal A_0$ is a function of the coupling 
and gain-loss coefficient:
\be
\mathcal A_0(\gamma,\C)=            \sqrt{\C}    \,   \mathcal H_0  \! \left(\frac{2 \sqrt{1-\gamma^2}}{\C} \right),  \label{ceiling}
\ee
with $h=\mathcal H_0(\sigma)$ being the inverse  of  the function $\sigma=-E_0(h)$. 
For each fixed value of $\C$ we will be referring to the function $\mathcal A_0(\gamma,\C)$ as the instability threshold.

Equation \eqref{ub} translates into a simple upper bound for the  threshold:
\be
 \mathcal A_0 (\gamma, \C) \leq (1-\gamma^2)^{1/4}.   \label{Aub}
\ee

More accurate estimates for the boundary of the stability domain are available in two opposite limits.
In the  limit $\sqrt{1-\gamma^2}/\C \to 0$, the approximate expression for $\mathcal A_0$ stems from equation \eqref{CL}:
\begin{align} 
\mathcal A_0=  \left( \frac{2}{3} \right)^{1/2} (1-\gamma^2)^{1/4} \left[ 1- \frac{1}{45}  \frac{\sqrt{1-\gamma^2}}{\C}    
 + \frac{7}{4050}   \frac{1-\gamma^2}{ \C^2}
+
O \left(  \frac{(1-\gamma^2)^{3/2} }{\C^3}  \right) \right].
\label{lar_cou}
\end{align}
(The first term in this expression reproduces the instability threshold for the solitons in the \PT-symmetric  coupler consisting of 
two parallel guiding planes \cite{Driben1,ABSK}.)

In the  weak coupling limit ($\C/ \sqrt{1-\gamma^2} \to 0$), 
an approximate expression for the instability threshold follows 
from equation \eqref{ACL}:
\begin{align}
\mathcal A_0= (1-\gamma^2)^{1/4}  \left[ 1
- \frac{\C}{\sqrt{1-\gamma^2}}  
 + \frac{5}{6} \frac{\C^2}{1-\gamma^2} 
+ O \left(  \frac{\C^3}{ (1-\gamma^2)^{3/2}}  \right)   \right]. 
\label{sma_cou}
\end{align}

\begin{figure}[t]
                        \includegraphics*[width=0.5\linewidth]{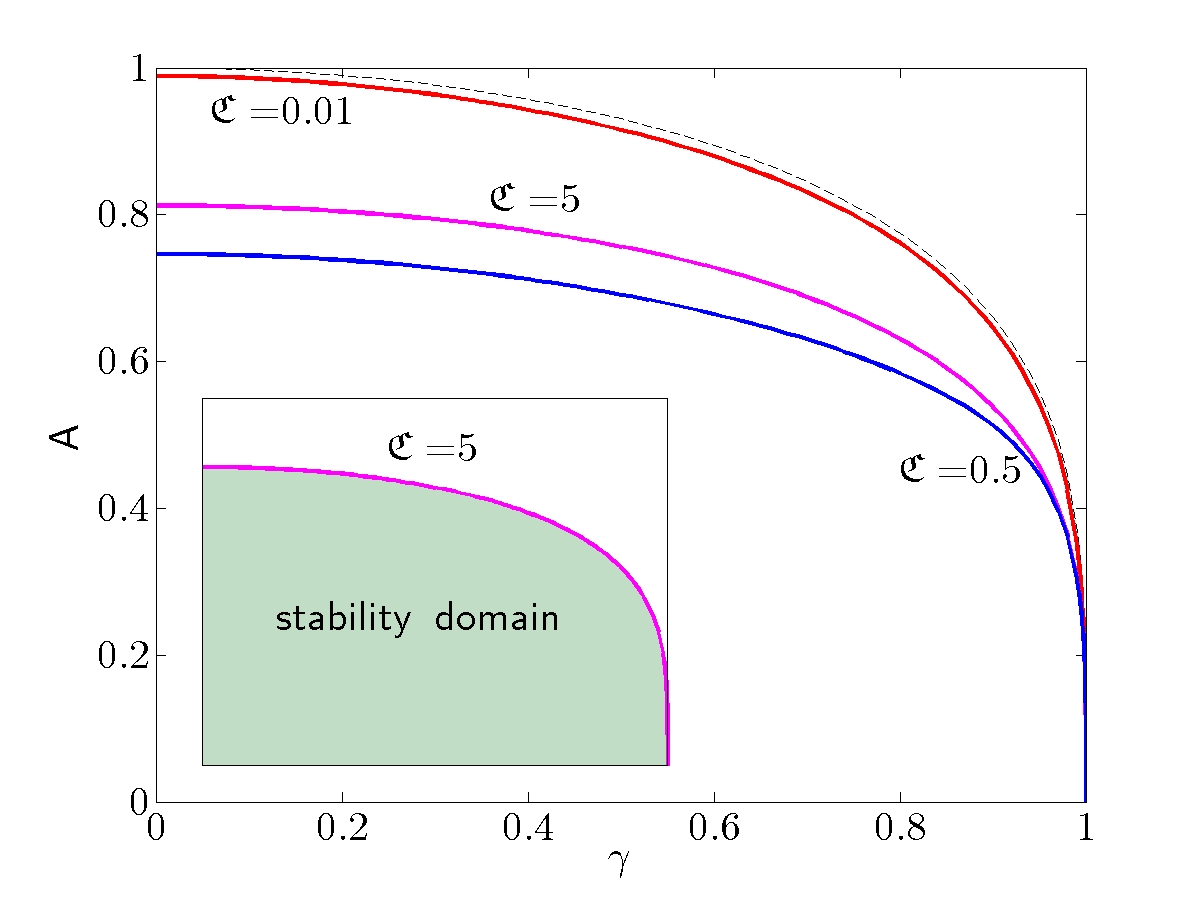} 
                              \caption{The in-phase soliton instability threshold for three different values of $\C$.
        Solid lines, top to bottom: $\C=0.01$, $\C=5$, $\C=0.5$.
  The dashed line demarcates the upper bound of the stability domain provided by equation \eqref{Aub}. 
  The inset illustrates the position of the stability domain relative to the threshold curve (here $\C=5$).
    }
 \label{stab_hf}
 \end{figure}


Plotting the curve \eqref{ceiling} with a variety of $\C$  (Fig \ref{stab_hf})
one observes that 
  the instability threshold is lowest when the coupling takes values near $\C=0.5$. 
 It is impossible to identify a single ``most unstable" value of  $\C$
  because
 small variations about $\C = 0.5$ raise either the small-$\gamma$ or large-$\gamma$  part of the curve \eqref{ceiling}
  while moving down the remaining part. 
  Instead of a single value, there is a finite (yet narrow) interval of  $\C$  encompassing $0.5$.
   The instability thresholds associated with  $\C$ in this interval form a thin 
  belt underlying instability thresholds associated with $\C$ further away from $\C=0.5$. 
  With this reservation in mind, we can refer to values close to $\C=0.5$ as ``most unstable".

 It is interesting to compare the ``most unstable" values of the
gain-to-gain and loss-to-loss coupling $\C$ to the value of the gain-to-loss  coupling in the ladder \eqref{B1}.
To make a fair comparison,  we
think of  each guide with gain as being connected to its counterpart with loss by 
{\it two\/} bonds, left and right,  with each coupling 
 coefficient being equal to  $1/2$. 
 (Geometrically, this corresponds to seeing each pair of gain and loss sites as a circular necklace 
 consisting just of two elements.)
 In this symmetric picture, each site of the ladder has four bonds: two horizontal and two vertical. 
 The maximum instability associated with $\C \approx 0.5$ implies that 
 the soliton's stability domain is  narrowest when the coupling coefficients of two
  horizontal and two vertical bonds are equal.

 As 
  $\C$ is raised from  $\C \approx 0.5$, the instability threshold is lifted.
  (See Fig \ref{stab_hf}.) 
  As $\C \to \infty$, it approaches the limit curve
 of $A= \sqrt{\frac23} (1-\gamma^2)^{1/4}$. 
 The instability threshold is also lifted 
 if we, instead, move $\C$  down from  $\C \approx 0.5$.
 As $\C \to 0$,  it approaches the limit curve in \eqref{Aub}: $A=\sqrt[4]{1-\gamma^2}$.

\subsection{Eigenvalue trajectories}

 \begin{widetext}
 
\begin{figure}
      \includegraphics*[width=0.4\linewidth]{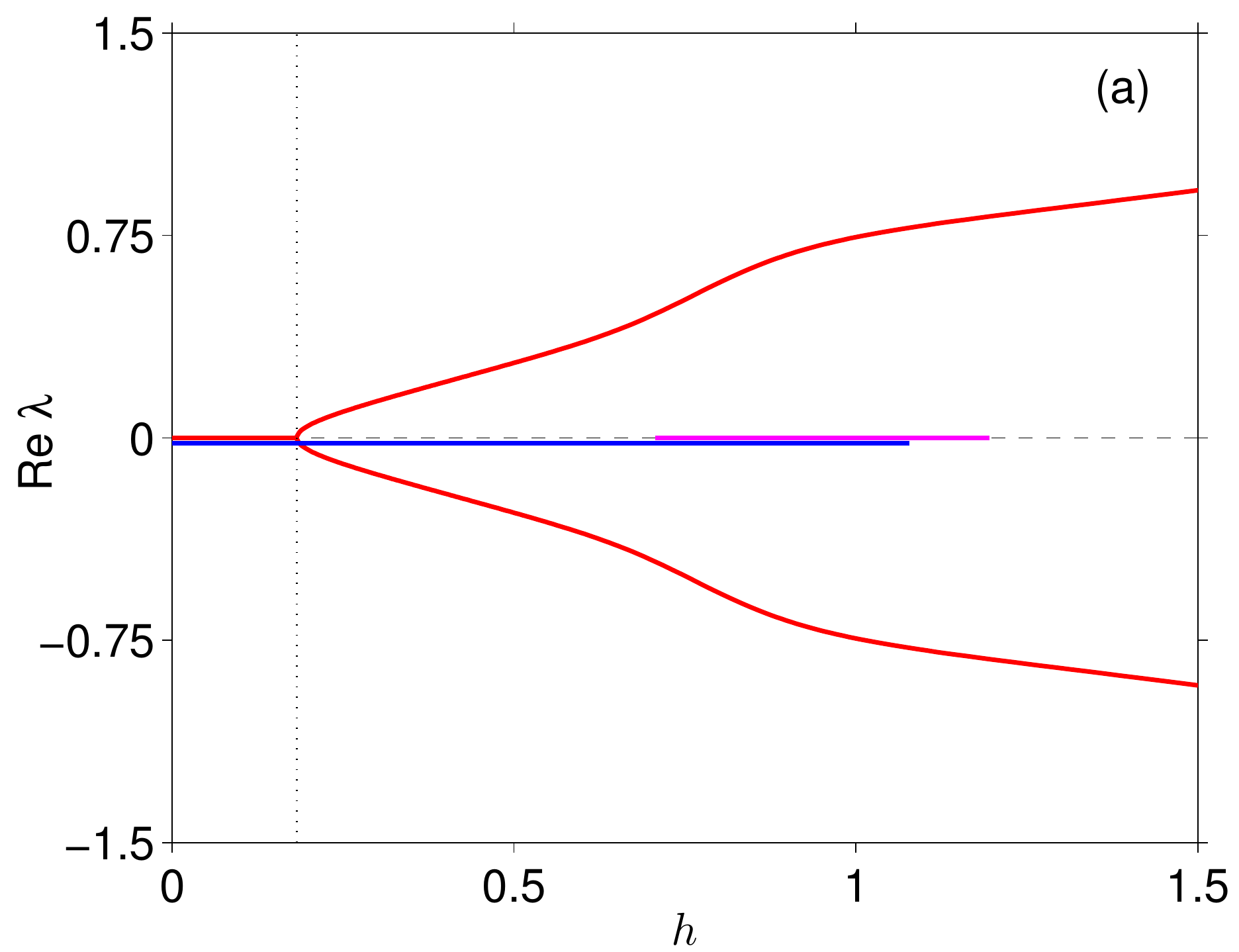}    
 \hspace*{5mm}  \includegraphics*[width=0.4\linewidth]{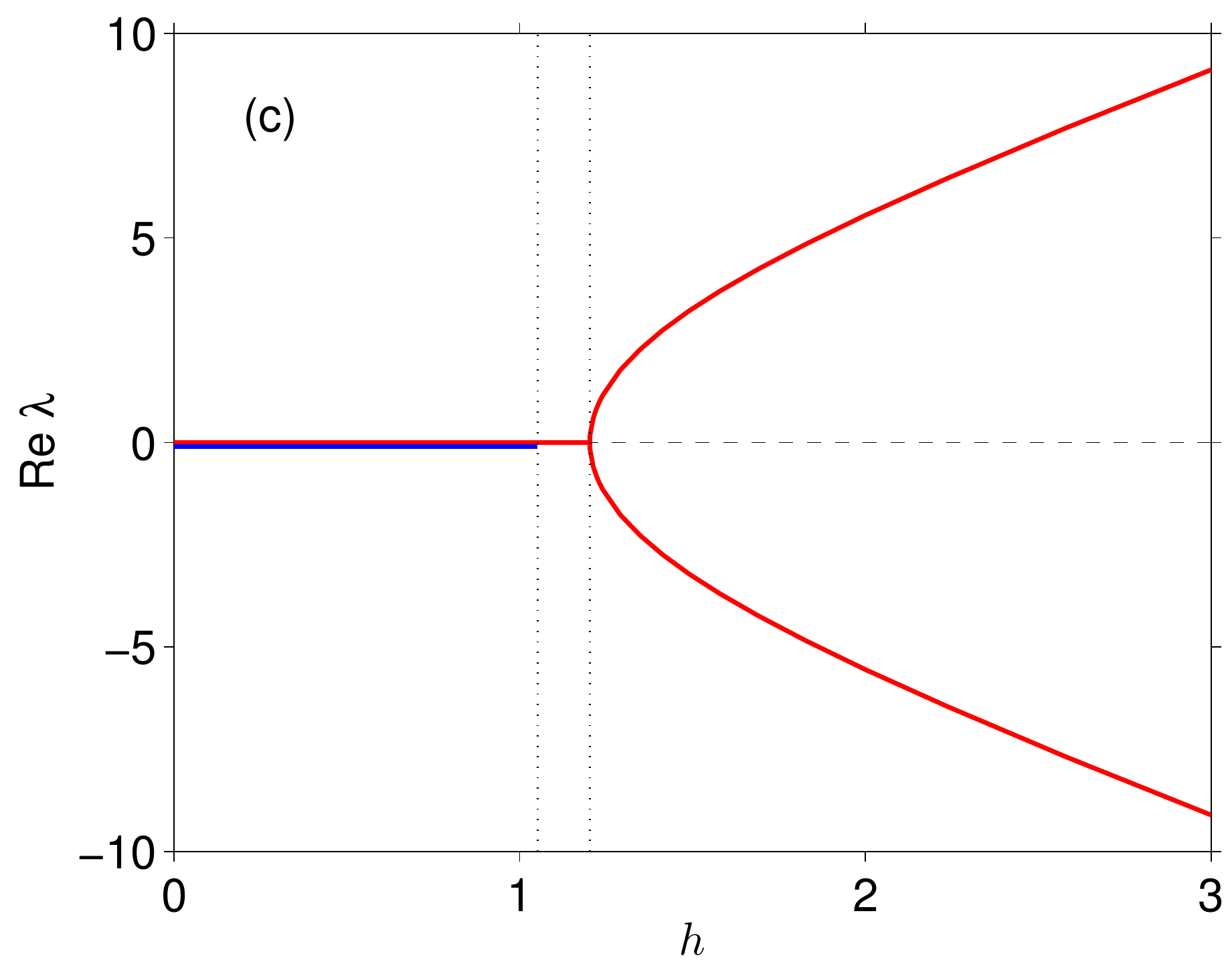}   
    
    \includegraphics*[width=0.4\linewidth]{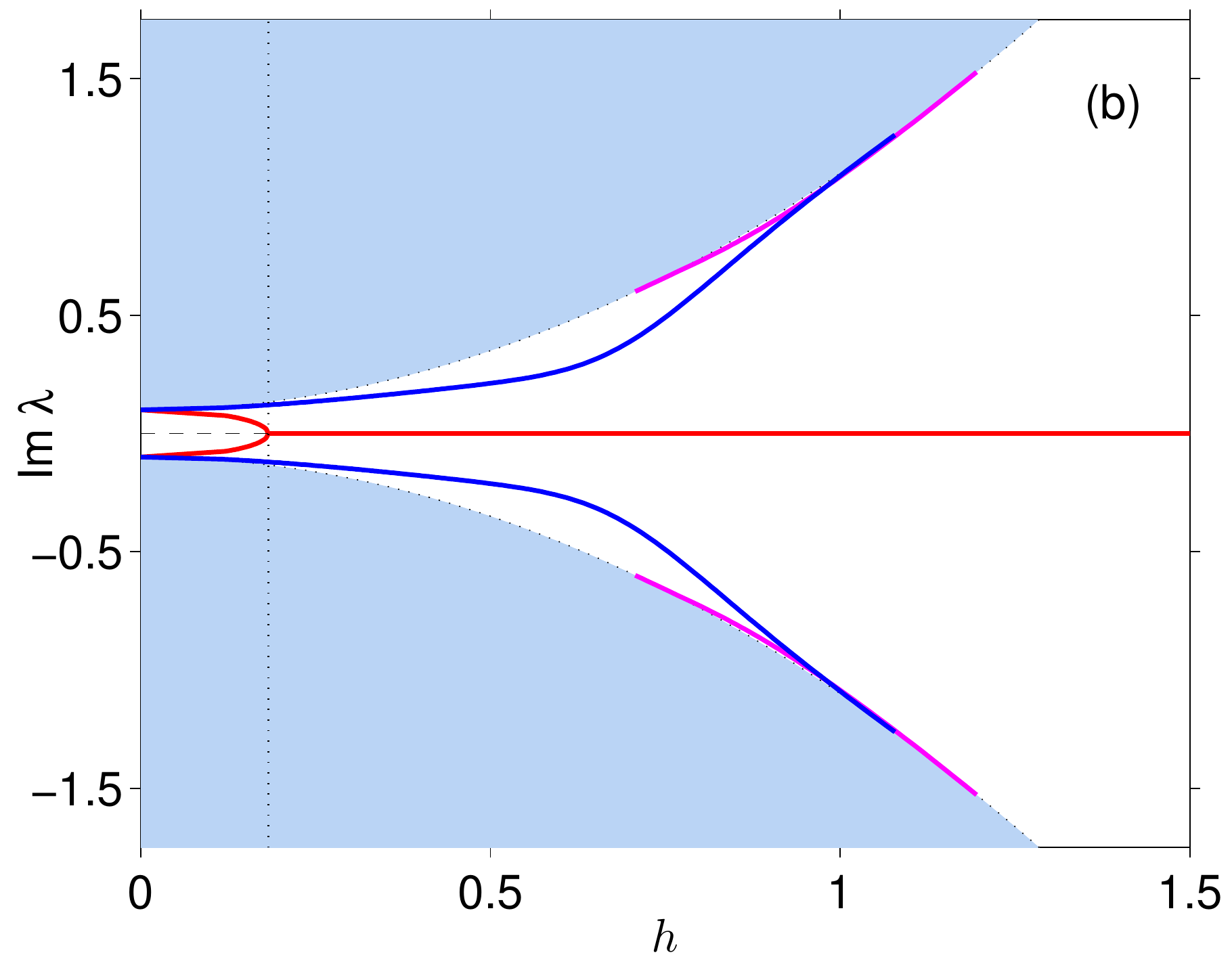}
   \hspace*{5mm} 
            \includegraphics*[width=0.4\linewidth]{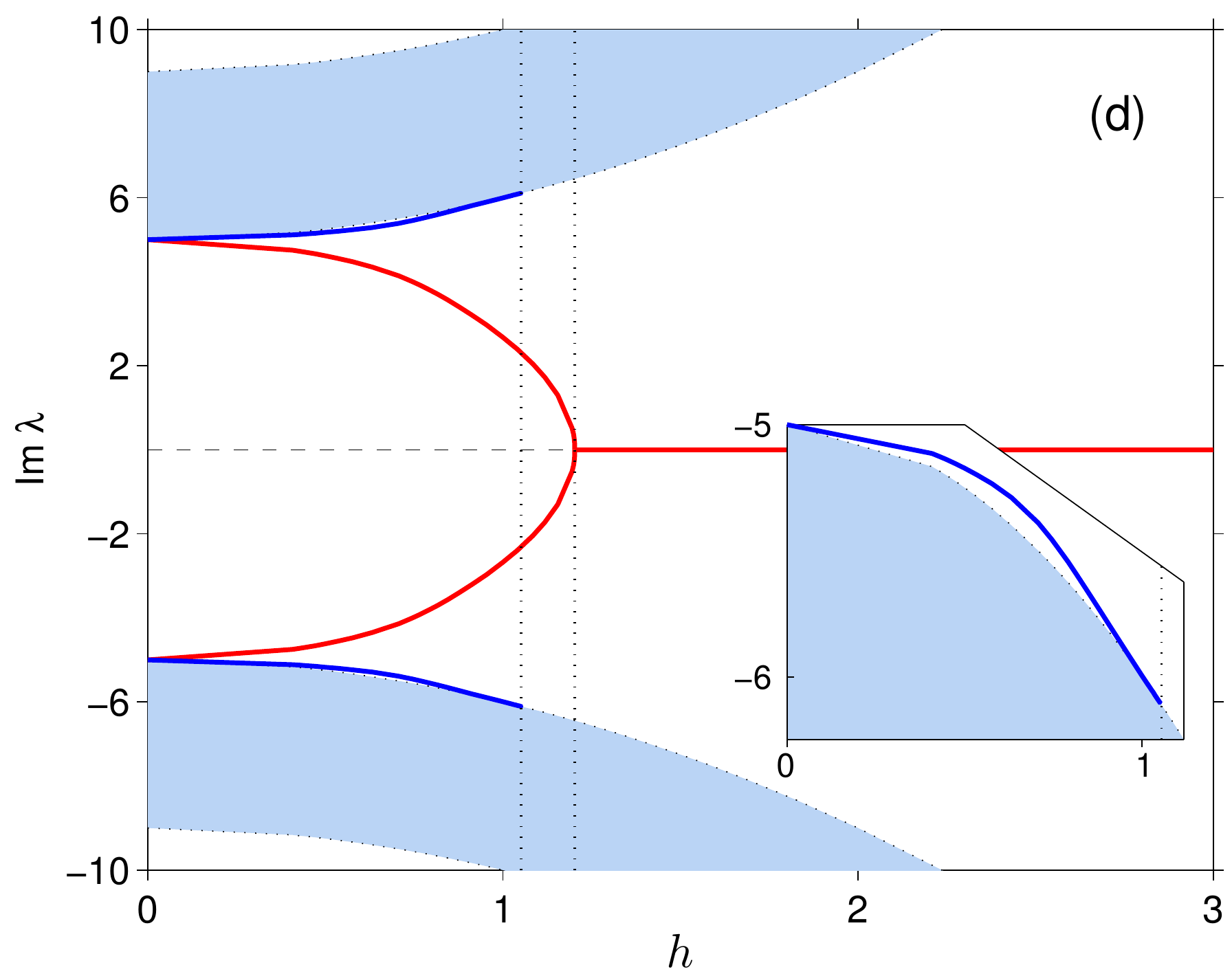}              
                        \caption{The  real  and imaginary part of eigenvalues $\lambda$ of the
   problem  \eqref{B8} as a function of the normalised amplitude $h$. In
   (a)-(b),  $\sigma=0.1$; in (c)-(d),  $\sigma=5$.
   A pair of  opposite imaginary eigenvalues $\pm i \omega_{\mathrm{even}}$  bifurcating from the continuous spectrum as $h$ is increased from 0,
   collides at the origin and diverges to infinities along the positive and negative real axis (shown by the red lines). 
   Also seen are pairs of imaginary eigenvalues $\pm i \omega_{\mathrm{odd}}$ and $\pm i {\tilde \omega}_{\mathrm{even}}$ which detach from the continuous spectrum
   but then immerse back in the continuum (deep blue and pink arcs, respectively.).   Note that in (a) and (c), the horizontal segments of the red, blue and pink
   curves have been shifted away from the  $\mathrm{Re} \, \lambda=0$ line for visual clarity.
   The inset in  (d) zooms in on the imaginary eigenvalue as it splits away from  the continuum and then returns to it. }
 \label{Plu}
 \end{figure}
  \end{widetext}

Although equations \eqref{thresh}-\eqref{ceiling}  provide the complete solution to the stability problem of the 
in-phase soliton,
they do not give any insight into the motion of the eigenvalues on the complex plane.
Yet the knowledge of stable and unstable eigenvalues can address a number of pertinent questions, in particular
classify the type of instability and identify modes of internal oscillation.
The aim of this subsection is to consider the eigenvalues: variationally, asymptotically and numerically.

The variational principle 
\eqref{A13} gives rise to simple upper and lower bounds of the 
 lowest eigenvalue $-\lambda^2$
of the scalar eigenvalue problem \eqref{A12} 
(see Appendix \ref{App2}).
The two bounds
allow one to follow the corresponding
pair of opposite symplectic eigenvalues
$\pm \lambda$,
as $h$ grows from 0 to $\infty$.

   For  $h$
smaller than $\mathcal H_0$, the set of symplectic eigenvalues may only consist of 
 pure imaginary pairs $\lambda = \pm i \omega$.
We denote $\pm i \omega_{\mathrm{even}}$ the pair with the smallest value of $\omega^2$; the notation will become clear later in this subsection.
Using \eqref{Q14}, the modulus-squared
 $\omega_{\mathrm{even}}^2$ is found to satisfy
  \be
 \omega_{\mathrm{even}}^2 \leq \sigma(\sigma-2h^2).
 \label{oms}
 \ee 
 Recalling  that the  continuous spectrum  of  symplectic  eigenvalues $\lambda= \pm i \omega$ in  \eqref{B8} 
  occupies the band $(\sigma+ h^2)^2 \leq \omega^2 \leq  (\sigma+ 4+h^2)^2$,
 we observe that the right-hand side in \eqref{oms} lies below the bottom edge of the continuous spectrum. Hence
 \be
 \omega_{\mathrm{even}}^2 < (\sigma +h^2)^2,
 \quad h>0.
  \label{inq}
 \ee
This implies that  the lowest eigenvalue $-\lambda_{\mathrm{even}}^2$ cannot
 bifurcate from the lower edge of the  continuum for any finite $h$.

 On the other hand, 
the inequality in \eqref{inq}  may become equality when $h=0$; hence the bifurcation {\it can\/} occur at this point. 
 Equation \eqref{Q17} indicates that for small $h$, the quantity
 $\omega_{\mathrm{even}}^2$ lies above $\sigma(\sigma-3h^2)$.  
 Taken together with the inequality \eqref{oms}, this fact implies that ``the innermost" 
 pair of imaginary eigenvalues (i.e. the pair with the smallest value of $\omega^2$)
 does indeed bifurcate from the continuum at the point $h=0$.

In fact, there are {\it two\/} pairs of pure imaginary
eigenvalues, $\lambda = \pm i \omega_{\mathrm{even}}$ and $\lambda= \pm i \omega_{\mathrm{odd}}$,
bifurcating from the edges of
 the continuous spectrum as $h$  is increased from zero.
  According to the asymptotic analysis of 
 Appendix \ref{App4}, one pair has even and the other one odd eigenfunctions; hence the choice of notation. 
 Equation \eqref{finlam} 
 gives 
\begin{align*}
 \omega_{\mathrm{even}}= 
\sigma- 1.438 h^2  + O(h^4),                                          &    \\   
 \omega_{\mathrm{odd}}=   \sigma+ 0.685 h^2 + O(h^4)      \ &
  \mbox{as} \ h \to 0.
\end{align*}
Note that the lowest value of $\omega^2$ is provided by the eigenvalue with the even eigenfunction.
This explains the choice of notation in \eqref{oms}-\eqref{inq}.

In the previous subsection we have shown that 
the
 pure imaginary  pair $\pm i \omega_{\mathrm{even}}$ (the pair with lowest $\omega^2$) converges at the origin and splits into
 the negative and positive real axis
as $h$ is increased through the critical value $\mathcal H_0$, 
When $h$ grows to infinity, 
equations \eqref{Q14}  and \eqref{Q15} give a corridor for the absolute value of the emerging real pair:
\be
\sigma(2 h^2-\sigma)
 \leq \lambda_{\mathrm{even}}^2  \leq
  \sigma(2 h^2 -\sigma +4- \frac{16}{3} h^{-2}).
\ee
Hence, $|\lambda_{\mathrm{even}}| = \sqrt{2 \sigma} h+O(h^{-1})$ as $ h \to \infty$. 
The anti-continuum limit (Appendix \ref{App5}) provides a more specific result:
\[
|\lambda_{\mathrm{even}}| = \sqrt{2 \sigma} \left[
h+ \left(1 -\frac{\sigma}{4} \right) h^{-1} + O(h^{-3}) \right].
\]

These variational 
and asymptotic considerations are corroborated by numerical solutions.
Figures \ref{Plu} show the evolution of the imaginary and real parts of eigenvalues as
$h$ grows from 0 to large values, with $\sigma$ being fixed. Clearly seen is
a pair of imaginary eigenvalues which detaches from the continuous spectrum at $h=0$,
moves on to the real axis, and diverges to infinity at a linear rate, as $h \to \infty$. 
\\

\section{Antiphase soliton}
\label{LFS}

To classify stability of the antiphase solitons, we consider the eigenvalue
problem \eqref{B8} with $\sigma<0$. 
Unlike equation \eqref{B8} with $\sigma >0$, this eigenvalue problem is amenable to 
analytical treatment only in two regions  on the $(\sigma, h )$ plane.
These are considered in subsections \ref{odd_mode} and \ref{wcsa}.
For other parameter values we have to rely upon the asymptotic and numerical eigenvalue analysis.

\subsection{Instability against an odd mode}
\label{odd_mode}

When ${\sigma}<0$, the matrix $L_-+\sigma I$ has a negative eigenvalue (equal to $\sigma$).
Hence, the lowest value of $-\lambda^2$ in the generalised eigenvalue problem \eqref{A12}
cannot be sought as a minimum of
the Rayleigh quotient \eqref{A13} over the entire $\ell^2$ space of square-summable bi-infinite sequences.
However, when $\sigma >-h^2$, we can use the Rayleigh quotient to establish the existence of 
eigenvalues with  odd (antisymmetric) eigenvectors ${\bf g}$: $g_{-n}=-g_n$, $n=0,1,2,...$.

Indeed,  
 the spectrum of the infinite matrix $L_-$ does not have eigenvalues other than ${\mathscr E}_0=0$ and 
 a continuous band  $h^2 \leq \mathscr{E}  \leq h^2+4$. 
(We have verified this by 
computing numerical eigenvalues of a large finite truncation of $L_-$
 over a representative sample  of $h$ values between 0 and $\infty$.)
The zero eigenvalue corresponds to an even (symmetric) eigenvector ${\bf z}_0= \textit{\textbf{R}}$. 
Therefore, the matrix $L_-+\sigma  I$ with $\sigma>-h^2$, is positive definite on the subspace $\frak S$ of $\ell^2$ 
consisting of {\it odd\/}  sequences $\bf g$. 
Consequently, the lowest eigenvalue of the generalised problem \eqref{A12} associated with an odd  eigenvector, 
is given by 
\be
-\lambda^2 = \min_{\frak S} \frac{\langle {\bf g} |L_++{\sigma} I| {\bf g} \rangle}{\langle {\bf g} |(L_-+{\sigma} I)^{-1}  |{\bf g}  \rangle}.
\label{G1}
\ee

Let $E_1$ denote the second lowest eigenvalue of the matrix $L_+$: 
\[
L_+ {\bf y}_1 =E_1 {\bf y}_1.
\]
The eigenvalue $E_1$ is positive \cite{LST,KK} and satisfies $0< E_1(h)<h^2$.  (See Fig.\ref{L1E1}(a)). The corresponding eigenvector ${\bf y}_1$ is odd 
and renders the quadratic form $\langle {\bf g} |L_++{\sigma}I| {\bf g} \rangle$ minimum in  $\frak S$.
Equation \eqref{G1} implies then that 
$-\lambda^2$ is positive respectively negative for $E_1+\sigma>0$ respectively $E_1+\sigma<0$. 
 
In the region $-h^2< \sigma < -E_1(h)$, the symplectic eigenvalue problem \eqref{B8} has a pair of real eigenvalues $\pm \lambda$ 
signifying the instability of the antiphase soliton  against an odd mode.
As $\sigma$ approaches $-E_1(h)$ from below, the two eigenvalues collide at the origin 
and move onto the imaginary axis. The  odd-mode instability is replaced with a mode of internal oscillation. 

In terms of the soliton's amplitude, $A$, the above odd-mode instability region is given by
\be
\sqrt{2} (1-\gamma^2)^{1/4} < A < \mathcal A_1(\C, \gamma),
\label{odd_inst} 
\ee
where
\be
\mathcal A_1= 
\sqrt{\C} \mathcal H_1 \left( {2\sqrt{1-\gamma^2}}/{\C} \right) 
\label{A1H1}
\ee
and
$h=\mathcal H_1(|\sigma|)$ is the function inverse to $|\sigma|=E_1(h)$. 
This function is shown in Fig \ref{L1E1}(c). 
The small- and large-$|\sigma|$ 
behaviour of $\mathcal H_1(|\sigma|)$ follow from the  corresponding asymptotics of the eigenvalue $E_1(h)$
(see the Appendix \ref{App1}):
\begin{align}
\mathcal H_1 = C \ln^{1/2} (|\sigma|^{-1}) \ \mbox{as} \ |\sigma| \to 0,         \nonumber \\
\mathcal H_1= |\sigma|^{1/2} + \mathrm{e.s.t.} \ \mbox{as} \ |\sigma| \to \infty. \label{ud}
\end{align}
Here $C$ is a positive constant and $\mathrm{e.s.t.}$  denotes a positive term that is smaller than $|\sigma|^{-n}$ with any $n>0$. 
The exponential approach of $\mathcal H_1$ to $|\sigma|^{1/2}$ as $|\sigma| \to \infty$, is  clearly visible in Fig. \ref{L1E1}(c).

\begin{figure}  
      \includegraphics*[width=0.5\linewidth]{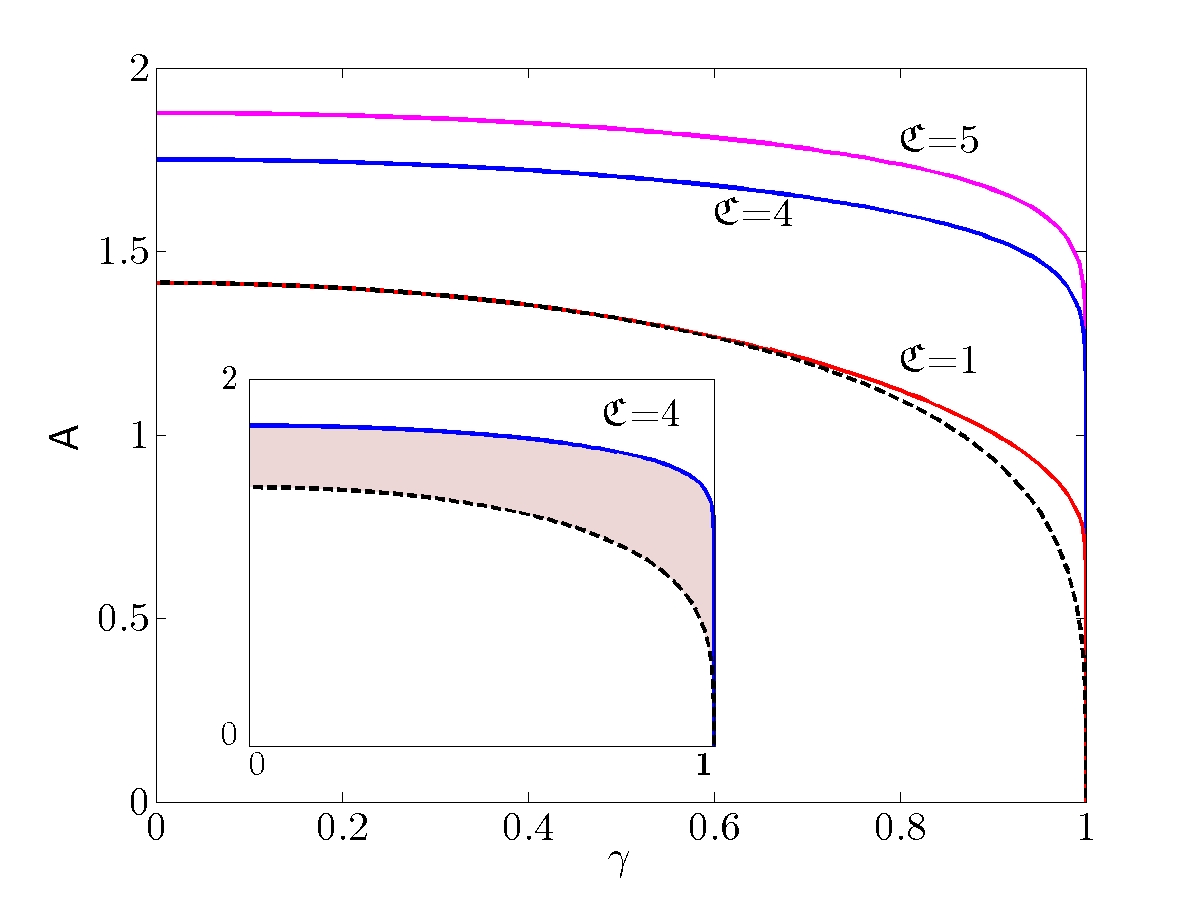}  
          \caption{The odd-mode instability band of the atiphase soliton established by 
       the analytical argument of section \ref{odd_mode}.           
    The  solid curves demarcate the upper boundary of the full odd-mode instability range, $A = \mathcal A_1(\C, \gamma)$,
   for three different values of $\C$.
 The dashed curve is given by $A= \sqrt 2 (1-\gamma^2)^{1/4}$; this curve is common for all $\frak C$.
  (Note that the  full  odd-mode instability domain extends below the dashed curve.)       
       The instability band \eqref{odd_inst} is tinted in the inset where only the solid curve with $\frak C=4$ has been retained.      
 }
 \label{H1_ga}
 \end{figure}

 The region \eqref{odd_inst} is displayed in 
Fig.\ref{H1_ga}   for three values of coupling $\C$: $\C=1$, $4$, and $5$.
It is important to emphasise here that the region   \eqref{odd_inst} constitutes 
 only {\it a part\/} of the full odd-mode instability range 
(namely,  the part where this instability can be established by analytical means).
  While the inequality
 $A < \mathcal A_1(\C,\gamma)$ does demarcate the top boundary 
 of the full odd-mode instability range, 
  the bottom boundary of the full range 
 lies somewhere below  the curve
$A= \sqrt{2} (1-\gamma^2)^{1/4}$ and can only be demarcated numerically. (See section \ref{evtr} below.)

As $\frak C \to 0$ and hence $|\sigma| \to \infty$, the equation \eqref{ud} indicates that 
 the width of the instability band  \eqref{odd_inst} becomes exponentially small.
(For instance, 
   the solid curve corresponding to $\C=0.1$
   would be indistinguishable from the dashed curve 
    in Fig.\ref{H1_ga}.)
   This argument alone does not yet imply that the full odd-mode instability domain on the $(\gamma, A)$-plane should shrink.
   However, the numerical study of symplectic eigenvalues demonstrates that this is exactly what happens:
   the width of the full odd-mode instability domain rapidly approaches zero as $\C \to 0$.

\subsection{Weakly coupled chain away from symmetry breaking: Analytical stability 
criterion}
\label{wcsa}

Defining
\[
P_+= -(L_-+ \sigma I) , \quad
P_-= - (L_+ + \sigma I),
\]
the eigenvalue problem \eqref{B8} is cast in the form 
 \be
 \label{bb8}
 \left( \begin{array}{cc}
 P_+  & 0 \\
 0 & P_-
 \end{array}
 \right)
 \left(
 \begin{array}{c}
 {\bf f} \\ {\bf g} \end{array}
 \right) = 
 \lambda J 
  \left(
 \begin{array}{c}
 {\bf f} \\ {\bf g} \end{array}
 \right). 
 \ee
 Let $|\sigma| > 4$.
 If $h^2<|\sigma|-4$, or, equivalently,
  \be
 A^2 < 2 \sqrt{1-\gamma^2} -4\C,    \label{W2}
 \ee
  both matrices $P_+$ and $P_-$ are positive
 definite. Hence the variational principle \eqref{A13} is valid
 and the lowest eigenvalue  $-\lambda^2$ is positive:
 \be
-\lambda^2 = \min \frac{\langle {\bf g} |P_+ | {\bf g} \rangle}{\langle {\bf g} |P_-^{-1}  |{\bf g}  \rangle}>0.
\label{A130}
\ee
  
 The upshot of this elementary consideration is that if the chain parameters satisfy
 \be
 \C < 1/2, \quad
  0 \leq  \gamma \leq  \sqrt{1-4 \C^2},      \label{W1} 
  \ee
the antiphase soliton has a range of stable amplitudes given by \eqref{W2}.


 It is important to emphasise that 
 the region described by the inequalities \eqref{W1} and \eqref{W2}
 constitutes only a part of the full stability domain
 of the antiphase soliton; see Fig \ref{C08}(a). The significance of this region  lies in the fact that 
  stability  is guaranteed by an accurate analytical argument here. Unlike the numerical eigenvalue analysis, 
  this argument would have not missed even a weak instability --- if there had been any.

  \subsection{Surprising stability of large-amplitude solitons}

\begin{figure}

       \includegraphics*[width=0.4\linewidth]{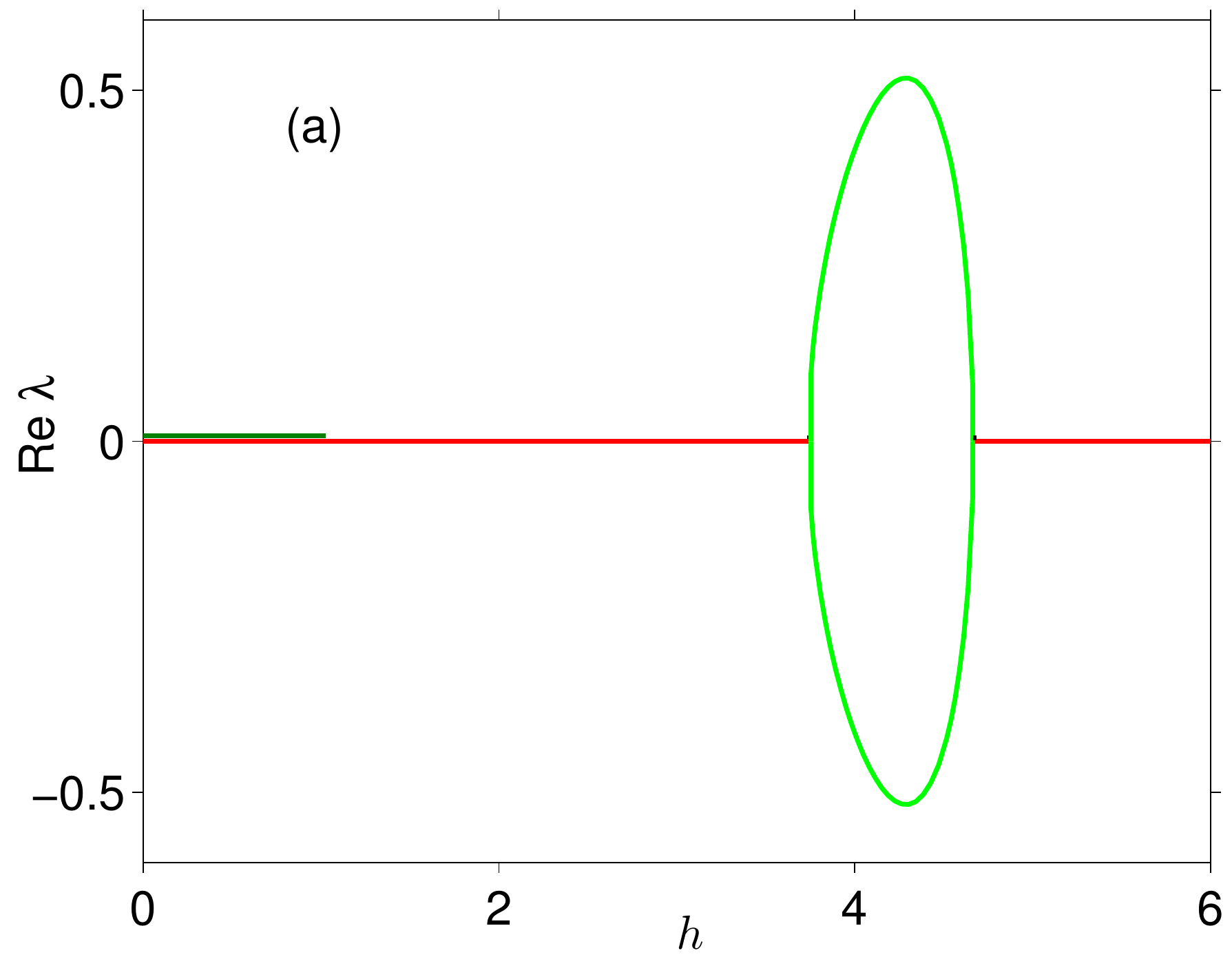}        
       \hspace*{5mm} 
  \includegraphics*[width=0.4\linewidth]{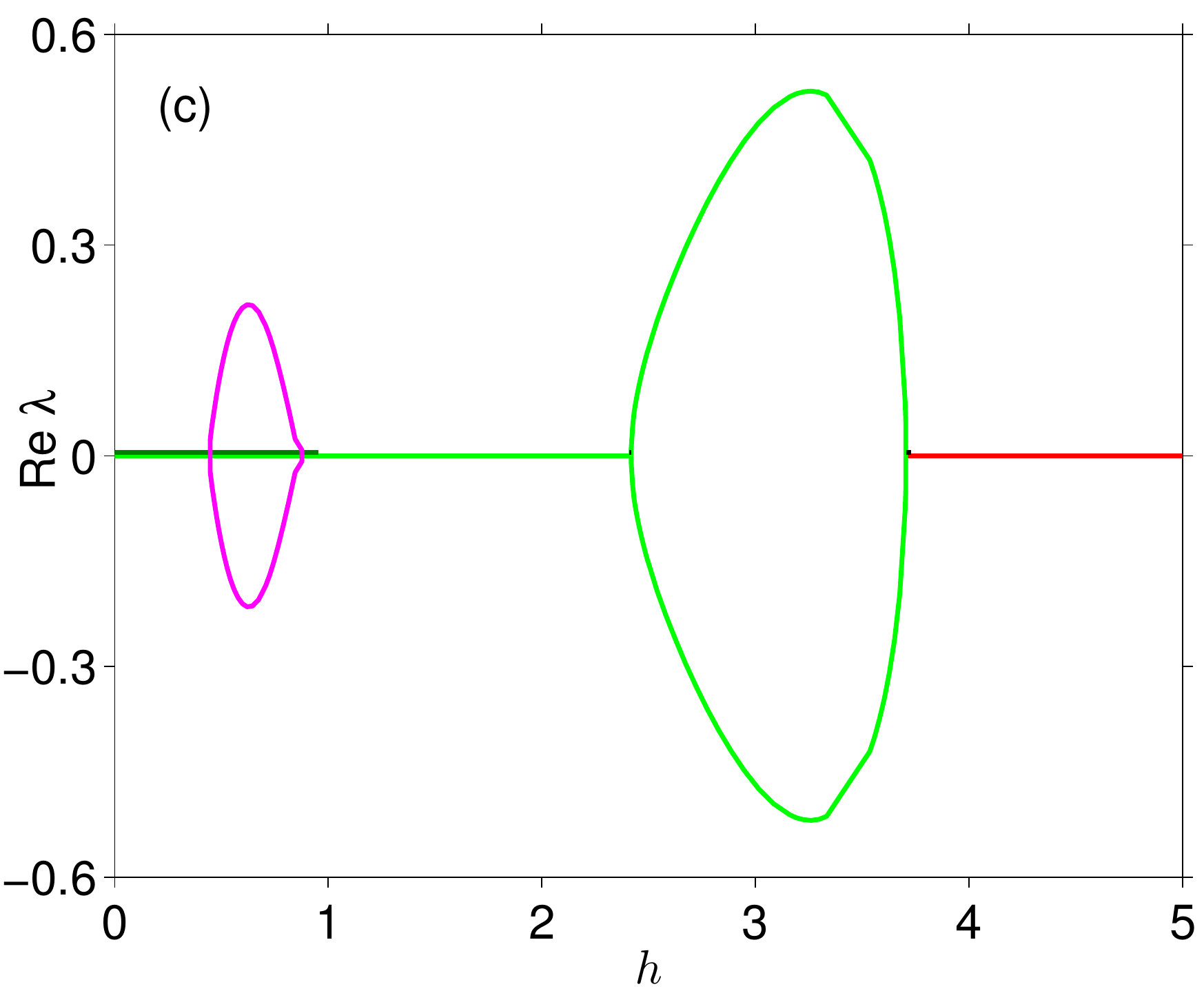}

    \includegraphics*[width=0.4\linewidth]{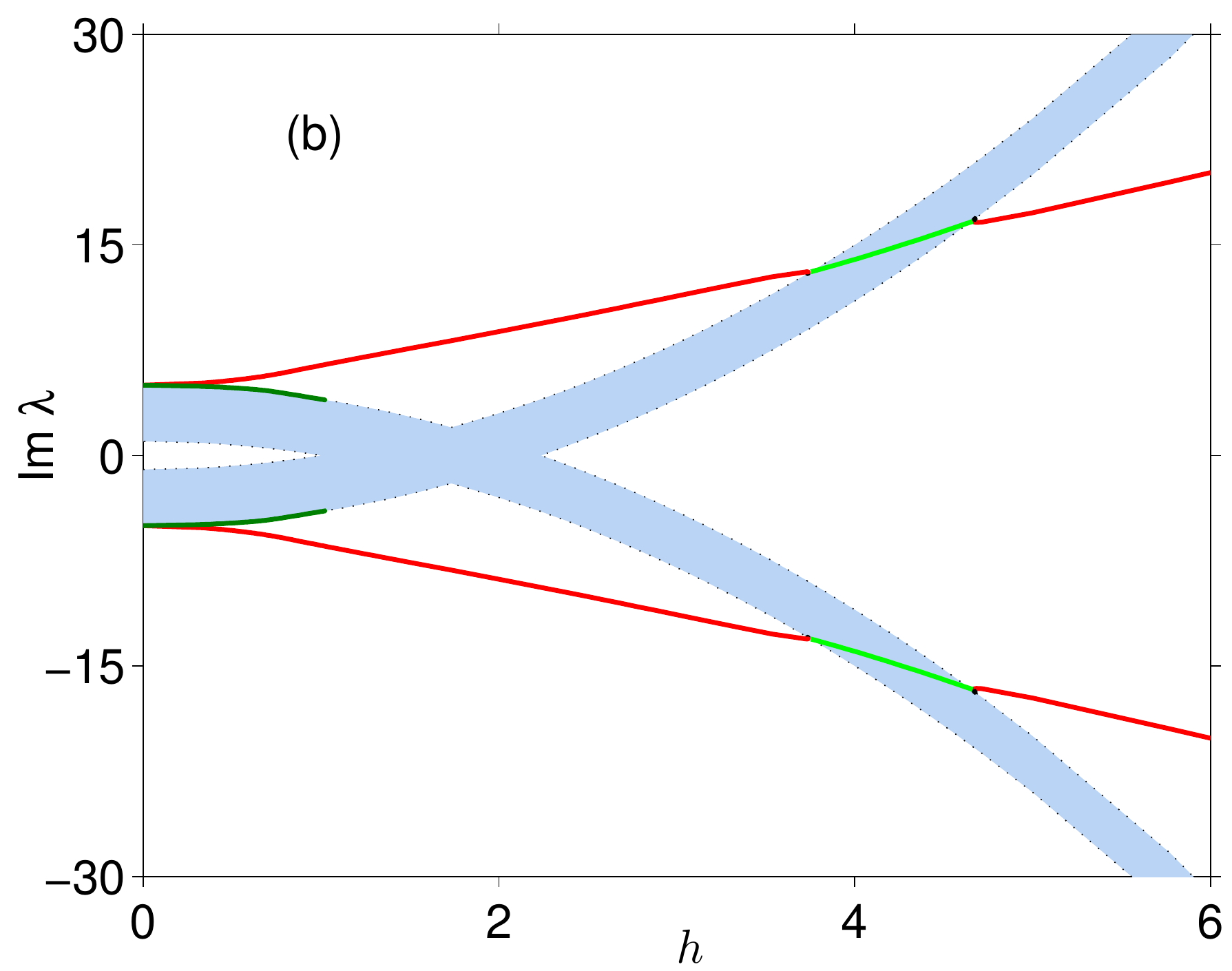}     
    \hspace*{5mm}
     \includegraphics*[width=0.4\linewidth]{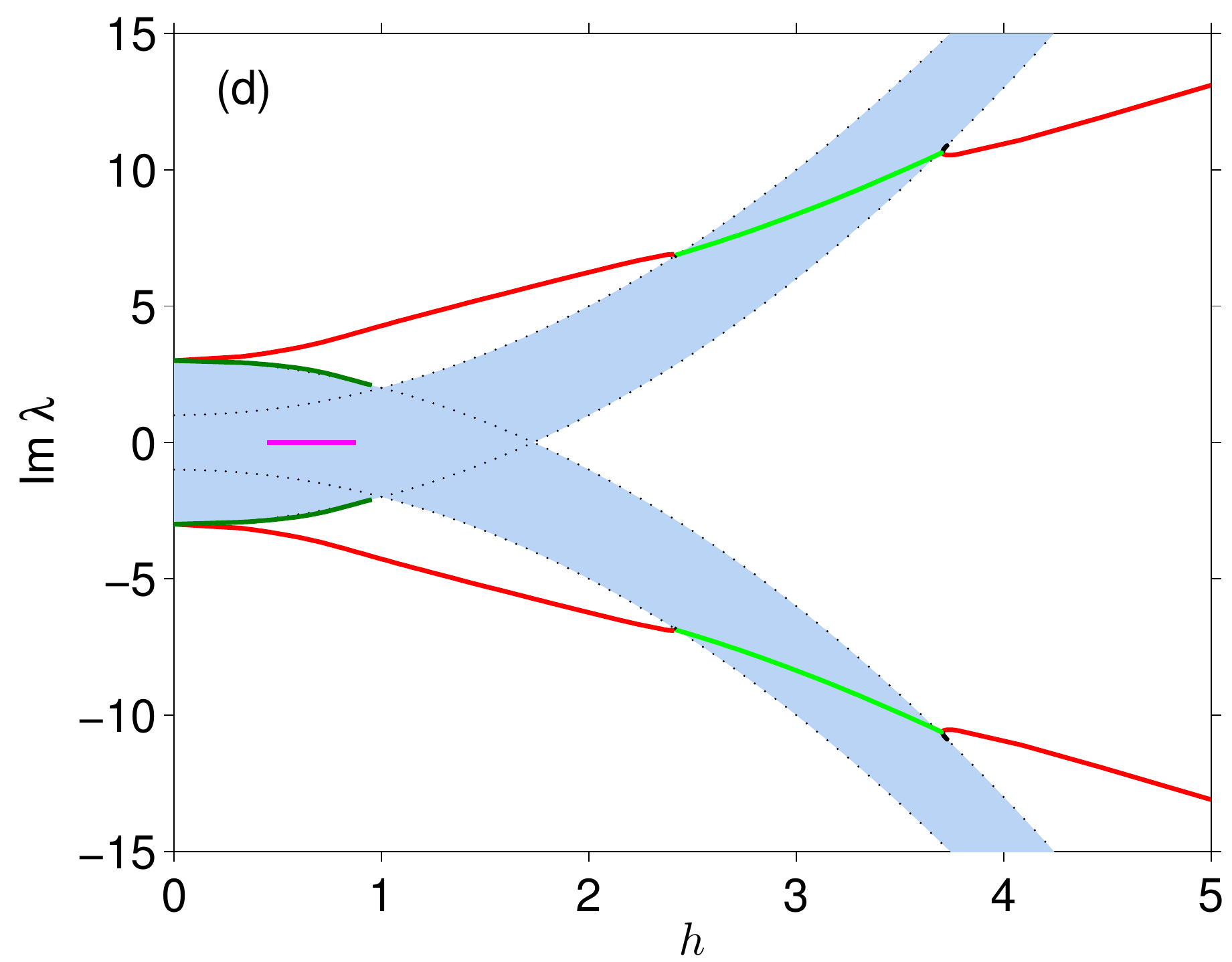} 
        
    \caption{The real (top) and imaginary (bottom) part of the symplectic eigenvalue $\lambda$   as a function of the normalised amplitude $h$. Left column: $\sigma=-5$;
   right column:      $\sigma=-3$. Each of the two red lines 
in (b,d)  follows an imaginary eigenvalue  bifurcating from a  continuous spectrum band at $h=0$.
The eigenvalue
   grows in absolute value and acquires a real part as it traverses the other continuum  band.
   One more pair of imaginary eigenvalues (shown in dark green)
   bifurcates from the continuous spectrum at $h=0$ but then returns to the continuum. 
    The pink oval in (c)  describes a  pair of real eigenvalues.
       }
 \label{Min5}
 \end{figure}

The two regions \eqref{odd_inst} and \eqref{W1},\eqref{W2}  exhaust all parameter values where the stability 
of the antiphase solitons can be classified using exact analytical criteria.
In the complementary part of the parameter space, one has to resort to numerical and asymptotic techniques.

Results of our numerical study of the eigenvalue problem \eqref{B8}
with negative $\sigma$  are summarised in Figs \ref{Min5} and \ref{Min01}.

 A striking difference between these two figures  on the one hand --- and stability properties of the corresponding continuous system \cite{ABSK} on the other --- is in the 
discrete soliton stabilisation as its amplitude becomes large enough.
(We remind the reader that once the coupling constant $\C$ has been fixed, the amplitude can be represented by 
the quantity  $h=A/ \sqrt{\C}$.)
 Indeed, all we see in Figs \ref{Min5} and \ref{Min01}  with a sufficiently large $h$ 
 is a pair of opposite pure imaginary eigenvalues.
 These eigenvalues are given by the power expansion  \eqref{J9} with coefficients as in 
\eqref{J10}:
\be
\lambda = \pm    i \sqrt{2 |\sigma|}  \left[ h+ \left( 1-\frac{\sigma}{4} \right) h^{-1} + O(h^{-3})  \right]. \label{div}
\ee

As $h$ is decreased from large values, each of the two  imaginary eigenvalues seems to collide with the corresponding branch of the continuous spectrum and acquire a
real part.
A scrutiny  of the collision area reveals that the imaginary eigenvalue approaching the continuum
collides not with the edge of the continuous-spectrum band 
 itself but with another pure imaginary eigenvalue that 
bifurcates from the edge just before the collision.
The distance in $h$ between the point of bifurcation and point of collision is so short
that this second eigenvalue is hardly discernible for large $|\sigma|$
(Fig \ref{Min5}). However  it is clearly visible when $|\sigma|$ is chosen smaller than $1$
(Fig \ref{Min01}).

The collision is nothing but the Hamiltonian Hopf bifurcation. It results in a quadruplet of complex eigenvalues $\pm \lambda, \pm \lambda^*$.
In what follows, the bifurcation value of $h$ is denoted  by $\mathcal H_3(|\sigma|)$.
Recalling that $h=A/\sqrt{\C}$ and $\sigma=-2\omega_0/\C$,
this gives the upper boundary of the soliton instability domain:
\be
 \mathcal A_3= \sqrt{\frak C} \mathcal H_3 \left( 2 \sqrt{1-\gamma^2}/{\frak C} \right),
\label{A3h3}  \ee
where the function $\mathcal H_3 (|\sigma|)$ admits a simple asymptote \cite{PY}:
\[
\mathcal H_3   \to     2|\sigma|^{1/2} + \left(
\frac{\sqrt{15}}{4} -\frac12 \right) |\sigma|^{-1/2} + O(|\sigma|^{-3/2})
 \ \mbox{as} \ |\sigma| \to \infty.
\]
The asymptotic behaviour for  the upper boundary of the soliton instability domain is then
\be
\mathcal A_3 \to   2 \sqrt{2}  (1-\gamma^2)^{1/4}  \left[ 1-\frac18 \left(1-\frac{\sqrt{15}}{2} \right) \frac{\C}{\sqrt{1-\gamma^2}}+
O\left( \frac{\C^2}{1-\gamma^2} \right) \right]
 \ \mbox{as} \ 
 \frac{\C}{\sqrt{1-\gamma^2}}  \to 0. \label{A_2} 
\ee

   \subsection{Small amplitude solitons: stable or long-lived}  
   \label{small_h}

\begin{figure}

   \includegraphics*[width=0.4\linewidth]{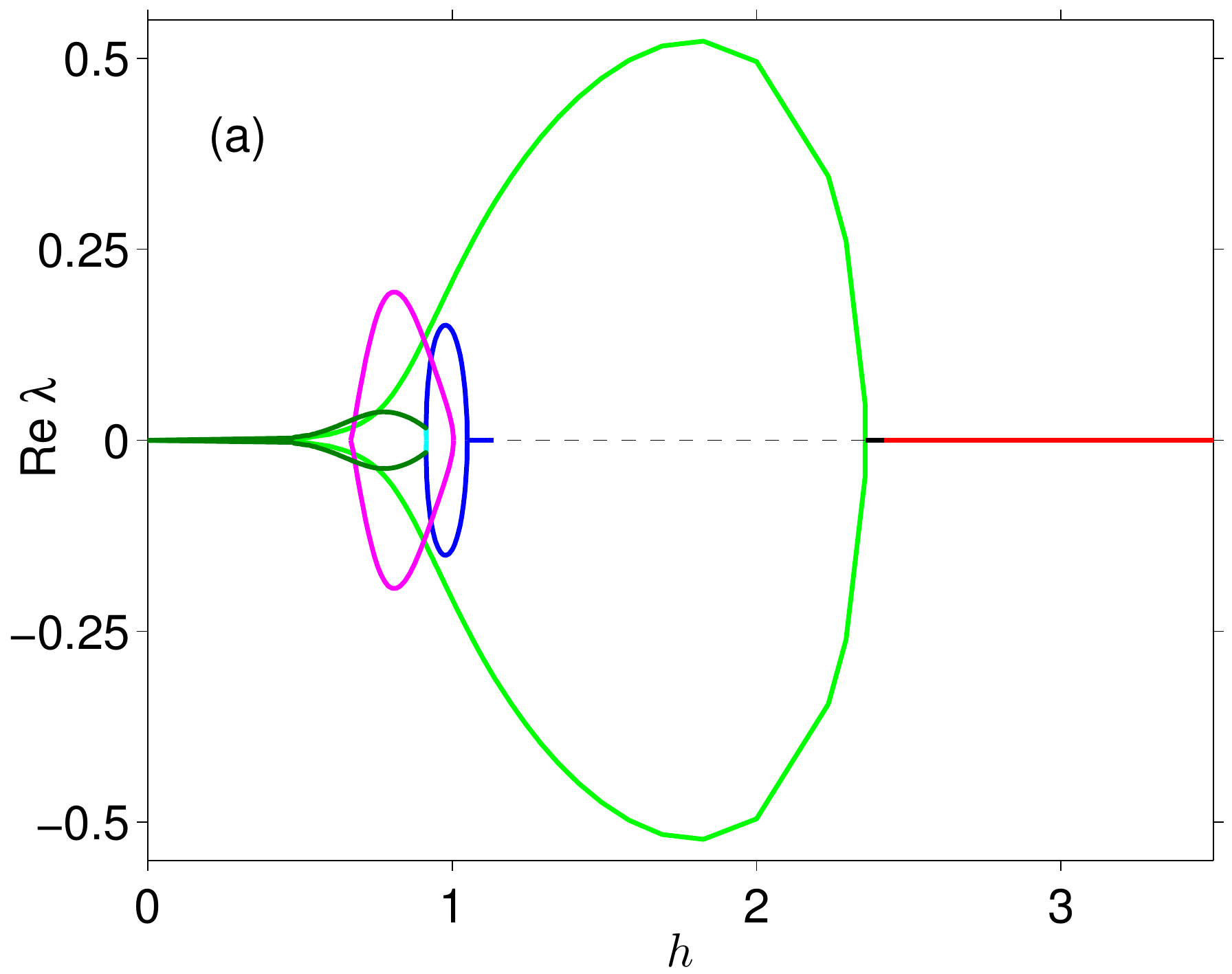} 
   \hspace*{5mm}
    \includegraphics*[width=0.4\linewidth]{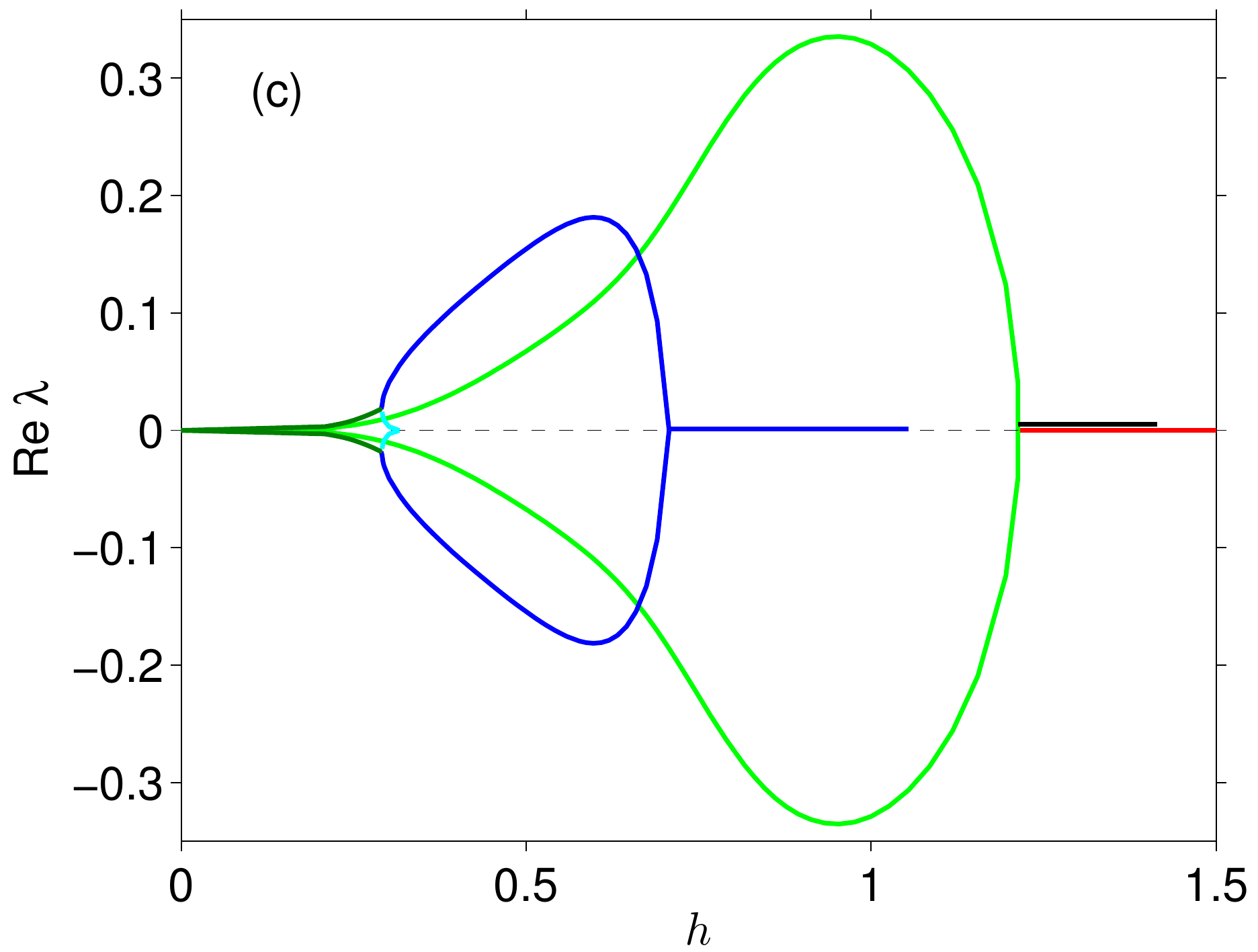}
      \includegraphics*[width=0.4\linewidth]{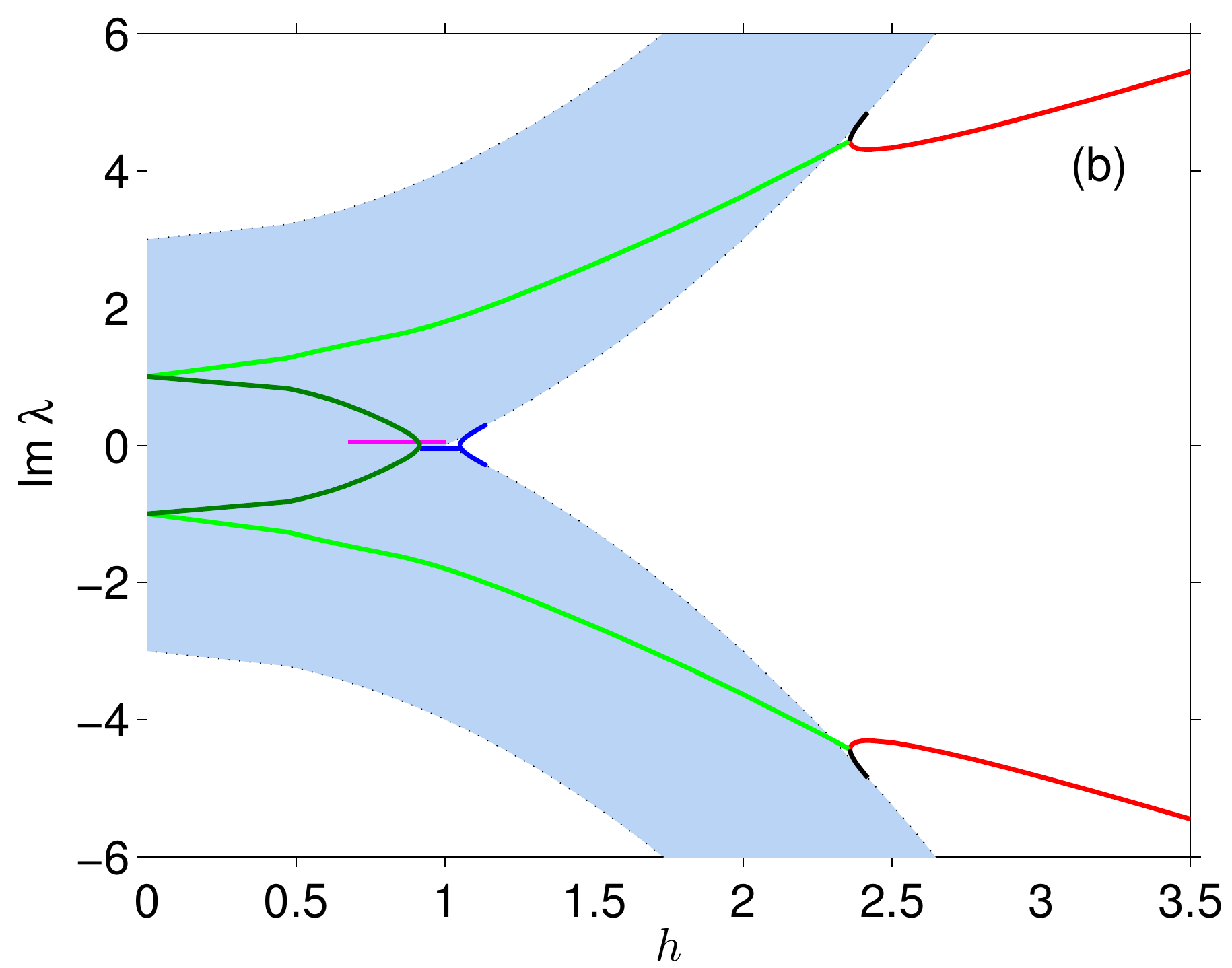}
      \hspace*{5mm}
        \includegraphics*[width=0.4\linewidth]{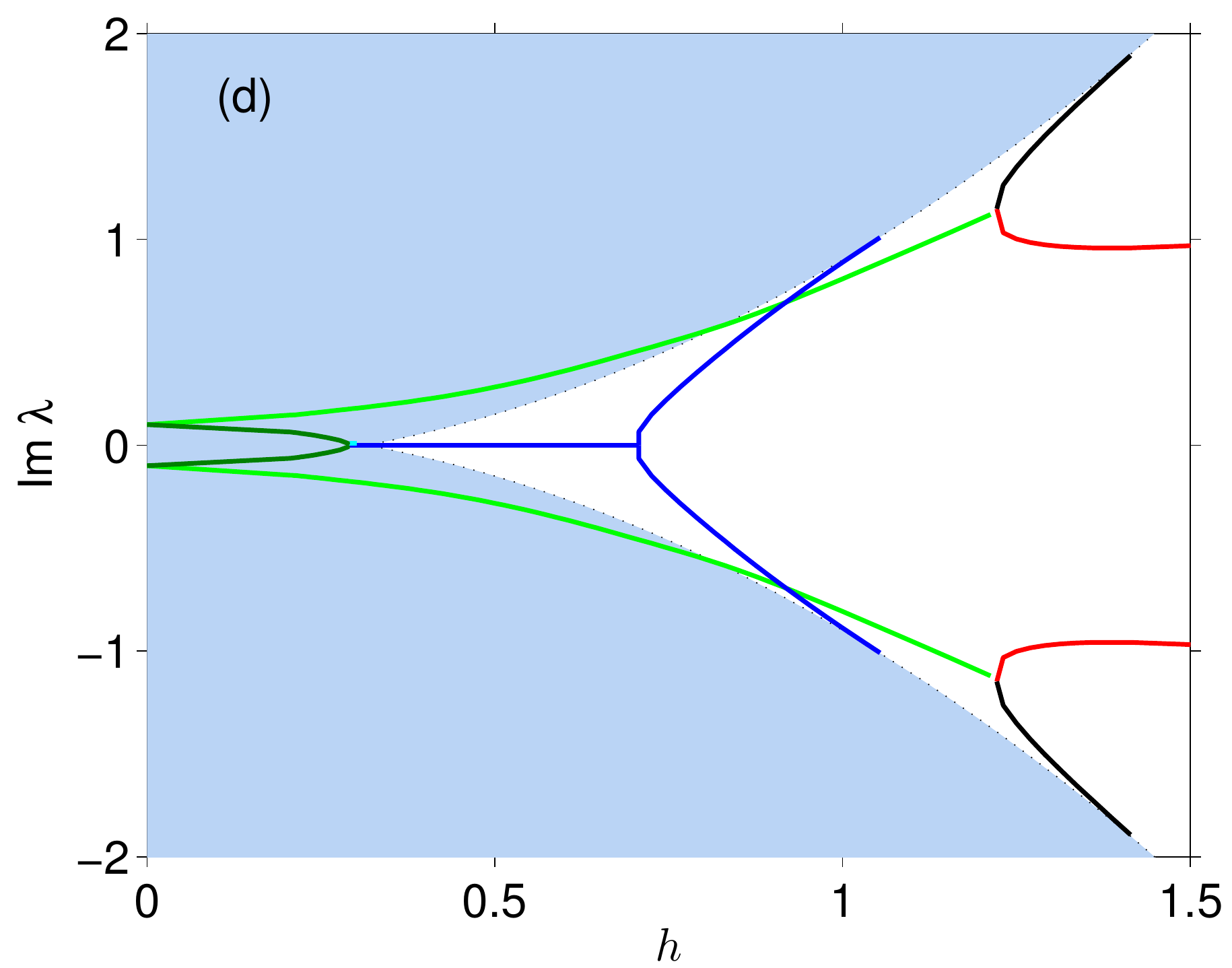}  
      \caption{The  real  and imaginary part of the eigenvalue $\lambda$ of the
   problem  \eqref{B8} as a function of the normalised amplitude $h$. Left column:  
   $\sigma=-1$;
 right column:  $\sigma=-0.1$. 
 Two light-green curves trace a complex quadruplet with even eigenvectors while 
 the dark-green arc marks complex eigenvalues with odd eigenvectors.
 As the imaginary parts of the ``even" quadruplet emerge from the continuous spectrum, the quadruplet dissociates into two pairs of
 imaginary eigenvalues
 described by the black and red curves.
 The ``odd" quadruplet also dissociates  as $h$ is increased; the products include two pairs of real eigenvalues shown in  light-blue and deep-blue, respectively. 
 The ``light-blue" eigenvalues converge on the origin and disappear in the gapless continuous spectrum. 
 The ``deep-blue" pair moves away from the origin but then reverses and crosses to the imaginary axis as well.
By the time the ``deep-blue" doublet reaches the origin, a gap will have opened  in the continuous spectrum.
The doublet reaches the boundaries of the gap
 and  immerses in the continuum. 
 The pink oval in (a)  represents a  pair of real eigenvalues with even eigenvectors.
}
 \label{Min01}
 \end{figure}

  The   limit $h \to 0$, with small or  finite $|\sigma|$ ($\sigma<0$),  is also  amenable to asymptotic analysis.
  See Appendix \ref{App6} and \ref{App4}, respectively.
As $h$ grows from zero, 
 either two pairs of pure imaginary or 
 two quadruplets of complex eigenvalues bifurcate from the endpoints of the continuous spectrum
 (more specifically, from the $\mathrm{Im} \, \lambda = \pm \sigma$ endpoints).
 Whether the bifurcating eigenvalues have real parts or not, 
 their  imaginary parts are given by equations \eqref{finlam} and \eqref{AD1}:
\begin{align*}
 \mathrm{Im} \, \lambda_{\mathrm{even}}= \pm 
 (\sigma- 1.438 h^2)+ O(h^4/ \sigma),  \\  
 \mathrm{Im}  \,  \lambda_{\mathrm{odd}}= \pm (\sigma+ 0.685 h^2) + O(h^4/\sigma).  
\end{align*}

 Numerically, the bifurcating imaginary eigenvalues are clearly visible 
 when $|\sigma| > 2$. See  e.g. Fig \ref{Min5}(b) for $\sigma=-5$ and Fig \ref{Min5}(d) for $\sigma=-3$.
  When $|\sigma| <2 $, by contrast, the point $\mathrm{Im} \, \lambda=  \sigma$ --- the bottom  of the upper  band of the continuous spectrum --- is 
 embedded in the lower band, which covers the interval $-\sigma -4 \leq \mathrm{Im} \, \lambda \leq -\sigma$. 
 (Compare the panel (b) of Fig \ref{cont} with panels (c) and (d).)
  In this case, the bifurcating   eigenvalues with small or zero real parts
 are concealed in the continuum bands
 and are much harder to  track. 

 The asymptotic analysis of the problem \eqref{B8} with a negative $\sigma$ of order one,
provides no clue on
 whether the eigenvalues 
 bifurcating from the edge of the continuum  have real parts or not.
 The only conclusion one can draw from it is that 
if an eigenvalue has a nonzero real part, this real part will have to lie beyond all orders of $h^n$.
(The proof of this fact is in Appendix \ref{App4}.)
  The asymptotic argument in Appendix \ref{App6} is of more use here as it
  guarantees the existence of an (exponentially small) real part 
when  $|\sigma|$ is small enough.
 Also useful is 
  the variational principle of section \ref{wcsa}; this principle ensures that  all eigenvalues of the operator \eqref{symplec}   with  small $h$ are pure imaginary if  $|\sigma|>4$.
Taken together, these two considerations suggest that there is a critical value $\sigma_c$ 
($|\sigma_c| \leq 4$) such that 
the bifurcating eigenvalues acquire an exponentially small real part when $|\sigma|< |\sigma_c|$
but remain pure imaginary when $|\sigma| \geq |\sigma_c|$.

Our numerics support this conjecture and indicate that $\sigma_c$ is close  to $-1.96$. 
For all  $|\sigma| \geq 1.96$ that we have examined, 
the two pairs of bifurcating eigenvalues  were  found to be pure imaginary.
They only develop nonzero real parts once their imaginary parts collide with the continuous spectrum bands at  some finite $h$.
(Note the square-root dependence of $\mathrm{Re} \, \lambda$  on $h$ in  the vicinity of the collision point in Fig \ref{Min5} (a) and (c).)

When $|\sigma| <1.96$, by contrast,  eigenvalues corresponding to small $h$ exhibit 
nonzero real parts. In this case, 
 the shape of the $\mathrm{Re} \, \lambda(h)$ curve is indicative of an exponential decay  as $h \to 0$.
 See  Fig \ref{Min01} (a,c) and note the difference in the behaviour of the curve from the 
square-root law near the collision point in Fig  \ref{Min5} (a,c).

The proximity of the critical value $\sigma_c$ to $-2$ sheds some light on the source of the weak oscillatory instability.
Unlike the case $|\sigma|>2$, the imaginary parts of  the  bifurcating  eigenvalues of  the operator \eqref{symplec}  with $|\sigma| \leq 2$
 are embedded in the continuous
spectrum right from the moment they were born, that is, as early as at  $h=0$.
Consequently, 
the emergence of a nonzero real part is  due to the resonance between the frequency corresponding to the 
imaginary part of the bifurcating eigenvalue, and frequencies of the continuous spectrum.

In summary, the small-amplitude antiphase solitons are stable if $|\sigma| \geq1.96$
or, equivalently, when $\frak C < 1.02$ and $\gamma \leq \sqrt{1-0.96 \frak C^2}$.
Conversely, in the region $|\sigma| < 1.96$ (that is, when $\frak C  > 1.02$ and $0 \leq \gamma \leq 1$  or when $\frak C \leq 1.02$ and $\gamma > \sqrt{1-0.96 \frak C^2}$),
these solitons manifest weak instability with growth rate exponentially small in $h$. 
In the latter case, solitons with sufficiently small amplitudes live long enough to be deemed {\it practically stable}.

\subsection{Weakly coupled chain away from $\mathcal{PT}$-symmetry breaking: disjoint stability domain}

When the structural parameter $|\sigma|$ is greater than   $4$ (that is, when $\C< 1/2$
  and $0 \leq \gamma \leq \sqrt{1-4 \C^2}$), 
the complex quadruplet associated with the  continuous-band crossing is the 
only set of unstable eigenvalues that arise as $h$ varies from $0$ to $\infty$. See Fig \ref{Min5} (a,b). In this case, 
as $h$ is decreased  from $\mathcal H_3$ (the point of collision with the ``inner" edge of the continuum), the imaginary part of the 
eigenvalue travels through the continuous band and crosses  through its ``outer" edge. At the point where the imaginary part
emerges from the continuum,
the real part of the eigenvalue appears to vanish. We denote this point  by $h=\mathcal H_2(|\sigma|)$
and the corresponding value of the amplitude  by $\mathcal A_2$:
\be
\mathcal A_2= \sqrt{\frak C} \mathcal H_2 \left( {2 \sqrt{1-\gamma^2}}/{\frak C} \right).
\label{A2h2}
\ee
The function $\mathcal H_2$ admits a simple asymptotic representation \cite{PY}:
\[
\mathcal H_2    \to   2|\sigma|^{1/2} - \left(
\frac{\sqrt{15}}{4} +\frac12 \right) |\sigma|^{-1/2} + O(|\sigma|^{-3/2})
 \ \mbox{as} \ |\sigma| \to \infty.
\]
This gives an asymptotic expression for 
the lower boundary of the instability region:
\be
\mathcal A_3 \to   2 \sqrt{2}  (1-\gamma^2)^{1/4}  \left[ 1-\frac18 \left(1 +\frac{\sqrt{15}}{2} \right) \frac{\C}{\sqrt{1-\gamma^2}}+
O\left( \frac{\C^2}{1-\gamma^2} \right) \right]
 \ \mbox{as} \ 
 \frac{\C}{\sqrt{1-\gamma^2}}  \to 0. \label{A_1} 
\ee

As with the inner-edge crossing, the
 numerical scrutiny confirms that this process is more complex than it seems.
 It involves two pairs of complex eigenvalues which converge on two opposite points on the imaginary axis
and dissociate into two  pure imaginary pairs. After that, one pair of opposite imaginary eigenvalues immerses in the continuum 
whereas the other pair moves away from it. 
The interval of $h$ where two imaginary pairs coexist is so tiny that 
 the complex eigenvalues appear to be stripped of their real parts as soon as  the imaginary parts have emerged out of the continuum.

  While considering diagrams pertaining to 
  large $|\sigma|$, 
a natural question  concerns the odd-mode instability of
section \ref{odd_mode}.
The domain of the odd-mode instability includes (though is not limited to)
the interval
 $|\sigma|^{1/2} < h < \mathcal H_1(|\sigma|)$.
Why is this instability not manifest 
 in the $\lambda(h)$ dependencies with large $|\sigma|$
 (Fig.\ref{Min5})?

 The answer is that since the function $\mathcal H_1(|\sigma|)$  approaches $|\sigma|^{1/2}$ exponentially fast 
as $|\sigma|$ grows, the above interval of unstable $h$ shrinks and becomes indiscernible (see Fig \ref{L1E1}(c)).
Furthermore, the numerical analysis indicates that the {\it full\/}  odd-mode instability domain also shrinks  as $|\sigma|$ grows.
Therefore, although the solitons with large negative $\sigma$ are prone to the odd-mode instability, 
the interval of   ``unstable" $h$ is minuscule.

The shrinking of the odd-mode instability band \eqref{odd_inst} 
 as the coupling $\C$ is reduced,
is clearly visible in Fig \ref{H1_ga}. 
In the weakly coupled chain of waveguides, the numerical detection of this band is only feasible 
if $\gamma$ is close to 1
(so that the argument of $\mathcal H_1$ in \eqref{A1H1} is small enough).

Our conclusions  can
be summarised as follows.
The stability domain in a  chain  with $\C< 1/2$
  and $0 \leq \gamma \leq \sqrt{1-4 \C^2}$
    consists of two disjoint parts. 
 The  soliton  is  unstable if its amplitude falls between $\mathcal A_2$ and $\mathcal A_3$,
 and stable when $A$ lies outside this band.
 Here $\mathcal A_2$ is as in \eqref{A2h2} and
 $\mathcal A_3$  as in \eqref{A3h3}, 
 with the corresponding asymptotic behaviours given by 
 \eqref{A_1} and
 \eqref{A_2}, respectively.
 (There is also a narrow band
  of the odd-mode instability immersed in the stability domain but it is of little significance due to its exponentially small width.)

 The disjoint stability domain  on the   $(\gamma, A)$-plane is illustrated in Fig \ref{C08}(a).  
  The 
dashed curve in  this panel demarcates  the region where the soliton's stability has been 
established using the variational principle \eqref{A130}.  It is obvious that the analytical argument secures only a small portion of the full stability domain.

\subsection{Numerical eigenvalue trajectories}
\label{evtr}

While the complex quadruplet crossing   the
continuous spectrum band is the only set of unstable eigenvalues for   $|\sigma| > 4$, chains with
smaller  values of $|\sigma|$ feature several more instabilities.
This subsection summarises results of our numerical study of the problem \eqref{B8}  with $0<|\sigma|<4$. 
As before in this section, we keep $\sigma<0$.

The complex quadruplet exhibited by the operator \eqref{symplec} with $|\sigma|>4$ persists for $|\sigma|<4$. 
In Figs \ref{Min5} and \ref{Min01}
the trajectories of  complex eigenvalues constituting the quadruplet in question, are delineated in light green. 
Similarly to the  $|\sigma|>4$ scenario,  the quadruplet  is born in a Hamiltonian Hopf bifurcation as $h$ is decreased through the point $\mathcal H_3(|\sigma|)$.
A further decrease of $h$   with a fixed $|\sigma| \geq 1.96$
 brings about an inverse Hopf bifurcation, at the point $h=\mathcal H_2(|\sigma|)$, 
where the quadruplet  dissociates into 
two opposite pairs of imaginary eigenvalues. 
When $|\sigma|<1.96$, by contrast, the quadruplet persists over the entire interval $0< h \leq \mathcal H_3$. 
(Its asymptotic behaviour as $h \to 0$ was discussed in section \ref{small_h}.)

We  note  that the eigenvectors associated with the ``light-green" quadruplet,
are symmetric
(that is, even in $n$). The left-right symmetry  of the eigenvectors distinguishes this quadruplet  from another quadruplet of
complex eigenvalues that is shown in dark green in Fig \ref{Min01}.
The eigenvectors of the ``dark-green" eigenvalues are  antisymmetric (odd in $n$). 
 
\begin{figure}
 \includegraphics*[width=0.49\linewidth]{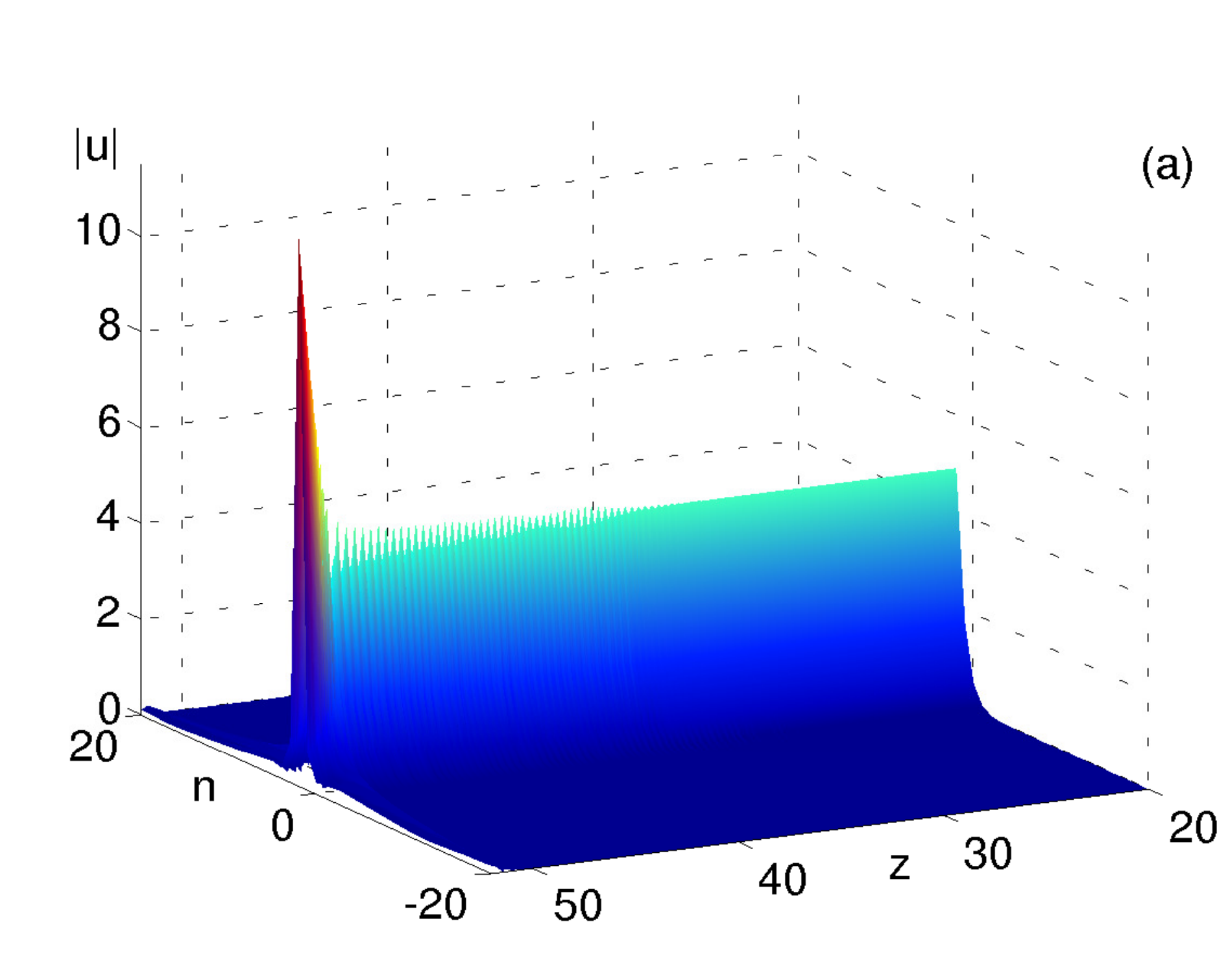} 
   \hspace*{1mm}
    \includegraphics*[width=0.49\linewidth]{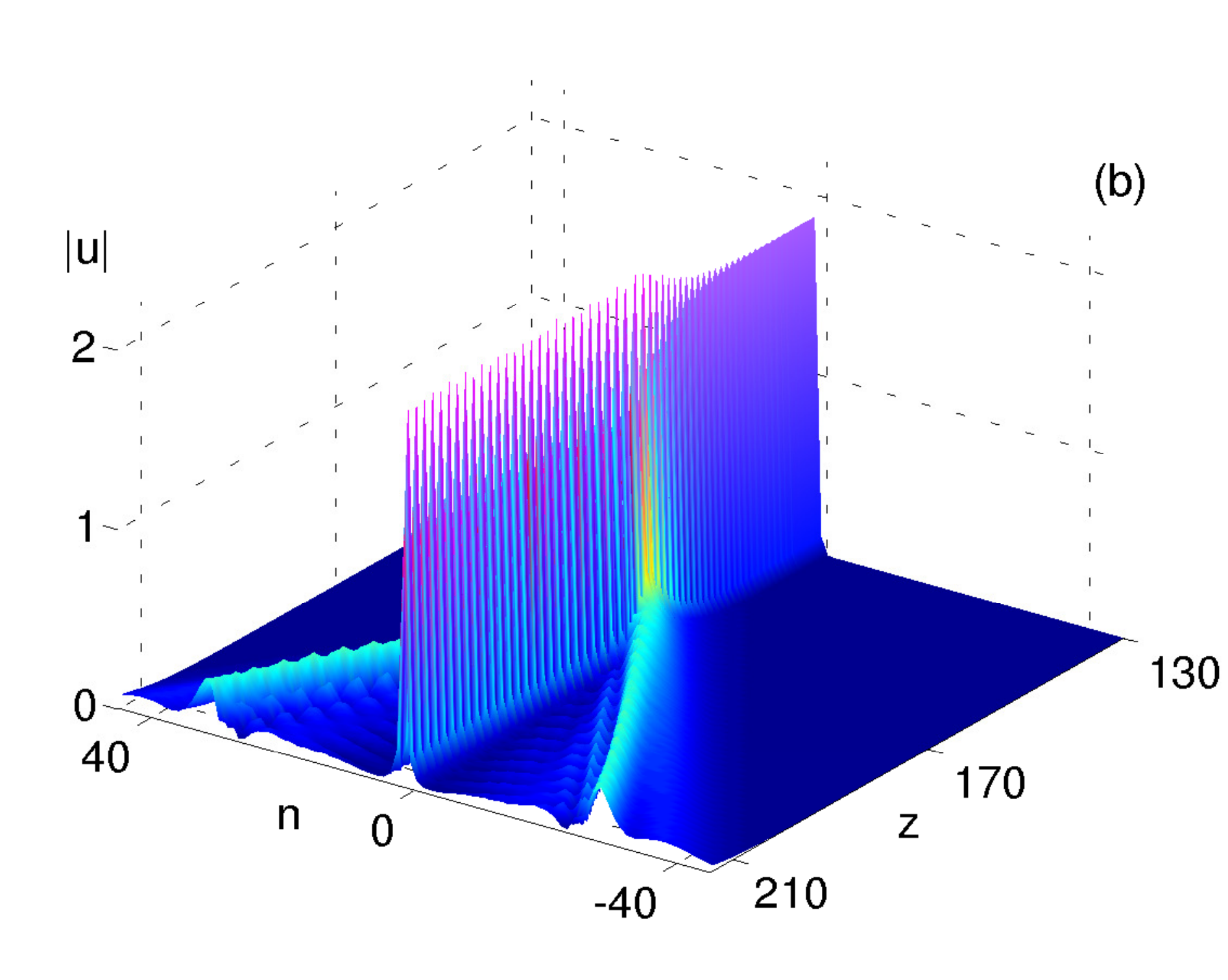} 
      \caption{The blow-up of the antiphase soliton (a) and formation of a breather (b).
            In (a), the parameters of the lattice are $\C=17.32$, $\gamma=0.5$, and the soliton's initial amplitude  is $A= 5.05$.
      These values correspond to the point $h=1.21$ in 
      Fig  \ref{Min01} (c,d)   (where $\sigma=-0.1$). 
      In (b), the lattice has $\C=0.366$ and $\gamma=0.4$; the initial amplitude of the decaying antiphase soliton is $A=2.51$.
      These parameters correspond to a point  $h=4.15$ in Fig \ref{Min5} (a,b) (where $\sigma=-5$).
      The instability responsible for the evolution shown in (a) and (b), is due to the complex quadruplet 
      marked by the light green lines in Fig \ref{Min01} and Fig \ref{Min5}, respectively.
}
 \label{blowbrea}
 \end{figure}

The left-right symmetry of the eigenvectors gives rise to a symmetric instability growth.
 Fig \ref{blowbrea} illustrates two possible outcomes of the decay of the antiphase soliton unstable against
 the ``light-green" complex quadruplet. In Fig \ref{blowbrea}  (a) the growing oscillations culminate in the 
 exponential blow-up of the soliton.
 In Fig \ref{blowbrea} (b), the oscillatory instability transforms the soliton into a breather ---
 a spatially localised structure with a steadily oscillating amplitude and width.

The pink loop in Fig \ref{Min5}(c) and \ref{Min01}(a) describes a pair of opposite real eigenvalues
(also with even eigenvectors).
These hail from the imaginary axis: as $h$ is decreased, a pair of pure imaginary eigenvalues 
bifurcates from the continuous spectrum, converges at the origin and  moves on to the real axis. 
As $h$ is decreased further, the eigenvalues 
reach their maximum absolute value and then return  to the origin.

The ``pink loop" is present in all diagrams with $|\sigma|$ between $4$ and approximately $0.4$. 
When $|\sigma| <1.96$, these real eigenvalues  
coexist with the complex quadruplet and do not produce any further reduction of the soliton
 stability domain. The stability domain 
 has a simple structure in this case: the  soliton is stable
when $h> \mathcal H_3(|\sigma|)$ and unstable otherwise.

\begin{figure}[t] 
   \includegraphics*[width=0.325\linewidth]{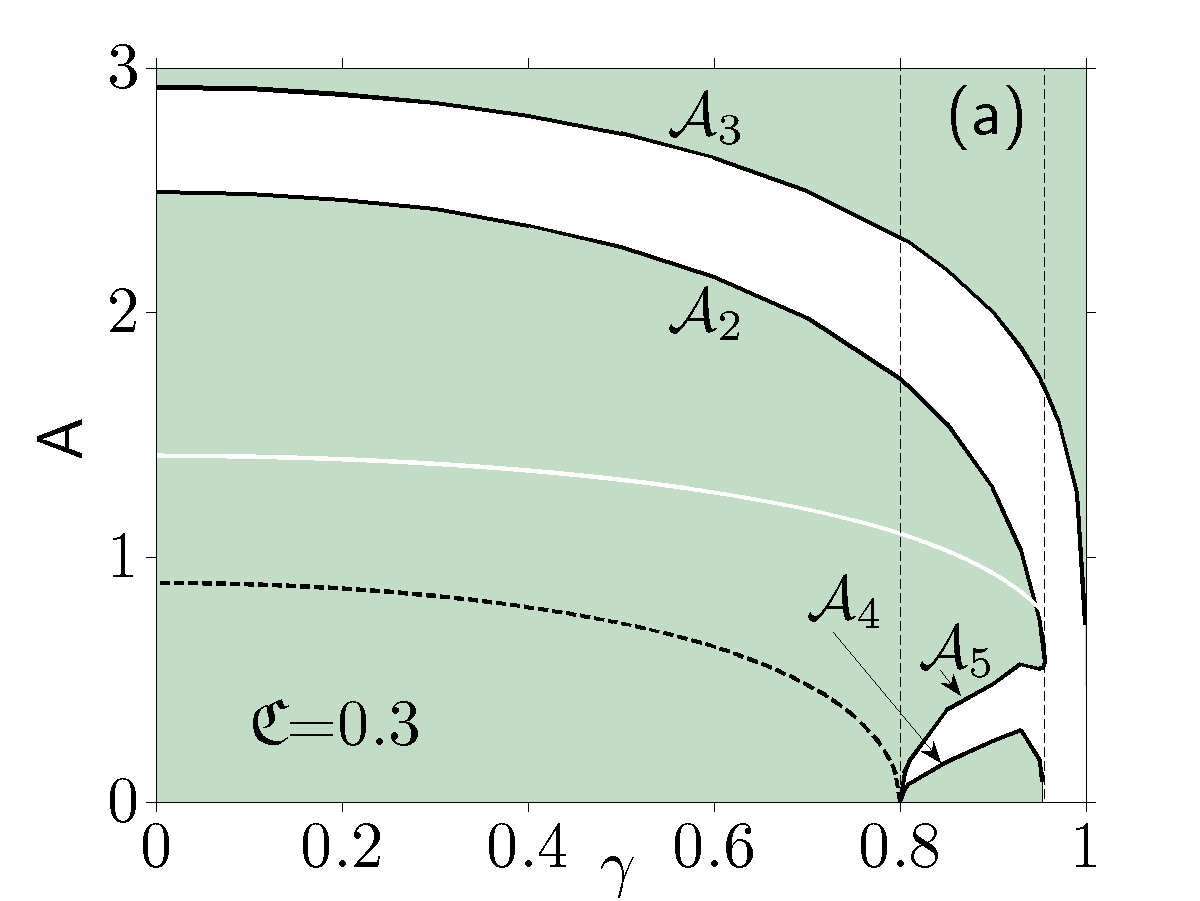}  
         \includegraphics*[width=0.325\linewidth]{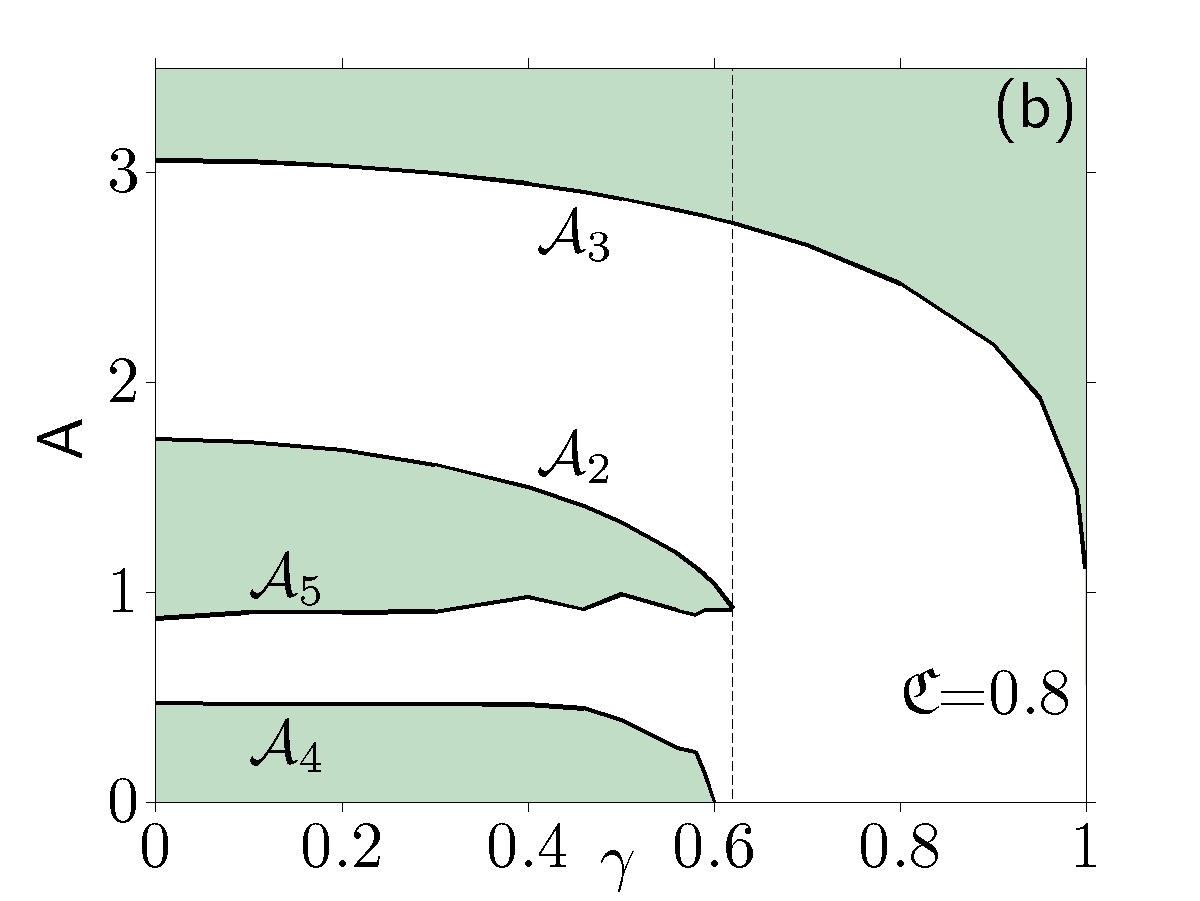}   
                 \includegraphics*[width=0.325\linewidth]{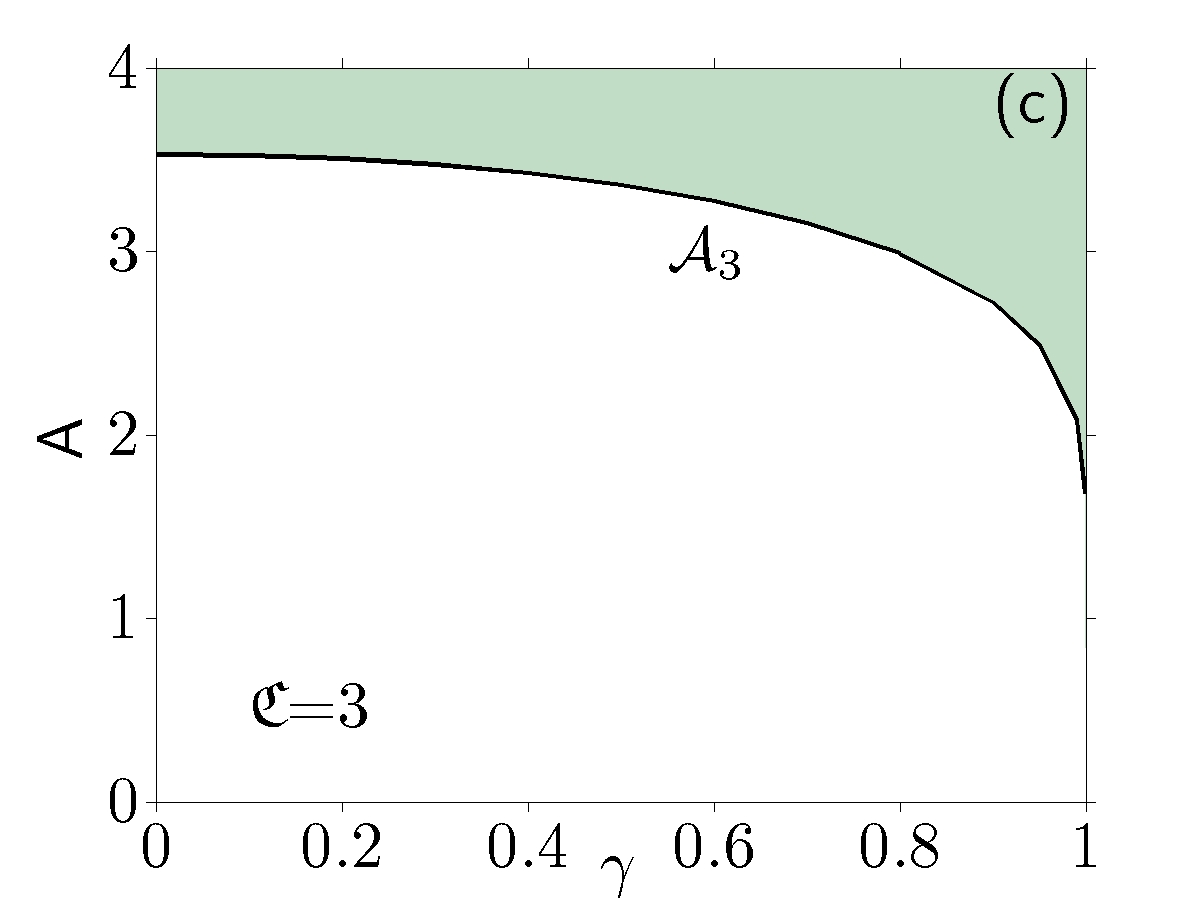}       
          \caption{
 Stability domain of the antiphase soliton in the chain with $\C<1/2$ (a),  $1/2< \C<1.02$ (b) 
          and $\C>1.02$ (c).   
              The stability domain is tinted green while  corridors of instability are left blank.
              The left vertical line in (a) is given by $\gamma= \sqrt{1-4 \C^2}$
              and the right one by $\gamma=\sqrt{1-0.96 \C^2}$.     The vertical line in (b) is  $\gamma=\sqrt{1-0.96 \C^2}$.
                               \newline      
     {\bf Panel (a):}  In the interval $0 \leq \gamma \leq \sqrt{1-4 \C^2}$,
  stable amplitudes lie on each side of  the band $\mathcal A_2< A < \mathcal A_3$.
  This corridor continues into 
   the interval $\sqrt{1-4\C^2} < \gamma \leq \sqrt{1-0.96 \C^2}$, where it is joined by the second instability band, $\mathcal A_4 < A< \mathcal A_5$.
In the interval $\sqrt{1-0.96 \C^2} < \gamma < 1$, stable amplitudes lie only above $\mathcal A_3$.
 What looks like 
   a white line through the stability domain is  a very narrow corridor of the odd-mode
   instability. The upper boundary of this band is given by $\mathcal A_1(\gamma)$; the lower boundary is determined numerically.
  The black dashed curve is given by \eqref{W2}; it bounds the region where the soliton is stable due to a rigorous analytical argument.        
    \newline
    {\bf Panel (b):}   
    When   $\gamma \leq  \sqrt{1-0.96  \C^2}$,  there are two instability bands: 
$\mathcal A_4<A< \mathcal A_5$ (lower) and $\mathcal A_2<A< \mathcal A_3$ (upper band). 
    In  the interval $\sqrt{1-0.96 \C^2} < \gamma < 1$,
         stable amplitudes lie only above $\mathcal A_3$.  
\newline
{\bf Panel (c):} 
stable amplitudes lie only above $ \mathcal A_3(\gamma)$.
 }
 \label{C08}
 \end{figure}

The inequality $|\sigma|<1.96$ describes a union of two regions on the $(\frak C, \gamma)$-plane:
\begin{subequations}  \label{conne1}
\begin{align} 
\frak C>1.02,   \quad  0 \leq  \gamma < 1;
\\
\frak C \leq 1.02,   
\quad 
  \gamma>  \sqrt{1- 0.96 \frak C^2}. 
 \end{align}
 \end{subequations}
In either case the soliton stability domain is connected:
$A \geq \mathcal A_3(\frak C, \gamma)$,
where $\mathcal A_3$ is as in \eqref{A3h3}. 
See Figs  \ref{C08}(a,b,c).

The real even (``pink") mode becomes more significant for $|\sigma|$ between $1.96$ and $4$.
When $|\sigma|$ is in this range,
 we have two separate intervals of ``unstable" values of $h$:
the interval $\mathcal H_2(|\sigma|)< h< \mathcal H_3(|\sigma|)$ bearing the complex quadruplet, and
an additional interval 
$\mathcal H_4(|\sigma|)< h< \mathcal H_5(|\sigma|)$, with $\mathcal H_5 < \mathcal H_2$,
 hosting the ``pink" pair of real eigenvalues.
 
\begin{figure}
 \includegraphics*[width=0.49\linewidth]{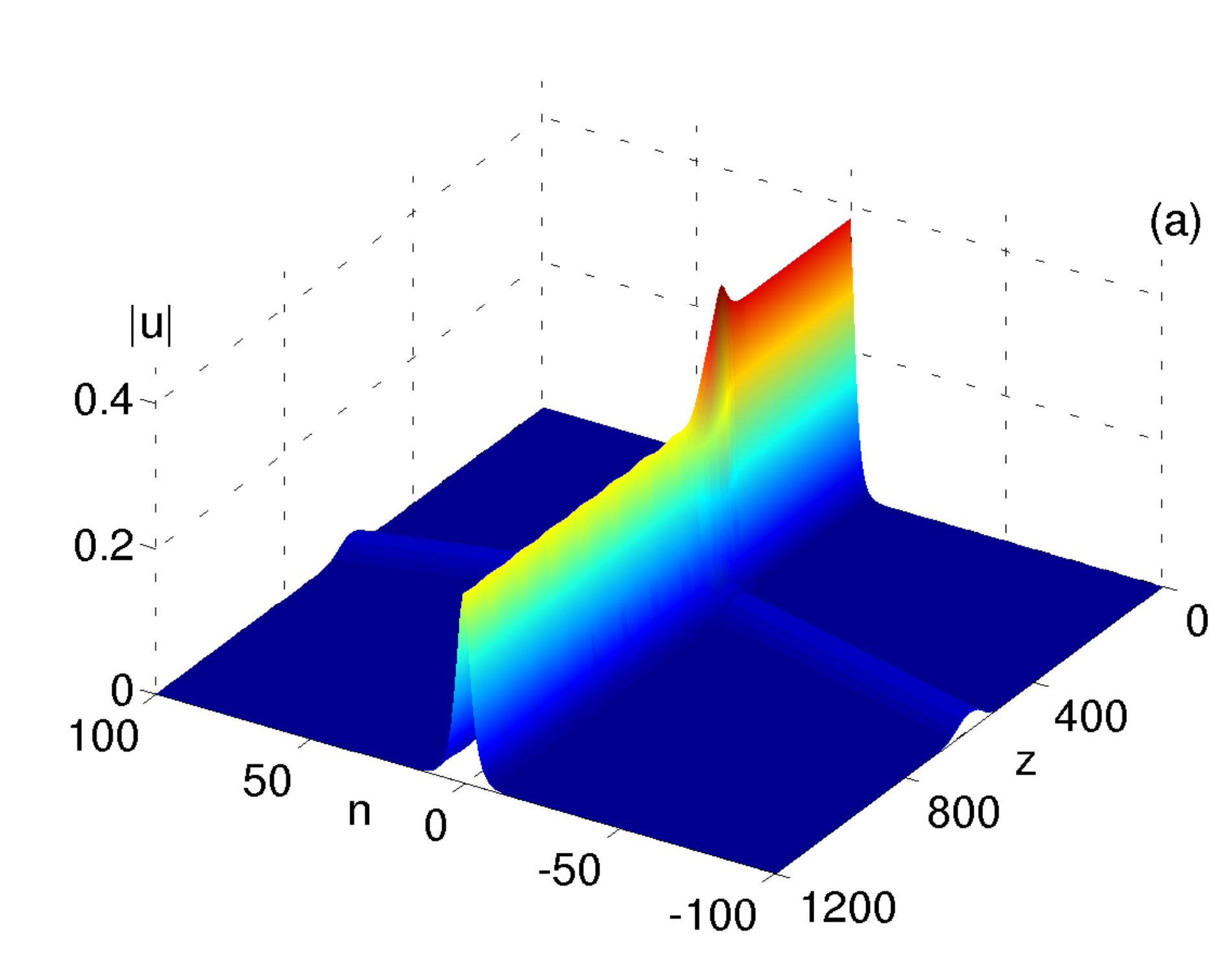} 
   \hspace*{1mm}
    \includegraphics*[width=0.49\linewidth]{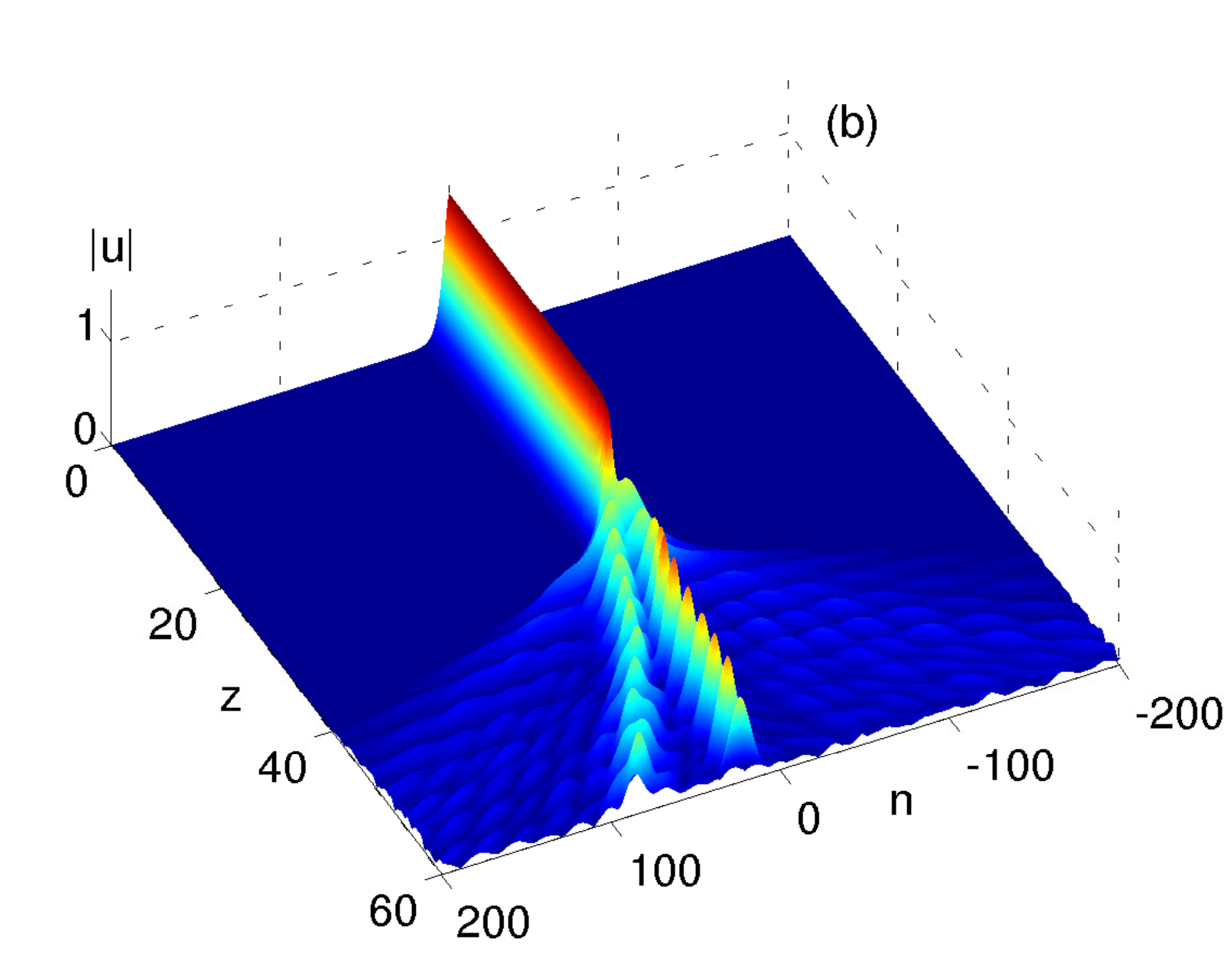} 
      \caption{Two more outcomes of the antiphase soliton instability growth. 
      (a): The instability caused by a real eigenvalue with an even eigenvector. In Figs \ref{Min5}(c,d) and \ref{Min01}(a,b), 
      this eigenvalue is marked by the pink line. The unstable soliton emits a pair of small-amplitude solitons with opposite
      velocities and transforms into a (stationary) antiphase soliton of smaller amplitude.
      Here the lattice coupling is  $\C=0.3$, gain-loss coefficient $\gamma=0.93$
      and the initial soliton's amplitude $A=0.37$. (These parameters fall into the ``white fang" region in Fig \ref{C08}(a).)
      The emerging soliton has $A=0.26$. (This belongs in the green region just under the white fang.)
      (b): The joint effect of the  real eigenvalue with an odd eigenvector (the deep-blue line in Fig \ref{Min01})
      and the complex quadruplet (light-green lines in Fig \ref{Min01}). The soliton breaks  into an eccentric bunch of breathers. Here
      $\C=19.9$, $\gamma=0.1$, and $A=1.41$. These values  correspond to the point $h= 0.32$ in Fig \ref{Min01}(c,d)
      (where $\sigma=-0.1$). 
            }
 \label{pinkblue}
 \end{figure}

Unlike the oscillatory instability discussed above, the monotonic growth 
  associated with the real mode cannot give rise to any breathing structures. A typical 
  evolution is illustrated in Fig \ref{pinkblue}  (a). The unstable antiphase soliton sheds some power and transforms into a stable soliton of the same variety.

The inequality $1.96 \leq |\sigma|<4$ selects two more regions on the $(\frak C, \gamma)$-plane:
\begin{subequations}
\begin{align} 
\frak C \leq 1/2,   
\quad
 \sqrt{1-4 \frak C^2}< \gamma  \leq \sqrt{1- 0.96 \frak C^2}; \\
 1/2 < \frak C <1.02,  \quad \gamma \leq  \sqrt{1-0.96 \frak C^2}.
  \end{align}
 \end{subequations}
 Each of these two regions features two non-overlapping  intervals  of unstable amplitudes, $\mathcal A_4< A< \mathcal A_5$
 and  $\mathcal A_2< A< \mathcal A_3$,
 where
 \be
 \mathcal A_n= \sqrt{\frak C} \mathcal H_n \left(  {2\sqrt{1-\gamma^2}}/{\frak C}  \right),
 \quad n=2,3,4,5.
 \label{An}
 \ee
  Two corridors of instability are clearly visible in the left part of 
   Fig \ref{C08}(b) and   middle section of Fig \ref{C08}(a).

Another instability that has a sizeable domain in chains with  $|\sigma|<2$, is due to a pair of real 
eigenvalues with odd eigenvectors. 
The inequality $|\sigma|^{1/2}< h< \mathcal H_1(|\sigma|)$ of section \ref{odd_mode}
 defines the upper part of the odd-mode instability domain; the lower part has to be determined numerically. 
The trajectory of the odd-mode eigenvalues on the $(h, \mathrm{Re} \, \lambda)$-plane is described by the deep-blue curve in Fig \ref{Min01} (a,c).
Like the even real mode discussed in the previous paragraphs, 
 this pair of real eigenvalues is  born as an imaginary pair which bifurcates from the continuous spectrum. [See the deep-blue curves in Fig \ref{Min01} (b,d).]
As $h$ is decreased to the value of $\mathcal H_1(|\sigma|)$, the two imaginary eigenvalues collide and move to the real axis
--- in agreement with the variational argument of section \ref{odd_mode}.

What the variational principle is powerless to determine though, is the trajectory of these eigenvalues on the complex plane as $h$ is  decreased
below $|\sigma|^{1/2}$. The numerical analysis, on the other hand,  reveals 
the arrival of yet another pair of real eigenvalues  from the imaginary axis via the origin
(shown in light blue in Fig \ref{Min01} (a,c)). Like the ``dark-blue" eigenvalues,  the newly born pair has odd eigenvectors.
At some $h< |\sigma|^{1/2}$, 
the ``light-blue" and ``deep-blue" eigenvalues collide, pairwise, and emerge into the complex plane.

The evolution of the resulting complex quadruplet 
depends on whether $|\sigma|$ is 
greater or smaller than 2.
When $|\sigma|<2$, the quadruplet persists all the way to $h=0$, with
the real parts of the complex eigenvalues decaying to zero. (See the stiletto-shaped dark-green curves in Fig \ref{Min01}(a)).
According to  section \ref{small_h}, the decay is exponentially  fast as $h \to 0$.
When $2<|\sigma|<4$, the complex eigenvalues converge, pairwise, at two opposite points on the imaginary axis
outside the continuous spectrum band. After that one pair of imaginary eigenvalues immerses in the continuum while
the other one remains outside the band, approaching the band edges only as $h \to 0$. (See the dark-green arcs in Fig \ref{Min5}(d).)


Regardless of the value of $\sigma$, 
the point $h=\mathcal H_1(|\sigma|)$ where the odd-mode instability sets in, is  to the left of $\mathcal H_3(|\sigma|)$,
the Hamiltonian Hopf bifurcation point where the even-mode (``light-green") complex quadruplet is born.
When $|\sigma|<2$, the even-mode quadruplet persists as $h$ is decreased from $\mathcal H_3$ to $0$.
Accordingly, in this  $\sigma$ range the ``deep-blue" real eigenvalues  and their descendent ``dark-green" complex quadruplet
coexist with the complex quadruplet shown in light green.
Therefore  although the odd-mode instability is detectable in direct computer simulations of the antiphase soliton,
its presence does not affect the soliton stability domain on the $(\gamma, A)$-plane.

Due to the coexistence of the real eigenvalues and the complex quadruplet, the odd-mode monotonic instability
and 
the even-mode oscillatory one, develop simultaneously.
A typical outcome of the soliton decay in the parameter range with $|\sigma|<2$  is
depicted in Fig \ref{pinkblue} (b). 
The antiphase soliton breaks into an asymmetric assembly of 
two or more travelling breathers.

When $2 < |\sigma|<4$, the point $h=\mathcal H_1(|\sigma|)$ is to the left of 
$\mathcal H_2(|\sigma|)$, the lower boundary of the 
``light-green"  quadruplet domain of existence. When $|\sigma|$ falls in this range, the intervals of the odd-
and even-mode  instability do not overlap. However the odd-mode instability band is so narrow here that this instability would be hard to observe
 in computer simulations of the soliton.
The odd instability band is marked as a white line in Fig \ref{C08}(a).

We close this subsection by acknowledging an earlier numerical study of the eigenvalue problem \eqref{B8}.
The authors of Ref \cite{PY} computed eigenvalues for three  fixed values of $h$
($h=0.50, 0.71$ and $3.16$) and varied $\sigma$. 
This is obviously not equivalent to the approach of the present paper where we fix $\sigma$ and vary $h$.

 \section{Concluding remarks}
\label{Conc}

\subsection{Conclusions}

In this paper we have considered stability of the site-  and bond-centred solitons in the chain of
\PT-symmetric  dimers with gain and loss.
Both classes of solitons come in two varieties: the in-phase and the antiphase solitons.
Our principal results are as follows.

1. We have shown that
 small perturbations about the soliton can be decomposed into a part tangent to the 
conservative invariant manifold containing the soliton, and a part 
that is transversal to that manifold. This decomposition 
proves instability of both varieties of the bond-centred solitons,
for all amplitudes and all values of the chain parameters.

2.
We have demonstrated that stability properties of a site-centred soliton 
with amplitude $A$ in the chain with the inter-dimer coupling $\C$ and 
gain-loss coefficient $\gamma$, are completely 
determined just by two combinations of $A$, $\C$ and $\gamma$.
These are the structural parameter $\sigma= \pm 2 \sqrt{1-\gamma^2} / \C$ 
(where the top and bottom sign correspond to the in-phase and antiphase soliton, respectively)
and the normalised amplitude $h= A/ \sqrt{\C}$. 

3.
We have proved that the in-phase site-centred soliton is stable if its
amplitude lies below a threshold value \eqref{ceiling} and unstable otherwise.
A simple upper bound for the threshold, equation \eqref{Aub},  has been established and 
 its
 explicit expressions in the small-coupling limit
($\C / \sqrt{1-\gamma^2} \to 0$) and in the limit where $\sqrt{1-\gamma^2}/\C \to 0$
have been provided.
(See equations \eqref{sma_cou} and \eqref{lar_cou}, respectively.)
The lowest threshold occurs in chains where the value of the dimer-to-dimer coupling $\C$ is
close to the value of the gain-to-loss coupling  within each dimer.

4.
With regard to the stability of the site-centered
antiphase  soliton, we have identified three characteristic coupling ranges: 
(a) $\frak C <1/2$,  (b) $1/2 < \frak C<1.02$, and  (c) $\frak C>1.02$.

4(a). In the weak-coupling range ($\frak C <1/2$), the structure of the stability
domain depends on whether the gain-loss coefficient 
(i) is smaller than $\sqrt{1-4\C^2}$;
(ii) falls between 
$\sqrt{1-4\C^2}$ and $\sqrt{1-0.96 \C^2}$, or
 (iii) is greater than $\sqrt{1-0.96 \C^2}$.
 (i) When $\gamma$ lies below $\sqrt{1-4 \C^2}$, the soliton is unstable if its amplitude satisfies $\mathcal A_2 < A< \mathcal A_3$ and stable otherwise. 
As $\frak C/ \sqrt{1-\gamma^2}  \to 0$, the functions $\mathcal A_2(\frak C, \gamma)$ and $\mathcal A_3(\frak C, \gamma)$
are given by the asymptotic expressions \eqref{A_1} and \eqref{A_2}, respectively.
Otherwise $\mathcal A_2$ and $\mathcal A_3$ are as in \eqref{A2h2} and \eqref{A3h3}, with $\mathcal H_2$ and $\mathcal H_3$ determined numerically. 
 (ii) When $\gamma$ falls between $\sqrt{1-4 \C^2}$ and $\sqrt{1-0.96 \C^2}$, there are two instability bands:  $\mathcal A_2 < A< \mathcal A_3$ 
and  $\mathcal A_4 < A< \mathcal A_5$, with $\mathcal A_n$ as in \eqref{An} and $\mathcal H_n$ determined numerically. 
(iii) When $\gamma$ is greater than  $\sqrt{1-0.96 \frak C^2}$, the  amplitudes 
above $\mathcal A_3$
 are stable and those  below $\mathcal A_3$ unstable.

 In addition to the above one or two instability bands, 
 we need to note an extremely thin corridor of odd-mode instability 
that crosses the  soliton stability domain when
$\C<1/2$ and $\gamma < \sqrt{1-0.96 \C^2}$. 
The upper part of this corridor is described by the inequalities in \eqref{odd_inst}, with 
the corresponding asymptotic expressions as in \eqref{ud}.

4(b). In the middle range ($1/2 < \frak C<1.02$), the  stability criteria depend on whether the 
gain-loss coefficient is greater or smaller than $\sqrt{1-0.96 \frak C^2}$. 
When $\gamma< \sqrt{1- 0.96 \frak C^2}$, the soliton is unstable if its amplitude falls in either of two bands, $\mathcal A_2<A< \mathcal A_3$ or
$\mathcal A_4<A< \mathcal A_5$ 
 --- and stable otherwise. When $\gamma> \sqrt{1-0.96 \frak C^2}$, the  amplitudes above  the threshold value $\mathcal A_3$
are stable
and those below the threshold are unstable.
 The boundaries $\mathcal A_n(\C, \gamma)$ ($n=2,...,5$)
 are as in \eqref{An}, with the functions $\mathcal H_n$ being determined numerically.

4(c). In the strong coupling range ($\frak C>1.02$),  solitons with amplitudes above $\mathcal A_3$  are stable and those below $\mathcal A_3$
 unstable.
The threshold amplitude is given  by equation \eqref{A3h3} where the function $\mathcal H_3(|\sigma|)$ is determined numerically.

4(d) 
In the parameter regimes where the small-amplitude solitons are unstable
(in the strong-coupling range $\frak C>1.02$ and in the range $\frak C<1.02$ with $\gamma > \sqrt{1-0.96 \C^2}$),
the instability growth rate becomes exponentially small as the soliton's amplitude tends to zero.
Hence unstable
 solitons with small amplitudes can live long enough to  be regarded as {\it practically stable}.

\subsection{Comparison with earlier work}


It is instructive to compare our approach and conclusions with those of earlier authors who
performed asymptotic analysis in the $\C, \gamma \to 0$
 limit  \cite{Susanto} 
and studied soliton stability numerically \cite{SMDK,Susanto}.

What we see as an advantage of our approach, is a particular choice of variables 
that leads to an eigenvalue problem in triangular form (see equations \eqref{B7} and \eqref{B70}).
The eigenvectors of the triangular matrices admit a natural decomposition into two disjoint classes: the
eigenvectors that belong to the scalar  reduction of the two-component  nonlinear Schr\"odinger equation 
\eqref{B1} and eigenvectors that lie outside the reduction manifold.

This decomposition alone was sufficient for us  to prove the instability of  all  bond-centred solitons, 
regardless of their amplitude and irrespective of the $\C$ and
$\gamma$   parameters of the chain. 
In contrast, the earlier approaches  could only establish the instability of the  
bond-centred soliton in the $\C \to 0$, $\gamma \to 0$ limit \cite{Susanto} --- and for several isolated parameter triplets 
that were  examined numerically \cite{SMDK,Susanto}. Numerical methods require particular caution 
as they
may not be accurate enough to discern 
small real parts of eigenvalues. This may mislead one into thinking that a weakly unstable soliton is stable
 (as in 
 the case of the in-phase bond-centred soliton with large $\C$ \cite{Susanto}.)

In the case of the site-centred soliton, the above decomposition 
reduces the eigenvalue problem to the form familiar from the theory of the  scalar nonlinear Schr\"odinger equation.
This reduction allowed us to use the standard analytical methods for the (discrete) Schr\"odinger operators.
The advantage of halving the 
 dimension of the vector space extends beyond the availability of analytics though.
 In particular, it simplifies power expansions in the anticontinuum limit 
 where, unlike \cite{Susanto}, we did not have to  assume that $\gamma$ is small.

 Another difference between our approach and previous work 
  is that we recognise that the stability properties of the soliton
 with amplitude $A$ in the chain with coupling $\C$ and gain-loss coefficient $\gamma$, 
 are completely determined just by two combinations of these three parameters. 
 Letting $h=A/\sqrt{\C}$ and $|\sigma| = 2 \sqrt{1-\gamma^2}/\C$ vary over their respective domains, we 
 cover the entire  $(A, \C, \gamma)$ space. 

Our numerical results on the  site-centred solitons in the limit $\C \to 0$, $\gamma \to 0$, 
are in agreement with those of \cite{Susanto}.
On the other hand, the  eigenvalue trajectories of the  site-centred antiphase soliton 
with larger $\C$ or $\gamma$ close to 1 --- the situation unexplored in \cite{Susanto} --- 
 are found to be much more complex than in the $\C, \gamma \to 0$ limit.

\section*{Acknowledgements}

We acknowledge useful discussions with  Georgy Alfimov, Sergej Flach, Panos Kevrekidis,  Andrei Sukhorukov, and Dmitry Zezyulin.
Special thanks go to Dmitry Pelinovsky for reading the manuscript and bringing Ref \cite{PY} to our attention. 
NA's stay in Canberra was supported via the Visiting Fellowship of the 
ANU.  IB's work in Bath was funded by
the European Union's Horizon 2020 research
and innovation programme under the Marie Sk{\l}odowska-Curie
Grant Agreement No. 691011.
These authors  were also supported by the National Research Foundation 
of South Africa
(grants 105835, 85751 and 466082).
Computations were performed at the UCT HPC Cluster.

\appendix
\section{The negative eigenvalue of  operator $L_+$ --- upper bound}
\label{App1}

The  discrete eigenvalues of the infinite-dimensional matrix \eqref{B10}, $E_0$ and $E_1$,   play a special role in our stability analysis. 
Unlike the eigenvalues of the  corresponding 
differential operator in \cite{ABSK}, the  matrix eigenvalues 
 do not admit analytic expressions.
 In this Appendix,  we derive a simple upper bound on the lower eigenvalue ($E_0$) valid for all $0<h< \infty$.

  The derivation starts with the scalar equation \eqref{B3}.
Multiplying it with ${\mathcal R}_n$ and summing over $n$,
we obtain
\be
2S_4 - S_2 = h^{-2} \sum ({\mathcal R}_{n+1}- {\mathcal R}_n)^2,
\label{a101}
\ee
where 
\be
S_2=  \sum {\mathcal R}_n^2,
\quad
S_4=
\sum {\mathcal R}_n^4.
\label{S2}
\ee
Eq.\eqref{a101} implies
\be
S_4  \geq \frac12  S_2.
\label{a1}
\ee

Next, noting that 
\be
(L_+)_{nm} = (L_-)_{nm} -4h^2 {\mathcal R}_n^2 \delta_{mn},
\label{a111}
\ee
and that ${\mathcal R}_n$ is the null eigenvector  of the matrix $L_-$, we have
\[
\sum_{n,m} {\mathcal R}_n (L_+)_{nm}  {\mathcal R}_m = -4 h^2 S_4.
\]
Using the inequality \eqref{a1}, this gives 
\be
\sum_{n,m} {\mathcal R}_n (L_+)_{nm}  {\mathcal R}_m  \leq -2 h^2  S_2.
\label{a2}
\ee

The lowest eigenvalue of the matrix $L_+$ is given by the minimum of the Rayleigh quotient
\[
E_0= \min_{\bf y} \frac{\sum_{n,m} y_n (L_+)_{nm}  y_m}{\sum_n y_n^2}.
\]
Letting here $y_n={\mathcal R}_n$ and using \eqref{a2},
we obtain an upper bound on the lowest eigenvalue:
\be
E_0 \leq -2h^2.
\label{a3}
\ee

\section{The negative eigenvalue of   $L_+$ in the continuum limit}
\label{App7}

In this and the next appendix, 
we establish the asymptotic behaviour of the discrete eigenvalues $E_0$ and $E_1$ as $h \to 0$.
 Although \eqref{B10}  is exactly the matrix arising in the linearisation of the
scalar discrete nonlinear Schr\"odinger soliton, these results are not in the 
existing literature on the subject.

In the  limit $h \to 0$,  the site-centred soliton solution of equation \eqref{B3} can be 
sought  as 
$\mathcal R_n = \mathcal R(X_n)$, where $\mathcal R(X)$ is a continuous function of its argument, and 
$X_n=nh$.
The function $\mathcal R(X)$ satisfies
\be
-2 \sum_{\alpha=0}^\infty \frac{h^{2\alpha}}{(2\alpha+2)!} \frac{d^{2\alpha+2}\mathcal R}{dX^{2\alpha+2}}+ \mathcal R - 2 \mathcal R^3=0,
\label{infdiff}
\ee
and can be
easily constructed as a power expansion
\be
\mathcal R=     \sum_{\alpha =0}^\infty  \mathcal R^{(\alpha)}h^{2\alpha} + \mathrm{e.s.t.}
\label{Rcont}
\ee
Here $\mathrm{e.s.t.}$ stands for exponentially small terms.
The first two terms in the expansion are 
\[
\mathcal R^{(0)}= 
\mathrm{sech\/} X, 
\]
and
\[
\mathcal R^{(1)}= \left(   \frac{1}{3} \mathrm{sech\/}^2 X
-\frac{7}{24}  + \frac{1}{24} X \tanh X
\right) \mathrm{sech\/} X.
\]

To determine the eigenvalues of the matrix $L_+$ in this limit 
we let $y_n= Y(X_n)$, where $Y(X)$ is a continuous function.
The system of linear algebraic equations
\be
L_+ {\bf y} = E {\bf y}
\label{405}
\ee
is transformed into a differential eigenvalue problem
\be
 -2 \sum_{\alpha =0}^\infty \frac{h^{2\alpha+2}}{(2\alpha +2)!} \frac{d^{2\alpha+2}Y}{dX^{2\alpha+2}}+ h^2(1 -6 \mathcal R^2)Y
 =EY.
 \label{412}
\ee
Expanding 
\be
Y(X)= \sum_{\alpha=0}^\infty \mathcal Y^{(\alpha)}(X) h^{2\alpha}+ \mathrm{e.s.t.}, \quad
E =  \sum_{\alpha=0}^\infty \mathcal E^{(\alpha)}  h^{2\alpha+2}+ \mathrm{e.s.t.},
\label{423}
\ee
and using \eqref{Rcont}, we equate coefficients of like powers of $h^2$ in \eqref{412}.

The coefficient of $h^{2 \alpha+2}$ gives us
\begin{align}
\left( \frak L_+        - \mathcal E^{(0)}  \right) \mathcal Y^{(\alpha)}= \sum_{\beta=1}^\alpha \frac{2}{(2\beta+2)!}
\frac{d^{2 \beta+2}}{dX^{2 \beta+2}} \mathcal Y^{(\alpha-\beta)} 
+ 6 \sum_{\beta=1}^\alpha \sum_{\mu=0}^\beta \mathcal R^{(\mu)} \mathcal R^{(\beta-\mu)} \mathcal Y^{(\alpha-\beta)}
+ \sum_{\beta=1}^\alpha \mathcal E^{(\beta)} \mathcal Y^{(\alpha-\beta)},
\label{414}
\end{align}
 where we have introduced the Sturm-Liouville operator 
  \be
{\frak L}_+= - d^2/dX^2+ 1 - 6 \sech^2 X.
\label{Lplus}
\ee
In particular, setting $\alpha=0$ we obtain from \eqref{414}:
\be
 {\frak L}_+ \mathcal Y^{(0)}= \mathcal E^{(0)} \mathcal Y^{(0)}.
 \label{417}
 \ee

Eigenvectors $y_n$ decaying to zero as $|n| \to \infty$ 
 correspond to eigenfunctions  $Y(X)$ decaying as $|X| \to \infty$; that is, 
discrete eigenvalues of \eqref{405} correspond to discrete eigenvalues of \eqref{417}. 
The 
 eigenvector 
 associated with $E_0$, the lowest eigenvalue of the matrix $L_+$, 
 is even in $n$:  $y_{-n}=y_n$. We normalise it by
$y_0=1$. 
The corresponding eigenfunction $\mathcal Y^{(0)}$ should also be even in its argument: $\mathcal Y^{(0)}(-X)=\mathcal Y^{(0)} (X)$. 
The eigenvalue of the operator 
 $\frak L_+$, associated with an even eigenfunction, is $\mathcal E^{(0)}= -3$, and
the eigenfunction is $\mathcal Y^{(0)}=\sech^2X$.

Equation \eqref{414} with $\alpha=1$ is
\[
( \frak L_+  -\mathcal E^{(0)}) \mathcal Y^{(1)}=\left( \frac{1}{12} \frac{d^4 }{dX^4} + 12 \mathcal R^{(0)} \mathcal R^{(1)} + \mathcal E^{(1)} \right)  \mathcal Y^{(0)},
\]
where the operator $\frak L_+ - \mathcal E^{(0)}$ in the left-hand side has a zero eigenvalue. 
A bounded solution of the above equation exists only if the right-hand side is orthogonal to the corresponding eigenvector, $\mathcal Y^{(0)}$.
This requirement fixes the coefficient $\mathcal E^{(1)}$: $\mathcal E^{(1)}=-1/5$. 
Consequently,  the asymptotic behaviour for the negative eigenvalue of $L_+$ as $h \to 0$, is
\be
E_0(h) = -3 h^2 -\frac15  h^4 +O(h^6).
\label{b1}
\ee
Higher order corrections to \eqref{b1} can be evaluated recursively, by considering equations
\eqref{414} with larger $\alpha$.

\section{The positive eigenvalue of   $L_+$ in the continuum limit}
\label{app10}

This appendix is a continuation of the previous one.  Here, we consider $E_1$, the second lowest eigenvalue of the matrix $L_+$.
(This eigenvalue is positive \cite{LST,KK}.) 
We show that as $h \to 0$,  $E_1$ becomes exponentially small in $h^2$.

The eigenvector associated with $E_1$,
is antisymmetric in $n$: 
$y_{-n}=-y_n$,  $y_0=0$. 
The antisymmetry in $n$ translates into the oddity of the functions $\mathcal Y^{(\alpha)}$ in \eqref{423}: $\mathcal Y^{(\alpha)}(-X)=- \mathcal Y^{(\alpha)}(X)$.
The odd eigenfunction of \eqref{417} is $\mathcal R^{(0)}_X= - \mathrm{sech\/} X \tanh X$; the
associated eigenvalue is $\mathcal E^{(0)}=0$. We prove, by induction, that 
$\mathcal Y^{(\alpha)}= \mathcal R^{(\alpha)}_X$ and $\mathcal E^{(\alpha)}=0$ for all $\alpha \geq 1$.

First, differentiating \eqref{infdiff} with respect to $X$ we observe that  
the derivative $\mathcal R_X$ is an eigenfunction of the operator
\[
 -2 \sum_{\alpha =0}^\infty \frac{h^{2\alpha+2}}{(2\alpha +2)!} \frac{d^{2\alpha+2}}{dX^{2\alpha+2}}+ h^2(1 -6 \mathcal R^2),
 \]
with the eigenvalue $E=0$. Accordingly, 
 equations \eqref{414} with
 $\mathcal Y^{(\alpha)}= \mathcal R^{(\alpha)}_X$ and  $\mathcal E^{(\alpha)}=0$
  become a  string of identities:
\begin{align}
\frak L_+  \mathcal R^{(\alpha)}_X= \sum_{\beta=1}^\alpha \frac{2}{(2\beta+2)!}
\frac{d^{2 \beta+2}}{dX^{2 \beta+2}} \mathcal R^{(\alpha-\beta)}_X 
+ 6 \sum_{\beta=1}^\alpha \sum_{\mu=0}^\beta \mathcal R^{(\mu)} \mathcal R^{(\beta-\mu)} \mathcal R^{(\alpha-\beta)}_X,
\quad \alpha=1,2,... .
\label{422}
\end{align}
Multiplying both sides 
 of \eqref{422} with the eigenfunction $\mathcal R^{(0)}_X$
 of ${\frak L}_+$ and integrating, these give
\begin{align}
 \sum_{\beta=1}^\alpha \frac{2}{(2\beta+2)!} \int \mathcal R^{(0)}_X
\frac{d^{2 \beta+2}}{dX^{2 \beta+2}} \mathcal R^{(\alpha-\beta)}_X  dX 
+ 6 \sum_{\beta=1}^\alpha \sum_{\mu=0}^\beta  \int \mathcal R^{(0)}_X
\mathcal R^{(\mu)} \mathcal R^{(\beta-\mu)} \mathcal R^{(\alpha-\beta)}_X  dX=0.
\label{420}
\end{align}

Assume that $\mathcal Y^{(\alpha)}= \mathcal R^{(\alpha)}_X$ and $\mathcal E^{(\alpha)}=0$ for all $0 \leq \alpha \leq \alpha_0-1$
with some $\alpha_0$,
and consider equation \eqref{414} with $\alpha=\alpha_0$:
\begin{align}
\frak L_+  \mathcal Y^{(\alpha)}= \sum_{\beta=1}^\alpha \frac{2}{(2\beta+2)!}
\frac{d^{2 \beta+2}}{dX^{2 \beta+2}} \mathcal R^{(\alpha-\beta)}_X 
+ 6 \sum_{\beta=1}^\alpha \sum_{\mu=0}^\beta \mathcal R^{(\mu)} \mathcal R^{(\beta-\mu)} \mathcal R^{(\alpha-\beta)}_X
+  \mathcal E^{(\alpha)} \mathcal R^{(0)}_X.
\label{419}
\end{align}
This equation admits a bounded solution only if the right-hand side is orthogonal to the eigenfunction of $\frak L$ associated with its zero eigenvalue --- that is, orthogonal to
 $\mathcal R^{(0)}_X$.
Making use of \eqref{420}
this solvability condition  reduces to
\[
\mathcal E^{(\alpha)} \int \left( \mathcal R^{(0)}_X \right)^2 dX=0,
\]
which implies $\mathcal E^{(\alpha)}=0$.
Comparing the right-hand side of  \eqref{419} to the equation \eqref{422}, we 
 obtain $\mathcal Y^{(\alpha)}$ with $\alpha=\alpha_0$: 
$\mathcal Y^{(\alpha)}= \mathcal R^{(\alpha)}_X+ C \mathcal R^{(0)}_X$, where $C$ is an arbitrary constant.
Without loss of generality, we can choose $C=0$; this is equivalent to 
dividing the eigenfunction $Y(X)$ by the normalisation factor $1+Ch^{2 \alpha}$.
 This completes the proof.

The bottom line of our analysis is that all $\mathcal E^{(\alpha)}$  in the expansion \eqref{423} are equal to zero and
so  $E_1$ is smaller than any power of $h^2$ in the $h \to 0$ limit.
Intuitively, this result is consistent with the exponential smallness of the 
obstacles to the translation motion of the discrete soliton 
(such as the value of the Melnikov function \cite{KK} or 
Peierls-Nabarro barrier \cite{JW}).

The function $E_1(h)$, computed numerically,  is shown in Fig \ref{L1E1}(b). 
The $e^{-C/ h^2}$-type of decay 
 of the eigenvalue  as $h \to 0$,  is clearly visible in the figure.

\section{Discrete eigenvalues of   $L_+$ in the anticontinuum limit}
\label{App8}

In this appendix, 
we establish the asymptotic behaviours of the eigenvalues $E_0$ and $E_1$ as $h \to \infty$.

To this end,  we use  the power-series
solution of the scalar equation \eqref{B3}. The fundamental mode (the single-hump solution even in $n$)
 centred on the zero site, has the following asymptotic expansion as $h \to \infty$:
\begin{align}
{\mathcal R}_0= \frac{1}{\sqrt{2}} \left[1+  \frac{1}{h^2} -  \frac{3}{2} \frac{1}{h^4}
+ O\left( \frac{1}{h^6} \right) \right],  \nonumber \\
{\mathcal R}_{\pm 1}=  \frac{1}{\sqrt{2}} \left[ \frac{1}{h^2} -  \frac{1}{h^4}    + O\left( \frac{1}{h^6} \right) \right],  \quad
{\mathcal R}_{\pm 2}=  \frac{1}{\sqrt{2}}  \frac{1}{h^4}   + O\left( \frac{1}{h^6}\right), 
\label{a4}
\end{align}
and
${\mathcal R}_{\pm n}=   O\left( h^{-2n} \right)$ for $n \geq 3$.

The eigenvalue $E_0$ can be sought in the form
\be
E_0= 
{\mathcal E}^{(0)}h^2+  \sum_{\alpha =0}^\infty 
{\mathcal E}^{(\alpha+1)}h^{-2\alpha},
\label{a6} \ee
and the corresponding eigenvector can also be written as a power series:
\be
y_n=  \sum_{\alpha=0}^\infty {\mathcal Y}_{n}^{(\alpha)} h^{-2(\alpha+n)}, \ n \geq 1.
\label{a5} \ee
Substituting \eqref{a4},          \eqref{a6}         and       \eqref{a5}  in \eqref{405}
and equating coefficients of like powers of $h$, 
we obtain the required asymptotic expansion
\be
E_0= -2 h^2-4+ \frac{16}{3} h^{-2} + O(h^{-4}),
\quad h \to \infty.
\label{a7}
\ee

Turning to 
 the eigenvalue
 $E_1$, associated with  an odd eigenvector, 
   we wish to show that 
\be
E_1= h^2 + \mathrm{e.s.t.} \ \mbox{as} \ h \to \infty,
\label{406}
\ee
where $\mathrm{e.s.t.}$ stands for the ``exponentially small terms" as $h \to \infty$. 
In \eqref{406}, the leading term  $h^2$ corresponds to  the lower edge of the continuous spectrum band.

It is convenient to introduce the distance of the eigenvalue to the continuum,
$E^\prime=E_1-h^2$.
 The solution of equation \eqref{405} with $E^\prime \neq 0$, 
decaying as $n \to \infty$,  is  $y_n=e^{-\kappa n}$. Here 
$\kappa>0$  is defined by 
\be
E^\prime= - 4 \sinh^2 \frac{\kappa}{2}.
\label{407}
\ee
Letting $y_n=e^{-\kappa n} z_n$, the equation \eqref{405} with $n>0$, is
cast in the form
\be
-e^{-\kappa} z_{n+1}- e^\kappa z_{n-1} + 2 z_n- 6 \epsilon^{-1} \mathcal R_n^2 z_n=  E^\prime z_n, 
\quad n>0,
\label{408}
\ee
where 
we have denoted  $\epsilon=h^{-2}$.
 As $n \to \infty$, the values $z_n$ approach a 
nonzero constant or  grow in proportion to $n$. 
We supplement \eqref{408} with the boundary condition corresponding to odd $y_n$:
\be
z_0=0.
\label{z0}
\ee

The eigenvalue and 
each component of the eigenvector in \eqref{407}-\eqref{z0}  are functions of $\epsilon$: $E^\prime=E^\prime(\epsilon)$, $z_n=z_n(\epsilon)$.
When $\epsilon =0$, the  eigenvector  is  $z_n(0)=n$, and the eigenvalue $E^\prime=0$.
When $\epsilon \neq 0$, we assume that an eigenvalue and eigenvector exist such that 
 $E^\prime(\epsilon) \to 0^-$ and $z_n(\epsilon) \to n$ as $\epsilon \to 0$.
We wish to  show that  such $E^\prime (\epsilon)$ has to be nonanalytic in $\epsilon$. 

To this end, we assume the opposite; that is, we assume that 
$E^\prime(\epsilon)$ admits a Taylor expansion 
\be
E^\prime = {\mathscr E}^{(\alpha)} \epsilon^\alpha +  {\mathscr E}^{(\alpha+1)} \epsilon^{\alpha+1} + ... 
\label{432}
\ee
with some integer $\alpha>0$. 
When $n$ is much greater than $ (\alpha+1)/2$, the potential  in \eqref{408} becomes much smaller than $E^\prime$ and
can be disregarded:
\be
-e^{-\kappa} z_{n+1}- e^\kappa z_{n-1} + 2 z_n =  - 4 \sinh^2 (\kappa/2)   z_n.
\label{431}
\ee

The only solution of \eqref{431} that grows slower than exponentially as $n \to \infty$,
is $z_n(\epsilon)=1$. (The other, linearly independent, solution is  $e^{2 \kappa n}$.)
However $z_n(\epsilon)=1$ is {\it not\/} a continuous perturbation of $z_n(0)=n$, and this fact 
 contradicts our assumptions.

This leads us to conclude that  $E^\prime(\epsilon)$ cannot have the Taylor expansion  \eqref{432}.
Instead, $E^\prime$ has to be smaller than any power of $\epsilon$. 
If that is the case,   the potential  in \eqref{408} will remain greater than $E^\prime$ even if $n \to \infty$
so that  the asymptotic behaviour of $z_n$ will not be governed by equation \eqref{431}.

The exponential approach of $E_1(h)$ to the value $h^2$, as $h \to \infty$,
is clearly visible in Fig \ref{L1E1}(b).

\section{Stability eigenvalues for the 
in-phase soliton --- upper and lower bounds}
\label{App2}

In this Appendix, we establish an upper and lower bounds on the lowest 
eigenvalue of the generalised eigenvalue problem \eqref{A12}. Here, $\sigma > 0$.

First, we evaluate the Rayleigh quotient \eqref{A13} using  $g_n=\mathcal{R}_n$ as a test function.
Equation \eqref{a111} gives
\[
 \frac{\langle {\bf g} |L_++{\sigma} I| {\bf g} \rangle}{\langle {\bf g} |(L_-+{\sigma}I)^{-1}  |{\bf g}  \rangle}=
 \frac{\sigma S_2 - 4 h^2 S_4}{\sigma^{-1} S_2},
\]
where $S_2$ and $S_4$ are as in \eqref{S2}.
Using \eqref{a1} this gives an upper bound on the ground-state eigenvalue $-\lambda^2$:
\be
-\lambda^2 \leq  \sigma (\sigma -2 h^2).
\label{Q14}
\ee

To  bound $-\lambda^2$ from below, we note two inequalities, valid for an arbitrary $\bf g$:
\begin{align*}
 \langle {\bf g} | L_++\sigma  I | {\bf g} \rangle \geq (E_0+\sigma)  \langle {\bf g}  | {\bf g} \rangle,
 \\
\langle {\bf g} | (L_-+\sigma I)^{-1} | {\bf g} \rangle \leq \sigma^{-1} 
 \langle {\bf g}  | {\bf g} \rangle.
   \end{align*}
   Substituting these in \eqref{A13} we obtain the lower bound:
   \be
-\lambda^2   \geq \sigma (E_0+\sigma). 
 \label{Q140}
\ee

Making use of
\eqref{b1}, the lower bound becomes, in the limit $h \to 0$,
\be
-\lambda^2 \geq 
\sigma [\sigma - 3 h^2 + O(h^4)].
\label{Q17}
\ee
In order to obtain the lower bound in the opposite limit $h \to \infty$, 
we use \eqref{a7} instead.
This gives
\be
-\lambda^2 \geq
\sigma \left[-2h^2-4+ \sigma + \frac{16}{3} h^{-2}  + O(h^{-4})\right] .
\label{Q15}
\ee

\section{Symplectic eigenvalues with $h \to 0$ and finite $\sigma$}
\label{App4}

The subject of this Appendix is the eigenvalue problem \eqref{B8} in the  limit 
 $h \to 0$, with finite $\sigma \neq 0$. This limit corresponds to solitons of small amplitude ($A \to 0$)
 in the chain with finite coupling $\frak C$ and gain-loss coefficient $\gamma$. 
 We show that there are only two  pairs of eigenvalues,
 with each eigenvalue being  pure imaginary to all orders in $h$.
 
The derivation simplifies if we transform to new variables \cite{ABSK}
\[
 {\mathscr A}_n=f_n-ig_n, \quad
{\mathscr B}_n= -(f_n+ig_n).
\]
In terms of ${\mathscr A}_n$ and ${\mathscr B}_n$, equations \eqref{B8} acquire the form
\begin{align}
-\Delta \iA_n +(h^2+\sigma -i \lambda- 4 h^2 \mathcal R_n^2) \iA_n - 2h^2 \mathcal R_n^2 \iB_n=0, 
\label{iA} \\
-\Delta \iB_n +(h^2+\sigma +i \lambda- 4 h^2 \mathcal R_n^2) \iB_n - 2h^2 \mathcal R_n^2 \iA_n=0.
\label{iB}
\end{align}

In the limit $h \to 0$, we   let  
\[
{\mathscr A}_n=  \sum_{\alpha=0}^\infty  {\sf A}^{(\alpha)}(nh)h^{2 \alpha}, \quad
{\mathscr  B} _n= \sum_{\alpha=0}^\infty  {\sf B}^{(\alpha)}(nh)h^{2 \alpha},
    \]   
where $\A^{(\alpha)}(X)$ and $\B^{(\alpha)}(X)$ are continuous functions of order 1, 
initially assumed to be complex. 
We also expand the eigenvalue:
\[
\lambda= i \sum_{\alpha=0}^\infty \nu_\alpha h^{ 2 \alpha},
\]
where  $\nu_\alpha =O(1)$ are complex coefficients.
The parameter $\sigma$ is  assumed to be $O(1)$. 

Substituting these expansions in \eqref{iA}-\eqref{iB} and equating the coefficient of $h^0$ to zero, we obtain
\[
(\nu_0+\sigma) \A^{(0)}=0, 
\quad
(\nu_0-\sigma)\B^{(0)}=0.
\]
 We choose 
\[
\nu_0=  \sigma,
\quad 
\A^{(0)}=0.
\]
(The alternative choice is  $\nu_0=-\sigma$  and $\B^{(0)}=0$; this simply leads to the other member of the  $\pm \lambda$-pair.)
The order $h^{2 \alpha}$, with $\alpha=1,2,...$, gives then 
a linear system
\begin{align}
2 \sigma {\sf A}^{(\alpha)}=  -\A^{(\alpha-1)}
- \sum_{\beta=1}^{\alpha-1} \nu_{\beta} \A^{(\alpha-\beta)}  
+ \sum_{\beta=1}^\alpha   \frac{2}{(2\beta)!} \frac{d^{2\beta}}{dX^{2\beta}} \A^{(\alpha-\beta)}  
+ \sum_{\beta=0}^{\alpha-1}   \frak R_\beta \left(2 \A^{(\alpha-\beta -1)}+ \B^{(\alpha-\beta -1)}\right),  \label{sA} \\
 (\frak L_0-\nu_1) \B^{(\alpha-1)}=
\sum_{\beta=2}^{\alpha}  \nu_{\beta} \B^{(\alpha-\beta)}  
+ \sum_{\beta=2}^\alpha \frac{2}{(2\beta)!} \frac{d^{2\beta}}{dX^{2\beta}} \B^{(\alpha-\beta)}  
+  \sum_{\beta=1}^{\alpha-1} \frak R_\beta  \left(2 \B^{(\alpha-\beta -1)}+ \A^{(\alpha-\beta -1)}\right) 
+   \frak R_0 \A^{(\alpha-1)},
\label{sB}
\end{align}
where 
\[
\frak R_\beta=  2 \sum_{\mu =0}^\beta \mathcal R^{(\mu)} \mathcal R^{(\beta-\mu)},
\quad \beta=0,1,... .
\]

Equation \eqref{sB} with $\alpha=1$ has the form of an eigenvalue problem 
\[
\frak L_0  \B^{(0)}= \nu_1  \B^{(0)},    
\]
where $\frak L_0$ is an operator of the  P\"oschl-Teller variety:
\be
\frak L_0= -d^2/dX^2+ 1 - 4 \, {\rm sech\/}^2 X.  \label{d2X} 
\ee
The  P\"oschl-Teller operator \eqref{d2X}  has only two discrete eigenvalues \cite{ABSK},
 $\nu_1^{\mathrm{(even)}}= \xi-3$ and $\nu_1^{\mathrm{(odd)}}=3 \xi-4$, with $\xi=(\sqrt{17}-1)/2$.    
  Numerically,   $\xi-3 \approx -1.438$ and $3 \xi-4 \approx  0.685$. The eigenfunction $\B^{(0)}= \mathrm{sech\/}^\xi X$ 
 corresponding to $\nu_1^{\mathrm{(even)}}$ 
is even and everywhere positive; the eigenfunction $\B^{(0)}= \mathrm{sech\/}^{\xi-1} X \tanh X$
associated with $\nu_1^{\mathrm{(odd)}}$ is odd and 
has one zero crossing. (That operator $\frak L_0$ does not have any other eigenvalues 
follows from the fact that its eigenvalue associated with the eigenfunction with $n$ zero crossings
cannot lie below the corresponding eigenvalue of the operator $\frak L_+$, equation \eqref{Lplus}. The operator $\frak L_+$ is 
well known to have only two discrete eigenvalues; hence the operator \eqref{d2X} cannot have more than two.)

Once $\B^{(0)}(X)$ has been chosen, equation \eqref{sA} gives $\A^{(1)}(X)$:
\[
\A^{(1)}= \frac{1}{2 \sigma} \frak R_0 \B^{(0)}.
\]

The conclusion of our analysis is that the symplectic operator \eqref{symplec} has only two pairs of eigenvalues $\lambda$ in 
the continuum limit $h \to 0$ with $\sigma=O(1)$. 
These are given by the following two-term asymptotic expansions:
\be
\lambda_{\mathrm{even}}=   \pm i \left[ \sigma + \frac12 (\sqrt{17}-7) h^2 +O(h^4) \right]. 
\quad
\lambda_{\mathrm{odd}}=   \pm i \left[ \sigma +   \frac12 (3 \sqrt{17}-11) h^2 +O(h^4) \right].  \label{finlam}
\ee
The notation reflects the fact that  the first pair  corresponds to  even and  the second one to odd eigenfunctions.

A simple recursion argument proves that the eigenvalues $\lambda_{\mathrm{even}}$ and $\lambda_{\mathrm{odd}}$ remain pure imaginary to all orders in $h$.

Indeed,  consider some $\alpha \geq 2$ and
assume we know $\A^{(\beta)}$, $\nu_\beta$  with $\beta=0,1,..., \alpha-1$ and $\B^{(\beta)}$ with $\beta=0,1,..., \alpha-2$.
Assume all these quantities are real. 
 The operator $\frak L_0-\nu_1 I$ in the left-hand side of equation \eqref{sB} is singular
 and the equation  admits a bounded solution only if the right-hand side is orthogonal to $\B^{(0)}$, the function spanning its kernel space. 
 This solvability condition determines $\nu_\alpha$, a real value.
  Once the orthogonality condition has been satisfied, we can solve  \eqref{sB}  for $\B^{(\alpha-1)}$. After that 
equation \eqref{sA} gives us $\A^{(\alpha)}$. Both $\B^{(\alpha-1)}$  and $\A^{(\alpha)}$ 
come out real.

It is fitting to note here that our conclusion does not rule out 
 the existence of an {\it exponentially small\/} real part of $\lambda$
 ($\mathrm{Re} \, \lambda \sim e^{-\kappa /h}$, $\kappa= \mathrm{const} >0$). 
 In particular, the exponentially small real part arises in the situation of small negative $\sigma$
 --- see the next appendix.

\section{Symplectic eigenvalues in the strong-coupling limit ($h \to 0$ and $\sigma \to 0$)}
\label{App6}

The limit $\C \to \infty$ corresponds to sending both 
$h$ and $\sigma$ to zero while keeping $\sigma/h^2$ of order one.
(Here we assume that the soliton's amplitude $A$ is finite.)
We write
\be
\sigma= \eta h^2, 
 \label{AD2}
\ee
where  $\eta=O(1)$. 

Letting $g_n=\G^{(0)}(nh)+ O(h^2)$ and $f_n= \F^{(0)}(nh)  +O(h^2)$, where 
$\G^{(0)}(X)$ and $\F^{(0)}(X)$ are continuous functions, 
 and developing $\lambda=\lambda_0+  \lambda_1 h^2 +O(h^4)$, we substitute these expansions,
 along with \eqref{AD2}, in the symplectic eigenvalue problem
 \eqref{B8}.
 Setting the coefficients of  $h^0$ to zero we obtain $\lambda_0=0$, while the order $h^2$ yields  a differential 
 eigenvalue problem
  \be
 \label{D1}
 \left( \begin{array}{cc}
 \frak L_+ + \eta & 0 \\
 0 &  \frak  L_-+ \eta
 \end{array}
 \right)
 \left(
 \begin{array}{c}
 \G^{(0)} \\ \F^{(0)}  \end{array}
 \right) = 
 \lambda_1 J 
  \left(
 \begin{array}{c}
 \G^{(0)} \\ \F^{(0)}  \end{array}
 \right).
 \ee
Here the Schr\"odinger operator $\frak L_+$ is as in \eqref{Lplus} and $\frak L_-$  is given by
\[
\frak L_-= -d^2/dX^2+ 1 -2  \, \mathrm{sech}^2 X.
\]

The symplectic eigenvalue problem \eqref{D1} with $\eta >0$
was studied numerically and asymptotically  \cite{ABSK}.
Smaller positive values of $\eta$ are characterised by a pair of opposite real eigenvalues and a pair of pure imaginary eigenvalues 
$\lambda_1= \pm i \omega_{\mathrm{odd}}$. As $\eta$ is increased, the  real eigenvalues grow in absolute value, but then decrease and, as $\eta$ goes
through $\eta_c=3$,  collide  and move onto the imaginary axis. The emerging imaginary pair $\lambda_1 = \pm i \omega_{\mathrm{even}}$ diverges to infinity --- along with the pair $\lambda_1= \pm i \omega_{\mathrm{odd}}$.
Asymptotically, as $\eta \to \infty$,  we have \cite{ABSK}
\be
\omega_{\mathrm{even}}= \eta + \frac12 (\sqrt{17}-7) + O(\eta^{-1}), \quad
\omega_{\mathrm{odd}}= \eta+\frac12 (3 \sqrt{17}-11)+O(\eta^{-1}).
\label{AD1}
\ee 
(Note that  the expansions \eqref{AD1} with
$\eta=\sigma/h^2$ and $\omega= \lambda /h^2$ reproduce our earlier result \eqref{finlam}.)

The eigenvalue problem \eqref{D1} with $\eta<0$ was originally encountered outside the context of \PT-symmetric waveguides.
(Specifically, it
 was derived in the stability analysis of one-dimensional solitons of the ``hyperbolic" nonlinear Schr\"odinger equation on the plane;
see \cite{Dima_Deconinck1} and references therein.)
Various regimes of eigenvalue trajectories have been analysed, numerically and theoretically, in \cite{Dima_Deconinck1,Jianke_book,ABSK,Dima_Deconinck2}.

When $\eta \to -\infty$,   there are two complex quadruplets of eigenvalues.
The imaginary parts are given \cite{ABSK}
 by the same equations \eqref{AD1} 
whereas the real parts undergo an exponential decay
(at  equal rates)  \cite{Dima_Deconinck2}:  
\be
  \mathrm{Re} \,  \lambda_1^{\mathrm{even}}
=  \mathrm{Re} \,  \lambda_1^{\mathrm{odd}}
 = 
 \mathrm{const\/} \times 
  \frac{      (|\eta|-1)^{\sqrt{17}/2}    }  
  {     \exp \left\{ \sqrt{2 \pi (|\eta|-1)       } \right\}     }.
\label{TMF}
\ee

The upshot of this analysis is that the symplectic problem \eqref{B8} with $\sigma<0$, 
where $h$ and $|\sigma|$ are  both asymptotically  small,
has two quadruplets of complex eigenvalues. 
As $h^2 / \sigma \to 0$,  the real parts of these eigenvalues
 tend to zero faster than any power of $h^2/\sigma$.

\section{Symplectic eigenvalues in the anti-continuum limit}
\label{App5}

In this Appendix we determine the asymptotic behaviour of a
symplectic eigenvalue  with an even eigenvector, as
 $h \to \infty$.
 The parameter $|\sigma|$ is considered to be finite in \eqref{B8},
 with $\sigma$ taking either sign.

 In the anticontinuum limit, equations \eqref{iA} and \eqref{iB} have the advantage over the symplectic eigenvalue problem in its original form 
 \eqref{B8}. Specifically, equations with $n \geq 1$ are diagonal to the leading orders in $h^2$:
 \begin{align*}
 -\Delta \iA_n + (h^2+ \sigma - i \lambda +O(h^{2-4n})) \iA_n +O(h^{2-4n}) \iB_n=0,       \\
  -\Delta \iB_n + (h^2+ \sigma +  i \lambda +O(h^{2-4n})) \iB_n +O(h^{2-4n}) \iA_n=0.
   \end{align*}
 Solutions satisfying $|\iA_n|, |\iB_n| \to 0$  as $n \to \infty$, are given by 
 \be
 \iA_n= a^n \iA_0, 
 \quad
 \iB_n= b^n \iB_0  \quad
( |a|, |b| <1),    
\label{ab}
 \ee
 where the multipliers $a$ and $b$ satisfy
 \begin{align}
 a+a^{-1}= h^2+\sigma +2 - i \lambda + O(h^{-2}),      \label{J3}\\
b+b^{-1}=  h^2+ \sigma+ 2 + i \lambda  +O(h^{-2}).   \label{J4} 
 \end{align}

The equations with $n=0$ give 
\be
\mathcal M \left( \begin{array}{c} \iA_0 \\ \iB_0 \end{array} \right) =0,  \label{MAB}
\ee
where 
\[
\mathcal M= \left(
\begin{array}{cc}
h^2-\sigma+2+ i \lambda + 2 a+O(h^{-2}) ) & h^2+2+O(h^{-2})  \\   
h^2+2+O(h^{-2})  &   h^2-\sigma+2- i \lambda + 2 b+O(h^{-2}) 
\end{array} \right).    
\]
In \eqref{MAB},  we have used  \eqref{ab} with $n=1$.
  Setting the determinant of $\mathcal M$ to zero, we observe that 
  $\lambda$ cannot grow faster than $h$ as $h \to \infty$. 
 Making use of this fact, equations \eqref{J3}-\eqref{J4} imply that 
  $a^{-1}$ and $b^{-1}$ are both $O(h^2)$. We expand
  \be
  \lambda=  i  \sum_{\alpha=0}^\infty  \nu_\alpha h^{1-2\alpha}
  \label{J9}
  \ee
  and
  \be
  a^{-1}= \sum_{\alpha=0} a_\alpha h^{2-\alpha}, 
  \quad
  b^{-1}= \sum_{\alpha=0} b_\alpha h^{2-\alpha}.
  \label{J11}
  \ee
  
  Substituting these expansions in $\det \mathcal M=0$ and equating coefficients of like powers of $h$, we obtain
 \be
 \nu_0= (-2 \sigma)^{1/2}, 
 \quad
 \nu_1=   \left(1- \frac{\sigma}{4} \right) \nu_0.   \label{J10}
 \ee
 The first few coefficients in \eqref{J11} are then straightforward from  \eqref{J3}-\eqref{J4}:
 \[
 a_0=b_0= 1, 
 \quad
 a_1=-b_1=\nu_0,
 \quad
 a_2=  b_2= \sigma+2.
 \]

The outcome of our analysis is that the anti-continuum limit  is
characterised by a pair of opposite eigenvalues, 
with $|\lambda|$ growing in proportion to $h$ as $h \to \infty$.
The eigenvalues are real if $\sigma>0$ and pure imaginary if $\sigma<0$.


\section*{References}

\begin{thebibliography}{99}

\bibitem{DD}     
I V Barashenkov and E V Zemlyanaya, Physica D {\bf 132}  (1999) 363;
I.V. Barashenkov, N.V. Alexeeva,  and E. V. Zemlyanaya, Phys. Rev. Lett.  {\bf 89} 104101  (2002);
I V Barashenkov and E V Zemlyanaya, SIAM J. Appl. Math. {\bf 64}  800 (2004);
I V Barashenkov, E V Zemlyanaya, and T. C. van Heerden, Phys. Rev.  E {\bf 83} 056609 (2011)



\bibitem{CGL} The 
soliton parameter selection in the complex Ginsburg-Landau equation
was discussed in the early work of
 B A Malomed, Physica D {\bf 29} 155 (1987);
O. Thual and S. Fauve, 
J. Phys. (Paris)
{\bf 49}, 1829-1833 (1988); 
S. Fauve and O. Thual, Phys. Rev. Lett. {\bf 64} 282 (1990);
B A Malomed, A A Nepolmnyashchy. Phys Rev A {\bf 42} 6009 (1990);  
V. V. Afanasjev, N. Akhmediev and J.M. Soto-Crespo,
 Phys. Rev. E {\bf 53}, 1931-1939 (1996). For comprehensive reviews, see 
 N.  Akhmediev, General Theory of Solitons. In: A. D. Boardman and A. P. Sukhorukov, eds. Soliton-driven Photonics.
 NATO science series, series II: Mathematics, Physics and Chemistry --- vol. 31. page 371 (Kluwer Academic Publishers, Dordrecht 2001);
N. Akhmediev and A. Ankiewicz,
 Three Sources and Three Component Parts
of the Concept of Dissipative Solitons. In: 
 N. Akhmediev and A. Ankiewicz (eds). Dissipative Solitons: From Optics to Biology and Medicine.
Lecture Notes in Physics 751. (Springer,  Berlin Heidelberg 2008)


\bibitem{Bender}
 C.M. Bender and S. Boettcher, Phys. Rev. Lett. {\bf 80}   5243 (1998);
 C M Bender, S Boettcher, and P N Meisinger,
  Journ Math Phys {\bf 40} 2201 (1999); 
  Bender C M, 
Contemp. Phys. {\bf 46}  277 (2005);
  C M Bender, Rep Prog Phys {\bf 70} 947 (2007)





\bibitem{Musslimani}
K.G. Makris,  R. El-Ganainy, D.N. Christodoulides, Z.H. Musslimani, 
Phys. Rev. Lett. {\bf 100} 103904 (2008)

\bibitem{Zheng} M.C. Zheng,
 D.N. Christodoulides, R. Fleischmann, T. Kottos, 
Phys. Rev. A {\bf 82} 010103 (2010)


\bibitem{Regensburger} A. Regensburger, C. Bersch,  M.-A. Miri, G. Onishchukov,
D. N. Christodoulides, and U. Peschel,
Nature (London) {\bf 488} 167 (2012).


\bibitem{Berry_Longhi} 
M.V. Berry, J. Phys. A {\bf 41} 244007 (2008); S. Longhi, Phys. Rev. A {\bf 81} 022102 (2010)


\bibitem{Nonreciprocal_propagation}
O. Bendix, R. Fleischmann, T. Kottos, and B. Shapiro,
Phys. Rev. Lett. {\bf 103}  030402 (2009);
H. Ramezani, T. Kottos, R. El-Ganainy, and
D. Christodoulides, Phys. Rev. A {\bf 82}  043803 (2010)

\bibitem{WGM}
Peng B, \"Ozdemir \c{S}K, Lei F, Monifi F, Gianfreda M, Long G, Fan S, Nori F, Bender CM, Yang L,  Nat. Phys. {\bf 10}, 394 (2014)

\bibitem{Shramkova} 
O.V. Shramkova and  G.P. Tsironis,  Scientific Reports {\bf 7}  42919 (2017)

\bibitem{Longhi} S. Longhi, Phys. Rev. Lett. {\bf 103}  123601 (2009);
M. Wimmer, M. A. Miri, D. N. Christodoulides, and U.
Peschel, Sci. Rep. {\bf 5} 17760 (2015)



\bibitem{Guo} 
A. Guo, G. J. Salamo, D. Duchesne, R.Morandotti, M. Volatier-Ravat, V. Aimez, G. A. Siviloglou, and D. N. Christodoulides,
Phys. Rev. Lett. {\bf 103} 093902 (2009)



\bibitem{single-mode_laser}
L. Feng, Z. J.Wong, R.-M.Ma, Y.Wang, and X. Zhang,  Science 346,
972 (2014);
H. Hodaei, M.-A. Miri, M. Heinrich, D. N. Christodoulides,
and M. Khajavikhan, Science {\bf 346}  975 (2014).

\bibitem{Chong}  Y. D. Chong, L. Ge, A. D. Stone, Phys. Rev. Lett. {\bf 106} 093902
(2011)
\bibitem{loss_lase} B. Peng,      \c{S}. K.    \"Ozdemir,
S. Rotter,  H. Yilmaz,  M. Liertzer,  F. Monifi,
C. M. Bender,  F. Nori,  L. Yang, 
Science {\bf 346} 328 (2014) 


\bibitem{Ramezani} H. Ramezani,
 T. Kottos, V. Kovanis, D. N. Christodoulides, 
Phys. Rev. A {\bf 85} 013818 (2012)

\bibitem{RKEC} H. Ramezani,
 T. Kottos,  R. El-Ganainy, D.N. Christodoulides, 
Phys. Rev. A {\bf 82} 043803 (2010)




\bibitem{SXK} A.A. Sukhorukov,
 Z.Y. Xu, Yu.S. Kivshar, 
Phys. Rev. A {\bf 82} 043818 (2010)




\bibitem{Miri2} M.-A. Miri,  P.  LiKamWa, and D. N. Christodoulides,
Optics Lett {\bf 37} 764 (2012)

\bibitem{standimer}
P G Kevrekidis, D E Pelinovsky and D Y Tyugin, J Phys A: Math Theor {\bf 46} 365201 (2013);
I V Barashenkov, G S Jackson and S Flach, Phys Rev A {\bf 88} 053817 (2013);
J Pickton and H Susanto, Phys Rev A {\bf 88} 063840 (2013)


\bibitem{Lin} 
Z. Lin, H. Ramezani, T. Eichelkraut, T. Kottos, H. Cao, and D. Christodoulides,
Phys. Rev. Lett. {\bf 106}  213901 (2011); M. Kulishov,
J. M. Laniel, N. BŽlanger, J. Aza–a, and D. V. Plant,
Opt. Express {\bf 13}  3068 (2005).

\bibitem{EPswitch} J  Doppler, 
 A  A  Mailybaev, 
J  B\"ohm, U Kuhl,  A  Girschik,
 F  Libisch, 
T  J  Milburn,
P Rabl,  N  Moiseyev and  S Rotter,
Nature {\bf 537} 76 (2016)



\bibitem{Feng} 
L Feng, Y-L Xu,  W. S. Fegadolli, M-H Lu, J E. B. Oliveira, V R. Almeida, Y-F Chen, and  A Scherer.
 Nature Mater. {\bf 12} 108 (2012)

\bibitem{Sanchez} L. L. S\'anchez-Soto and J. J. Monzon, Symmetry  {\bf 6} 396
(2014)

\bibitem{plasmonics} H. Benisty,
A. Degiron, A. Lupu, A. De Lustrac,
 S. ChŽnais,  S. Forget, M.  Besbes, G. Barbillon, A. Bruyant,
S. Blaize, and G. LŽrondel,
Opt. Express {\bf 19}, 18004 (2011);
M. Mattheakis, Th. Oikonomou, M. I. Molina and G. P. Tsironis,  IEEE J. Sel. Top. in Quantum Electronics {\bf 22} 5000206 (2015)

\bibitem{Lambda} C. Hang,
  G. Huang, and V. V. Konotop,  Phys Rev Lett {\bf 110} 083604 (2013)

\bibitem{OM}
H. Jing,     \c{S}. K.  \"Ozdemir, Z. Geng,  J. Zhang,  X.-Y. L\"u,      B. Peng,
L. Yang, and F. Nori,  Sci. Rep. {\bf 5}  9663
(2015).




\bibitem{Kepesides}
K V Kepesidis,  T J Milburn, J Huber,  K G Makris, S Rotter, and
P Rabl,  New J. Phys. {\bf 18}  095003 (2016) 



\bibitem{Lazarides}  N. Lazarides and G. P. Tsironis,
Phys. Rev. Lett. {\bf 110}   053901 (2013).


\bibitem{Kottos} T. Kottos, Nat. Phys. {\bf 6} 192 (2010)





\bibitem{Rueter}
C.E. R\"uter, 
 K.G. Makris, R. El-Ganainy, D.N. Christodoulides, M. Segev, D. Kip, 
Nat. Phys. {\bf 6} 192 (2010)

\bibitem{Regensburger2} 
A. Regensburger, M.-A. Miri, C. Bersch, J. N\"ager, G. Onishchukov, D.N. Christodoulides, and U. Peschel, Phys. Rev. Lett. {\bf 110} 223902 (2013) 

\bibitem{Miri1} 
M. Wimmer, A. Regensburger, M.-A. Miri, C. Bersch, D. N. Christodoulides and  U. Peschel,
Nature Comm {\bf 6} 7782 (2015) 


\bibitem{Christo_2016}
Zhang Z Y, Zhang Y Q,  Sheng J T, Yang  L, Miri M A,  Christodoulides D N,  He B,  Zhang Y P,  Xiao M,
Phys. Rev. Lett. {\bf 117} 123601 (2016) 





\bibitem{oligomer}
K. Li and P. G. Kevrekidis,  Phys. Rev. E {\bf 83} 066608 (2011);
D. A. Zezyulin and V.V. Konotop, Phys. Rev. Lett.  {\bf 108} 213906 (2012);
K Li, P G Kevrekidis, B A Malomed, and U G\"unther, Journ. Phys. A: Math. Theor. {\bf 45} (2012)  444021;
J D'Ambroise, P G Kevrekidis, and S Lepri, 
J. Phys. A: Math. Theor.  {\bf 45} (2012) 444012


\bibitem{Liam}
I.V. Barashenkov, L. Baker, and N.V. Alexeeva, Phys Rev A {\bf 87} 033819 (2013)

\bibitem{Izrailev}
O. V\'azquez-Candanedo,  J. C. Hern\'andez-Herrej\'on, F. M. Izrailev,  and D. N. Christodoulides,
Phys. Rev. A {\bf 89} 013832 (2014)

\bibitem{Turitsyn} 
A J Mart\'inez, M I Molina, S K Turitsyn, and Y S Kivshar, Phys Rev A {\bf 91} 023822 (2015)

\bibitem{binary} S. V. Dmitriev, A. A. Sukhorukov, and Yu. S. Kivshar, Opt. Lett. {\bf 35}  (2010) 2976 



 \bibitem{SMDK} S V Suchkov, B A Malomed, S V Dmitriev, Y S Kivshar,  Phys Rev E {\bf 84}, 046609 (2011)

\bibitem{KPZ}
V. V. Konotop, D. E. Pelinovsky,  and D. A. Zezyulin, EPL {\bf 100} 56006  (2012);
D E Pelinovsky, D A Zezyulin, V V Konotop, Journ Phys A: Math Theor {\bf 47} 085204 (2014)

\bibitem{staggered}
J D'Ambroise,  P  G  Kevrekidis, and B  A  Malomed, Phys Rev E {\bf 91}  033207 (2015)


\bibitem{Chernyavsky}
A Chernyavsky and D E Pelinovsky, J. Phys. A: Math. Theor. {\bf 49}      475201      (2016);
Symmetry  {\bf 8}  59  (2016)


\bibitem{Miri3}
M.-A. Miri and A. Al\`u,  New J. Phys. {\bf 18} 065001 (2016)


\bibitem{Driben1} 
R. Driben and B.A. Malomed, Opt. Lett. {\bf 36} 4323 (2011)

\bibitem{Driben2}
R. Driben and B.A. Malomed, EPL {\bf 96} (2011) 51001


\bibitem{ABSK} N V Alexeeva, I V Barashenkov, 
A A Sukhorukov, and Y S Kivshar,  Phys. Rev. A {\bf 85} 063837 (2012)




\bibitem{BSSDK} 
I.V. Barashenkov, 
S. V. Suchkov,  A. A. Sukhorukov, S. V. Dmitriev, and Yu. S. Kivshar,
Phys. Rev. A {\bf 86} 053809 (2012)


\bibitem{Driben3}
R. Driben and B.A. Malomed, EPL {\bf 99} (2012) 54001




\bibitem{reviews} 
V V Konotop, J Yang, and D A Zezyulin, Rev. Mod. Phys. {\bf 88} 035002 (2016);
S V Suchkov, A A Sukhorukov, J H Huang, S V Dmitriev, C Lee, and Y S Kivshar,
Laser and Photonics Reviews {\bf 10} 177 (2016) 


\bibitem{Susanto}
O B  Kirikchi,    A A Bachtiar, and H Susanto, Adv. in Math. Phys.
 9514230 (2016) 



\bibitem{Dima_Deconinck2}
D. E. Pelinovsky,    E. A. Rouvinskaya, O. E. Kurkina, and B. Deconinck,
Theor. Math. Phys. {\bf 179} 452 (2014)


\bibitem{HT} D Hennig and G Tsironis, Phys Reports {\bf 307} (1999) 333



\bibitem{KC} Yu S Kivshar and D K Campbell, Phys Rev E {\bf 48} 3077 (1993)



\bibitem{LST} E W Laedke, K H Spatschek,  S K Turitsyn.  Phys Rev Lett  {\bf  73}  1055 (1994);
 E W Laedke,  O Kluth, K H Spatschek.  Phys Rev E  {\bf  54}  4299 (1996)


\bibitem{KK} T Kapitula and P Kevrekidis, Nonlinearity {\bf 14} 533 (2001)


\bibitem{PKF} D E Pelinovsky, P G Kevrekidis, D J Frantzeskakis. Physica D {\bf 212} 1 (2005)



\bibitem{Herrmann} M Herrmann. Discrete and Cont Dyn Systems A {\bf 31} 737 (2011)

\bibitem{ABK} G L Alfimov, V A Brazhnyi, and V V Konotop. Physica D {\bf 194} 127 (2004)





\bibitem{Arnold}
V. I. Arnold, Mathematical Methods of Classical Mechanics
(Graduate Texts in Mathematics). Springer Verlag, New York (2010)



\bibitem{JW} M Jenkinson and M I Weinstein. arXiv: 1601.04598 [nlin.PS]






\bibitem{Dima_Deconinck1}
B. Deconinck, D. Pelinovsky, and J. D. Carter, Proc. Roy. Soc. London Ser. A {\bf 462} 2039 (2006)


\bibitem{Jianke_book}
J. Yang, Nonlinear Waves in Integrable and Nonintegrable Systems (Math. Modeling Computation, Vol. 16),
SIAM, Philadelphia, Penn. (2010)


\bibitem{PY}
D E Pelinovsky and J Yang, Physica D {\bf 255} 1 (2013) 


\end{thebibliography}
\end{document}